\documentclass[fleqn,usenatbib]{mnras}

\usepackage{newtxtext,newtxmath}

\usepackage[T1]{fontenc}
\usepackage{ae,aecompl}

\usepackage{graphicx}	
\usepackage{amsmath}	
\usepackage{amssymb}	
\usepackage{xspace}		
\usepackage[usenames,dvipsnames]{color}
\usepackage{siunitx}
\usepackage{threeparttable}
\usepackage{longtable}
\usepackage{booktabs}
\usepackage{arydshln}
\usepackage{multirow}

\usepackage{rotating}

\usepackage{subcaption}
\newcommand{\gaia}{{\it Gaia}\xspace}
\newcommand{\jktebop}{{\tt JKTEBOP}\xspace}
\newcommand{\sbop}{{\tt SBOP}\xspace}
\newcommand{\python}{{\tt Python}\xspace}
\newcommand{\lk}{{\tt Lightkurve}\xspace}

\title[The age and metallicity of NGC~2506]{Extremely precise age and metallicity of the open cluster NGC~2506 using detached eclipsing binaries}

\author[Knudstrup et al.]{
E.~Knudstrup\hyperlink{https://orcid.org/0000-0001-7880-594X}{\includegraphics[scale=0.08]{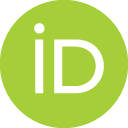}},$^1$\thanks{E-mail: emil@phys.au.dk (EK)} 
F.~Grundahl\hyperlink{https://orcid.org/0000-0002-8736-1639}{\includegraphics[scale=0.08]{figures/ORCIDiD_icon128x128.png}},$^1$
K.~Brogaard\hyperlink{https://orcid.org/0000-0003-2001-0276}{\includegraphics[scale=0.08]{figures/ORCIDiD_icon128x128.png}},$^{1,2}$ 
D.~Slumstrup\hyperlink{https://orcid.org/0000-0003-4538-9518}{\includegraphics[scale=0.08]{figures/ORCIDiD_icon128x128.png}},$^{3,1}$
\newauthor
J.~A.~Orosz\hyperlink{https://orcid.org/0000-0001-9647-2886}{\includegraphics[scale=0.08]{figures/ORCIDiD_icon128x128.png}},$^4$ 
E.~L.~Sandquist\hyperlink{https://orcid.org/0000-0003-4070-4881}{\includegraphics[scale=0.08]{figures/ORCIDiD_icon128x128.png}},$^4$
J.~Jessen-Hansen\hyperlink{https://orcid.org/0000-0003-4985-7606}{\includegraphics[scale=0.08]{figures/ORCIDiD_icon128x128.png}},$^1$ 
M.~N.~Lund\hyperlink{https://orcid.org/0000-0001-9214-5642}{\includegraphics[scale=0.08]{figures/ORCIDiD_icon128x128.png}},$^1$
\newauthor
T.~Arentoft\hyperlink{https://orcid.org/0000-0002-4696-6041}{\includegraphics[scale=0.08]{figures/ORCIDiD_icon128x128.png}},$^1$
R.~Tronsgaard\hyperlink{http://orcid.org/0000-0003-1001-0707}{\includegraphics[scale=0.08]{figures/ORCIDiD_icon128x128.png}},$^5$
D.~Yong\hyperlink{https://orcid.org/0000-0002-6502-1406}{\includegraphics[scale=0.08]{figures/ORCIDiD_icon128x128.png}},$^6$
S.~Frandsen,$^1$
\& H.~Bruntt$^7$
\\
$^1$Stellar Astrophysics Centre, Department of Physics and Astronomy, Aarhus University, Ny Munkegade 120,\\ DK-8000 Aarhus C, Denmark\\
$^2$Astronomical Observatory, Institute of Theoretical Physics and Astronomy, Vilnius University, Sauletekio av. 3, 10257 Vilnius, Lithuania\\
$^3$European Southern Observatory, Alonso de Cordova 3107, Vitacura, Santiago de Chile, Chile\\
$^4$Astronomy Department, San Diego State University, 5500 Campanile Drive, San Diego, CA 92182-1221, USA\\
$^5$DTU Space, National Space Institute, Technical University of Denmark, Elektrovej 328, 2800 Kgs. Lyngby, Denmark\\
$^6$Research School of Astronomy and Astrophysics, The Australian National University, Canberra, ACT2611, Australia \\
$^7$Aarhus Katedralskole, Skolegyde 1, DK-8000 Aarhus C, Denmark
}

\date{Accepted XXX. Received YYY; in original form ZZZ}

\pubyear{2020}

\begin{document}
\label{firstpage}
\pagerange{\pageref{firstpage}--\pageref{lastpage}}
\maketitle

\begin{abstract}
Accurate stellar parameters of stars in open clusters can help constrain models of stellar structure and evolution. Here we wish to determine the age and metallicity content of the open cluster NGC~2506. To this end we investigated three detached eclipsing binaries (DEBs; V2032, V4, and V5) for which we determined their masses and radii, as well as four red giant branch stars for which we determined their effective temperatures, surface gravities, and metallicities. 
Three of the stars in the DEBs have masses close to the cluster turn-off mass, allowing for extremely precise age determination. Comparing the values for the masses and radii of the binaries to BaSTI isochrones we estimated a cluster age of $2.01 \pm 0.10$~Gyr. This does depend on the models used in the comparison, where we have found that the inclusion of convective core-overshooting is necessary to properly model the cluster. From red giant branch stars we determined values for the effective temperatures, the surface gravities, and the metallicities. From these we find a cluster metallicity of $-0.36 \pm 0.10$~dex. Using this value and the values for the effective temperatures we determine the reddening to be E$(b-y) = 0.057 \pm 0.004$~mag. Furthermore, we derived the distance to the cluster from {\it Gaia} parallaxes and found $3.101 \pm 0.017$~kpc, and we have performed a radial velocity membership determination for stars in the field of the cluster. Finally, we report on the detection of oscillation signals in $\gamma$~Dor and $\delta$~Scuti members in data from the TESS mission, including the possible detection of solar-like oscillations in two of the red giants.

\end{abstract}



\section{Introduction}
\label{sec:int}
Age and metallicity determination of open clusters is of great interest since; i) it allows us to test stellar evolution theory by comparing the observed cluster sequence in a colour-magnitude diagram (CMD) to theoretically calculated {\it isochrones}, ii) by combining the ages and chemical compositions with the kinematical properties of the clusters, they can be used in a much grander scheme to decipher the formation and evolution of the Galaxy in the field of Galactic Archaeology. In the latter context NGC~2506 is particularly interesting as it belongs to a group of metal-deficient clusters located just beyond the solar circle in the galactic anticenter \citep{art:twarog2016}. 

\begin{figure}
	\centering
    \includegraphics[width=\columnwidth]{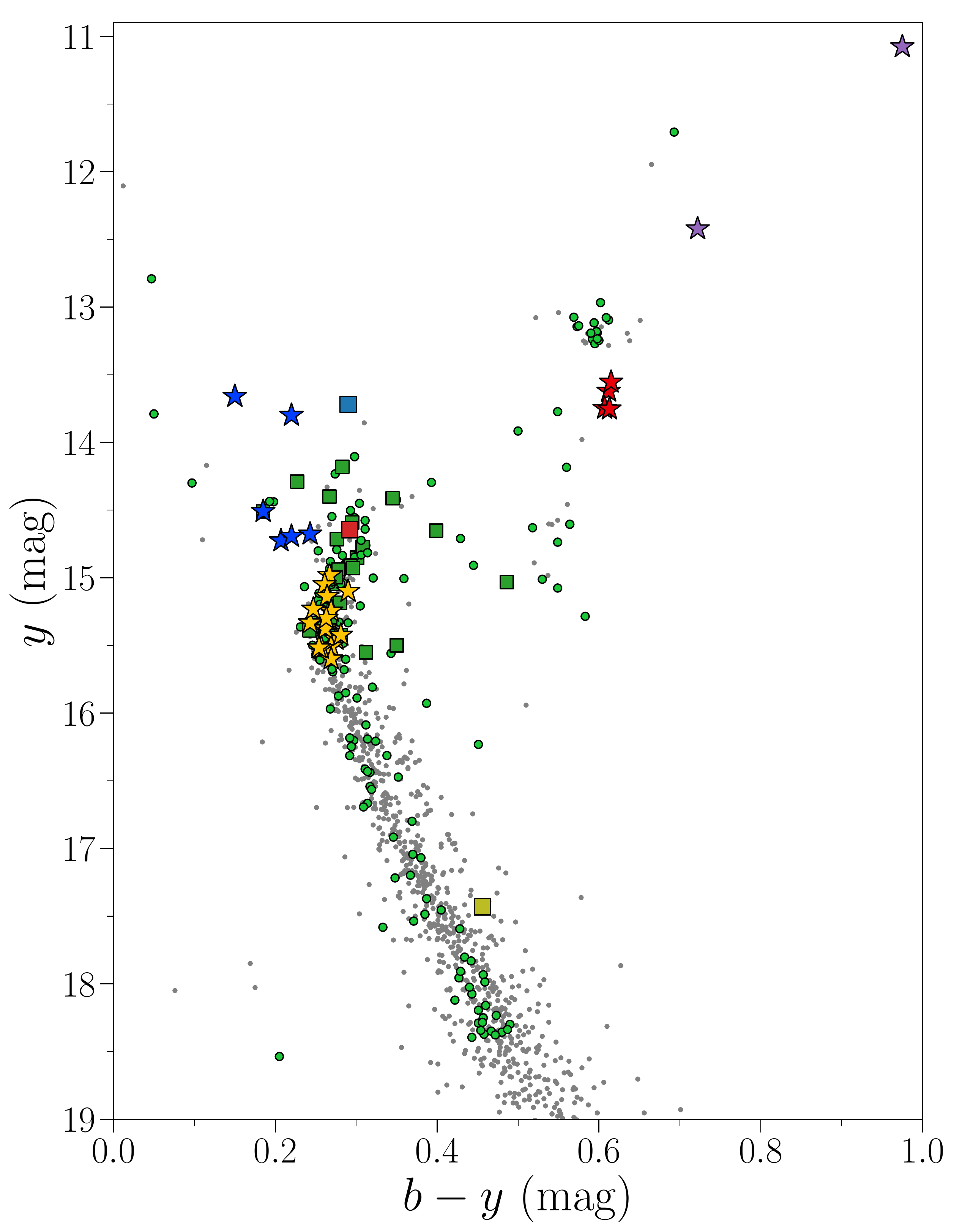}
    \caption{Cleansed colour-magnitude diagram of NGC 2506. Grey dots are {\it Gaia} proper motion members of the cluster (see Section~\ref{sec:gaia}) and green dots and squares mark the confirmed radial velocity members from spectroscopy of single and multiple systems, respectively; we have thus removed all stars we deemed non-members (see Section~\ref{sec:rv_members}). Yellow and blue stars denote respectively the $\gamma$~Dor stars and BSs reported in \citet{art:arentoft2007}. The blue, red, and yellow squares denote V2032, V4, and V5 listed in Table~\ref{tab:targets} alongside the RGB stars marked with red and purple stars in this figure. We performed a spectroscopic analysis of the RGB stars marked with red stars and we report on the possible detection of solar-like oscillations for the stars marked with purple.}
    \label{fig:cmd}
\end{figure}

In the context of stellar evolution and probing the interior of stars, NGC~2506 is an extremely promising cluster as it harbors a multitude of stellar oddballs. \citet{art:arentoft2007} reported on the discovery of three oscillating blue stragglers (BSs) bringing the total in the cluster up to six, as well as the discovery of no less than 15 $\gamma$~Doradus ($\gamma$~Dor) stars. BSs are stars residing in a brighter and bluer region of the main sequence turn-off in a cluster (see Figure~\ref{fig:cmd}). The origin of BSs is still debated, but viable formation scenarios involve binary mass transfer and/or the merging of two stars, either by a direct collision or the merging of the components in a binary \citep[e.g.,][]{art:chatterjee2013,art:simunovic2014,art:brogaard2018}. The blue stragglers are situated in the instability strip and we detect $\delta$~Scuti-like oscillations (see Section~\ref{sec:astero}) in all of the blue stragglers. We will therefore use the terms blue stragglers and $\delta$~Scuti stars interchangeably. $\gamma$~Dor stars are a type of variable stars, which as seen in Figure~\ref{fig:cmd} can be found at or just above the main sequence turn-off, depending on the cluster. $\gamma$~Dor stars show photometric variations of up to $0.1$~mag, which are caused by non-radial g-mode pulsations that allow for probing of the stellar interior. $\gamma$~Dor stars can therefore be used to constrain convective core-overshooting and rotation in stellar models \citep{art:lovekin2017}. Precise age and metallicity determination of NGC~2506 is therefore valuable as it means constraining the parameters for these stars.

The proposed ages of NGC~2506 ranges from more than $3$~Gyr in one of the earliest studies \citep{art:mcclure1981} to just below $2$~Gyr in the more recent ones \citep{art:netopil2016,art:twarog2016}. The literature seems to agree that NGC~2506 is a metal-deficient cluster with a reported upper limit of around $-0.2$~dex \citep{art:netopil2016}, but there is no clear consensus on the metallicity. 

It is possible to determine the masses and radii of the components in detached eclipsing binaries (DEBs) with great precision. Should one or both of the components turn out to have a mass close to the cluster turn-off mass, it is possible to place a tight constraint on the age of the binary system and therefore the cluster \citep[e.g., as for NGC~6791 in][]{art:grundahl2008,art:brogaard2011,art:brogaard2012}.

We aim to constrain the age and metallicity of NGC~2506 by analyzing three DEBs, meaning that we will measure the masses and radii of six stars in the cluster. To supplement our age and metallicity estimates, we will perform a spectroscopic analysis of four red-giant branch (RGB) stars. These will allow us to constrain the metallicity of the cluster and will allow us to firstly check if the metallicity is consistent with what is suggested by the DEBs, and secondly we might then choose models within a small range of this metallicity to further constrain the age.

The paper is structured as follows. In Section~\ref{sec:targets} we briefly introduce our target stars. In Sections~\ref{sec:spec} and \ref{sec:phot} we respectively present our spectroscopic and photometric data. Section~\ref{sec:orb.anal} contains our orbital analysis of the DEBs and the stellar parameters deduced therefrom. In Section~\ref{sec:clpar} we report on the derived cluster parameters. In Section~\ref{sec:gaia} we present our derived distance to the cluster and membership determination using data from the \gaia mission \citep{art:gaia2016}. The discussion is given in Section~\ref{sec:disc} and finally we draw our conclusions in Section~\ref{sec:conc}.



\begin{table}
	\centering
    \caption{Names, WEBDA ID, and coordinates of the target stars. The detached eclipsing binaries are above the solid line and the red-giant branch stars are below. The red-giant branch stars above the dashed lines are the ones for which we perform a spectroscopic analysis and the two listed below are the ones in which we possibly detect solar-like oscillations. The index for these (RGBXXX) refers to their index in our $uvby$ photometry (Table~\ref{tab:rv_members}).}
   \begin{threeparttable}
    \begin{tabular}{c c c c c c c}
    \toprule
    Name/WEBDA & $\alpha_{2000}$ & $\delta_{2000}$ & $y$ & $(b-y)$ \\
    \midrule
    V2032/4132 & 08 00 00.6 & -10 45 38 & 13.719 & 0.290 \\
    V4/1136 & 08 00 08.2 & -10 45 50 & 14.645 & 0.292 \\
    V5/1335 & 08 00 10.3 & -10 43 17 & 17.430 & 0.456 \\
    \hline
    RGB231/7108\tnote{\textdagger} & 08 00 23.3 & -10 48 48 & 13.622 & 0.612 \\
    RGB433/2375 & 08 00 11.5 & -10 50 19 & 13.555 & 0.615 \\
    RGB913/2255 & 08 00 09.4 & -10 48 13 & 13.748 & 0.607 \\
    RGB2358/4274 & 08 00 00.8 & -10 44 04 & 13.753 & 0.613 \\
   \hdashline
   RGB383/2402 & 08 00 20.1 & -10 49 59 & 12.422 & 0.722 \\
   RGB526/-\tnote{$\chi$} & 08 00 18.2 & -10 49 21 & 11.077 & 0.975 \\
    \bottomrule
    \end{tabular}
		\begin{tablenotes}
        	\item[\textdagger] Identifier from \citet{art:anthony-twarog2018}.
            \item[$\chi$] No identifier found.
		\end{tablenotes}
\end{threeparttable}
    \label{tab:targets}
\end{table}

\section{Targets}\label{sec:targets}
The names, WEBDA ID\footnote{\url{https://webda.physics.muni.cz/cgi-bin/ocl_page.cgi?cluster=ngc+2506}}, and coordinates of the targets are listed in Table~\ref{tab:targets}. Displayed in Figure~\ref{fig:cmd} is the $b-y,y$ \citep[data from][]{art:grundahl2000} CMD of NGC~2506 with the targets highlighted. Also shown in Figure~\ref{fig:cmd} is the position of the confirmed $\gamma$~Dor stars and BSs. V4 was discovered by \citet{art:kim2001} and V5 by \citet{art:arentoft2007}. It was only very recently we detected an eclipse in V2032 and as such nothing about the system has been published yet.

From our analysis we have found that the binary V4 has an outer companion on a much wider orbit. The most massive component in V4 is close to the turn-off mass of the cluster, which is around $1.5$~$\rm M_\odot$, making it one of the systems that allow for precise age determination. In this sense V2032 is an even more auspicious system as both components seem to be located on the subgiant branch -- an evolutionary phase of rapid expansion making the isochrones almost completely vertical in the mass-radius diagram (see, e.g., Figure~\ref{fig:MR}). Precise determination of the masses of these components will therefore completely lock the age of the cluster. The components of V5 are somewhat lower in terms of mass than the cluster turn-off mass with the lowest of the two having a mass of around $0.7$~$\rm M_\odot$. This means that the masses of all the components in the binaries span a range in mass that covers the transition between stars above $\sim 1.2$~$\rm M_\odot$ with a convective core and stars below with a radiative core, which will help anchor the isochrones.

In addition to the DEBs we have spectra of four RGB stars. These will provide us with a firm grip on the metallicity of the cluster. Furthermore, they will allow us to probe a more evolved stage of stellar evolution in a different parameter space, namely $\log g$ and $T_{\rm eff}$. Finally, the \gaia mission is providing precise parallaxes and proper motion for billions of stars, which is extremely useful in cluster studies as this allows for not only distance determination, but also membership determination.



\section{Spectroscopic observations}\label{sec:spec}

Here we present our spectroscopic observations of stars in the cluster, where we first discuss the membership based on radial velocities (RVs). We then present our measurements of the chemical composition of the cluster through an analysis of spectroscopic measurements of RGB stars, with a subsequent derivation of the colour excess of the cluster. In Section~\ref{sec:spec.rad} we describe how we obtained RVs for the DEBs. Finally, we present measurements for the luminosity ratios of V2032 and V4, both from the spectroscopic measurements, but also from measuring the spectral energy distribution (SED). 

\subsection{Radial velocity members from spectroscopy}\label{sec:rv_members}
We obtained 15 epochs of GIRAFFE spectroscopy (ESO programme 075.D-0206(B); this is the same programme as the data for the DEB V4, see Table~\ref{tab:spec.V4}, and the RGB stars in Section~\ref{sec:spec.rgb}) from ESOs Very Large Telescope (VLT) for NGC~2506 in order to define membership near the cluster turn-off region and RGB. The setting (HR14A) with a central wavelength near 6515~{\AA} and a resolution of 18000 (Medusa mode) was utilized. All spectra were recovered from the \url{http://giraffe-archive.obspm.fr} site which provided a re-reduction of the ESO GIRAFFE data. We note, however, that at the time of writing this webpage is no longer active. 

To derive the velocities we cross-correlated each obtained stellar spectrum with a solar template and calculated the average velocity, standard deviation of the individual velocities as well as the width of the fitted gaussian. This resulted in a histogram of velocities for 122 objects, with a clear peak in the distribution at $v_{\rm rad} = 83.8$~km/s with a Full Width Half Maximum (FWHM) of 4.7~km/s. We then assigned membership by requiring that an object has an average velocity within two FWHM of the cluster mean. Following this we inspected the 15 epochs of RVs for each target to make sure that binaries would be correctly assigned as members or non-members.   

In Table~\ref{tab:rv_members} the basic information for each target is provided; ID (from the $uvby$ photometry), $y$ and $b-y$ in the Str{\"o}mgren system, average velocity, standard deviation of the 15 RVs, and the Gaussian $\sigma$ from the fit to the cross-correlation function (CCF). The two second to last columns indicate whether a significant epoch-to-epoch variability was found (0 = RV constant, 1 = RV variable) and the membership status (1 = member, 0 = non-member) based on the RV. This forms the basis for the colour coding used in Figure~\ref{fig:cmd}. In the very last column we list both a cross-match with the catalog created by \citet{art:cantat2018} available in the VizieR Online Data Catalog \citep{art:cantat2019}, where they assessed cluster membership based on the \gaia proper motions and parallaxes, and the spectroscopic membership by \citet{art:anthony-twarog2018}, where we have adopted their membership classification. The values listed in Table~\ref{tab:rv_members} are the probabilities for membership they provide. As a sanity check we also did a cross-match between our target stars in Table~\ref{tab:targets} and \citet{art:cantat2019} -- all stars, with the exception of V5, were found to be members. This could be due to the faintness of the system as the RV curves in Figure~\ref{fig:lcs_rvs} clearly suggest that V5 is a member of the cluster. Likewise we cross-matched our targets with the catalog by \citet{art:anthony-twarog2018}, where again all targets were listed as members, with the exception of V5 and RGB525 for which we could not find a match. A version of Table~\ref{tab:rv_members} is available online containing magnitudes from all four Str\"omgren filters with associated uncertainties.

\subsection{Spectroscopic analysis of red-giant branch stars}\label{sec:spec.rgb}

\begin{table}
    \centering  
    \caption{Atmospheric parameters of the four RGB stars. The uncertainties are only internal.}
    \begin{tabular}{c c c c c c c c c c c c}
    \toprule
    & $T_\mathrm{eff}$ & $\log g$ & $v_{\rm mic}$ \\
    & (K) & (cgs;dex) & (km/s) \\
    \midrule
    RGB231 & 4870~$\pm$~30 & 2.65~$\pm$~0.03 & 1.10~$\pm$~0.04 \\
    RGB433 & 4840~$\pm$~30 & 2.60~$\pm$~0.05 & 1.15~$\pm$~0.03 \\
    RGB913 & 4920~$\pm$~30 & 2.70~$\pm$~0.05 & 1.10~$\pm$~0.05 \\ 
    RGB2358 & 4970~$\pm$~70 & 2.80~$\pm$~0.10 & 1.00~$\pm$~0.10 \\
    \midrule
    & $\rm [Fe/H]$ & $\rm [\alpha/Fe]$ & $\rm [Mg/Fe]$ \\
    & (dex) & (dex) & (dex) \\
    \midrule
    RGB231 & -0.36~$\pm$~0.01 & 0.10~$\pm$~0.02 & 0.12~$\pm$~0.02 \\
    RGB433 & -0.37~$\pm$~0.01 & 0.10~$\pm$~0.02 & 0.13~$\pm$~0.01 \\
    RGB913 & -0.36~$\pm$~0.01 & 0.09~$\pm$~0.02 & 0.08~$\pm$~0.02 \\
    RGB2358 & -0.34~$\pm$~0.03 & 0.06~$\pm$~0.05 & 0.10~$\pm$~0.06 \\
    \midrule
    & $\rm [Si/Fe]$ & $\rm [Ca/Fe]$ & $\rm [Ti/Fe]$ \\
    & (dex) & (dex) & (dex) \\
    \midrule
    RGB231 & 0.13~$\pm$~0.02 & 0.14~$\pm$~0.05 & -0.01~$\pm$~0.01 \\
    RGB433 & 0.13~$\pm$~0.05 & 0.13~$\pm$~0.01 & 0.00~$\pm$~0.02 \\
    RGB913 & 0.16~$\pm$~0.05 & 0.11~$\pm$~0.01 & 0.01~$\pm$~0.01 \\
    RGB2358 & 0.05~$\pm$~0.09 & 0.09~$\pm$~0.05 & 0.02~$\pm$~0.03 \\
   \midrule
   & SNR & SNR & \\
   & @\SI{5000}{\angstrom} & @\SI{6000}{\angstrom} & \\
   \midrule
    RGB231 & 105 & 230 & \\
    RGB433 & 110 & 230 & \\
    RGB913 & 100 & 220 & \\
    RGB2358 & 100 & 215 & \\
    \bottomrule
    \end{tabular}
    \label{tab:rgbs}
\end{table}

The spectra for the RGB stars were obtained using UVES under the programme with ID 075.D-0206(B). We used UVES/FLAMES in the 580~nm setting, resulting in a spectral resolution of 47,000. The atmospheric parameters of the four RGB stars presented in Table~\ref{tab:rgbs} were determined spectroscopically from an equivalent width analysis of Fe lines using DAOSPEC \citep{Stetson2008a} to measure line strengths. The line list is from \citet{art:slumstrup2019} and the methodology follows that of \citet{art:slumstrup2017,art:slumstrup2019}, who has derived the metallicities for giant stars in NGC~188, M67, NGC~6819, and NGC~6633 as well as in the Hyades \citep{art:arentoft2019} in a self-consistent way. Using this method \citet{art:slumstrup2017,art:arentoft2019} finds the ``canonical'' values for the metallicity of M67 and the Hyades. Compared to previous studies of NGC~2506 \citep[e.g.,][]{art:friel1993,art:carretta2004} the data presented here have significantly higher spectral resolution and spectral range as well as a higher signal-to-noise ratio (SNR), which is comparable to that of \citet{art:slumstrup2017,art:slumstrup2019}.

The atmospheric parameters were determined with the auxiliary program \texttt{Abundance with SPECTRUM} \citep{Gray1994} using ATLAS9 stellar atmosphere models \citep{Castelli2004} and solar abundances from \citet{Grevesse1998}. Non-LTE (local thermodynamic equilibrium) effects have been shown to be small for Fe in this parameter range \citep[of the order of \SI{0.1}{dex};][]{Asplund2005a,Mashonkina2011} and we therefore assume LTE. The effective temperatures were determined by requiring that the Fe abundance of each absorption line has no dependency on the excitation potential, i.e., excitation equilibrium. Likewise, the microturbulent velocity was determined by requiring that the Fe abundances show no trend with the reduced equivalent width of the lines ($\log \left( \frac{\rm EW}{\lambda} \right)$). The surface gravities were determined by invoking ionization equilibrium - requiring that the mean abundances of the two ionization stages FeI and FeII are in agreement, because FeII lines are much more sensitive to pressure changes than FeI lines in this parameter range. This is, however, also sensitive to the effective temperature and heavy element abundance and several iterations were realized to reach agreement on every parameter.

The metallicity of NGC~2506 has been determined several times in the literature and different values have been obtained. The higher determinations are from, e.g., \citet{Mikolaitis2011} and \citet{Reddy2012} with values of $\rm [Fe/H]= -0.24 \pm 0.05$~dex and $\rm [Fe/H] = -0.19 \pm 0.06$~dex, respectively. These are significantly higher than our mean cluster metallicity of $-0.36$~dex, which is in slightly better agreement with results on the lower end of determinations as, e.g., the study of many open clusters presented by \citet{Friel2002} that gives a mean cluster metallicity of $-0.44$~dex. The $\alpha$ abundances in Table~\ref{tab:rgbs} are calculated as $\rm [\alpha/Fe] = \frac{1}{4} \cdot ([Mg/Fe] + [Ca/Fe] + [Si/Fe] + [Ti/Fe])$. We also provide the individual elemental abundances because there are interesting systematic differences in the abundances of the standard $\alpha$ elements, with $\rm [Ti/Fe]$ showing no $\alpha$ enhancement, whereas the other three elements show slight $\alpha$ enhancement for all stars. The two studies by \citet{Mikolaitis2011} and \citet{Reddy2012} do not find this same significant difference between Titanium and the other three $\alpha$ elements used here.

\subsubsection{Reddening from RGBs}\label{sec:red}

The intrinsic spectroscopic parameters for the RGB stars in Table~\ref{tab:rgbs}, i.e., $T_{\rm eff}$, $\log g$, and $\rm [Fe/H]$, allow us to determine the reddening, E$(B-V)$, of the cluster. This was done by calculating the bolometric corrections for the \gaia filters, ${\rm BC}_{G_{\rm BP}}$ and ${\rm BC}_{G_{\rm RP}}$, using the spectroscopic parameters and compare these to the observed \gaia colour, since ${\rm BC}_{G_{\rm RP}} - {\rm BC}_{G_{\rm BP}}=G_{\rm BP}-G_{\rm RP}$. Any discrepancy between the two should be due to the reddening. We used the bolometric corrections from \citet{art:casagrande2018a,art:casagrande2018b} with $\rm [\alpha/Fe]=0.0$~dex.

To incorporate the uncertainties on the spectroscopic parameters, our approach was to do a Markov chain Monte Carlo (MCMC) analysis using the program \texttt{emcee} \citep{pack:foremanmackey2013}, where we drew from Gaussian distributions for the spectroscopic parameters (Table~\ref{tab:rgbs}) and a uniform distribution for the reddening, in the sense $\mathcal{N}(\mu,\sigma)$ ($\mu$ being the mean and $\sigma$ the uncertainty) and $\mathcal{U}(a,b)$ ($a=0.0$ and $b=0.4$), respectively. We then determined ${\rm BC}_{G_{\rm RP}}$ and ${\rm BC}_{G_{\rm BP}}$ for each draw and calculated the corresponding maximum likelihood, or rather the logarithm of the maximum likelihood: 
\begin{equation*}
    \log \mathcal{L} = -\frac{1}{2} \sum^{4}_{i=1} \log (2 \pi \sigma^2_i) + \frac{(y_{{\rm BC},i} - y_{{\rm Obs.},i})^2}{\sigma_{{\rm Obs.},i}^2} \, ,
    \label{eq:logl}
\end{equation*}
where $y_{\rm BC} = {\rm BC}_{G_{\rm RP}} - {\rm BC}_{G_{\rm BP}}$, $y_{\rm Obs.} = G_{\rm BP}-G_{\rm RP}$, and $\sigma_{\rm Obs.}$ is the uncertainty on the observed \gaia colour. This yielded a value of E$(B-V)=0.080^{+0.005}_{-0.006}$~mag, corresponding to E$(b-y)=0.057 \pm 0.004$~mag. 

This value is a bit higher than the values found in \citet{art:carretta2004} of E$(b-y) = 0.042 \pm 0.012$~mag (from E$(b-y)=0.72 \cdot $E$(B-V)$) and E$(b-y) = 0.042 \pm 0.001$~mag found in \citet{art:twarog2016}. The value we have found for the reddening can be used to calculate the effective temperatures for the stars in the binaries, which is discussed in Section~\ref{sec:sed}.

\subsection{Radial velocities for the detached eclipsing binaries}\label{sec:spec.rad}

VLT was also used to obtain all of the spectroscopic data of V4 and V5 as well as part of the spectroscopic data of V2032, where both the Ultraviolet and Visual Echelle Spectrograph \citep[UVES;][]{book:dekker2000} and the GIRAFFE spectrographs have been used for V4, but only GIRAFFE has been used for V5 and V2032. The data from UVES were acquired in 2005 by feeding the spectrograph by the Fibre Large Array Multi Element Spectrograph \citep[FLAMES;][]{art:pasquini2002} resulting in a medium resolution of $R=47,000$. When UVES is fed by FLAMES the spectrum is imaged onto two beams hitting two separate CCDs -- a lower CCD covering $4777-5750$~\r{A} and an upper one covering $5823-6819$~\r{A}. Likewise, the GIRAFFE spectrograph was also fed by FLAMES resulting in a resolving power of $R=33,700$. The GIRAFFE spectra were obtained in 2009 and 2010. All the spectroscopic data from VLT are summarised in Table \ref{tab:spec.V4}. The second batch of spectroscopic data for V2032 was acquired using the Nordic Optical Telescope (NOT) covering epochs from 2012 to 2015. The spectra were obtained at a resolution of $R=46,000$ using the FIbre-fed Echelle Spectrograph \citep[FIES;][]{art:telting2014}. This is summarised in Table~\ref{tab:spec.V2032}.

\begin{table*}
    \centering
    \footnotesize
    \caption{The spectroscopic data taken with ESOs VLT located on Cerro Paranal, Chile. Listed are the 31 spectra (two of which have been excluded) of V4 and the subset of 17 spectra of V2032 and V5. Shown are the dates, exposure times, barycentric velocity corrections (BVCs), and the radial velocities of the primary and secondary components. The spectra taken with UVES (above the dashed horizontal line) have the programme ID 075.D-0206(B), whereas the spectra taken with GIRAFFE have the programme ID 084.D-0154(A). 
    }
    \begin{threeparttable}
    \begin{tabular}{c c c c c c c c c c}
        \toprule
        & & & & \multicolumn{2}{c}{V4} & \multicolumn{2}{c}{V5} & \multicolumn{2}{c}{V2032} \\ \cmidrule(r){5-6} \cmidrule(r){7-8} \cmidrule(r){9-10}
        yyyy-mm-dd & BJD & Exp. & BVC & $v_{\rm rad}^{\rm p}$ & $v_{\rm rad}^{\rm s}$ & $v_{\rm rad}^{\rm p}$ & $v_{\rm rad}^{\rm s}$ & $v_{\rm rad}^{\rm p}$ & $v_{\rm rad}^{\rm s}$ \\        
         &  & (s) & (km/s) & (km/s) & (km/s) & (km/s) & (km/s) & (km/s) & (km/s) \\        
        \midrule
\multicolumn{10}{c}{UVES}  \\ 
\cmidrule(r){4-7} 
2005-03-28 & 2453458.5395 & 2450 & -22.704 & $180.2 \pm 1.7$ & $-29.8 \pm 0.3$ & - & - & - & - \\
2005-04-02 & 2453463.5118 & 2450 & -23.653 & $35.7 \pm 1.0$ & $144.0 \pm 0.8$ & - & - & - & - \\
2005-04-03 & 2453463.5412 & 2450 & -23.734 & $43.2 \pm 1.1$ & $133.8 \pm 0.4$ & - & - & - & - \\
2005-04-11 & 2453471.5461 & 2700 & -24.875 & $-2.0 \pm 1.2$ & $189.0 \pm 0.8$ & - & - & - & - \\
2005-04-11 & 2453471.5785 & 2700 & -24.955 & $-4.6 \pm 1.5$ & $191.7 \pm 0.8$ & - & - & - & - \\
2005-04-15 & 2453475.5097 & 748 & -25.189 & $175 \pm 6$ & $-19.6 \pm 0.3$ & - & - & - & - \\
2005-04-16 & 2453476.5384 & 2600 & -25.310 & $92.0 \pm 1.9$ & $78.2 \pm 0.5$ & - & - & - & - \\
2005-04-16 & 2453476.5696 & 2600 & -25.385 & $85 \pm 2$ & $82.9 \pm 0.7$ & - & - & - & - \\
2005-05-05 & 2453496.4841 & 1800 & -25.192 & $107.3 \pm 0.9$ & $56.7 \pm 0.7$ & - & - & - & - \\
2005-05-05 & 2453496.5061 & 1800 & -25.241 & $106.6 \pm 0.9$ & $60.2 \pm 0.9$ & - & - & - & - \\         
2005-05-11 & 2453502.4878 & 2600 & -24.617 & $70.6 \pm 1.8$ & $101.4 \pm 0.6$ & - & - & - & - \\
2005-05-12 & 2453502.5197 & 2600 & -24.679 & $67.4 \pm 1.1$ & $107.4 \pm 0.2$ & - & - & - & - \\
2005-05-13 & 2453504.4964\tnote{$\chi$} & 2450 & -24.387 & $177 \pm 2$ & $-25.6 \pm 0.4$ & - & - & - & - \\
2005-05-14 & 2453504.5259 & 2450 & -24.441 & $177 \pm 3$ & $-23.8 \pm 1.1$ & - & - & - & - \\ \hdashline
\multicolumn{10}{c}{GIRAFFE}  \\ 
\cmidrule(r){4-7} 
2009-12-14 & 2455179.7583\tnote{\textdagger} & 3600 & 17.614 & - & - & $18.1 \pm 0.9$ & $173 \pm 3$ & $55.4 \pm 0.6$ & $110.8 \pm 0.8$  \\ 
2009-12-18 & 2455183.7744 & 3600 & 16.151 & $-7.7 \pm 1.9$ & $199.0 \pm 0.5$ & $38 \pm 2$ & $145.1 \pm 1.4$ & $49.1 \pm 0.2$ & $117.8 \pm 0.3$  \\ 
2010-01-03 & 2455199.7369 & 3600 & 9.883 & $129.5 \pm 1.4$ & $35.6 \pm 0.4$ & $29.5 \pm 0.8$ & $158.2 \pm 1.4$ & $82.67 \pm 0.18$ & $82.82 \pm 0.18$  \\ 
2010-01-04 & 2455200.7519 & 3600 & 9.409 & $-2 \pm 5$ & $190.8 \pm 0.8$ & $58.4 \pm 0.5$ & $123.5 \pm 0.6$ & $78.4 \pm 0.8$ & $87.6 \pm 1.6$  \\ 
2010-01-05 & 2455201.8243\tnote{\textdagger} & 3600 & 8.770 & - & - & $155.0 \pm 0.6$ & $-17 \pm 5$ & $73.0 \pm 0.4$ & $92 \pm 2$  \\ 
2010-01-06 & 2455202.7917 & 3600 & 8.426 & $103 \pm 3$ & $66.2 \pm 5$ & $66.5 \pm 0.8$ & $117.0 \pm 1.0$ & $69.44 \pm 0.16$ & $98.2 \pm 0.7$  \\ 
2010-01-07 & 2455203.7635 & 3600 & 8.072 & $-14 \pm 14$ & $200 \pm 50$ & $24.0 \pm 0.8$ & $164 \pm 3$ & $65.8 \pm 0.4$ & $99.7 \pm 0.6$  \\ 
2010-01-10 & 2455206.6635 & 3600 & 7.043 & $-11.7 \pm 1.3$ & $199.0 \pm 0.5$ & $16.8 \pm 0.4$ & $176.7 \pm 1.3$ & $58.7 \pm 0.3$ & $108.6 \pm 0.3$  \\
2010-01-11 & 2455207.6647 & 1698 & 6.591 & $182.3 \pm 3$ & $-27.29 \pm 0.03$ & $83.8 \pm 0.6$ & $84.0 \pm 0.6$ & $55.80 \pm 0.19$ & $110.81 \pm 0.13$  \\ 
2010-01-14 & 2455210.6097 & 3600 & 5.386 & $183.9 \pm 3$ & $-31.6 \pm 1.0$ & $34.5 \pm 1.0$ & $156 \pm 2$ & $50.1 \pm 0.4$ & $116.2 \pm 0.2$  \\
2010-01-15 & 2455211.6858 & 3600 & 4.713 & $55 \pm 3$ & $119.4 \pm 0.7$ & $152.44 \pm 0.15$ & $-7.1 \pm 0.9$ & $49.4 \pm 0.3$ & $117.66 \pm 0.09$  \\ 
2010-01-28 & 2455224.7121 & 3600 & -1.357 & $169.8 \pm 1.6$ & $-7.6 \pm 0.8$ & $115.1 \pm 0.5$ & $41.4 \pm 1.4$ & $98.9 \pm 0.5$ & $66.4 \pm 0.4$ \\
2010-01-29 & 2455225.6235 & 3600 & -1.546 & $115 \pm 3$ & $49.6 \pm 0.2$ & $143.2 \pm 0.6$ & $3 \pm 2$ & $91.1 \pm 0.2$ & $72.2 \pm 0.3$  \\ 
2010-01-31 & 2455227.6399\tnote{\textdaggerdbl} & 3600 & -2.507 & $171.0 \pm 0.7$ & $-15.7 \pm 0.3$ & $61.4 \pm 0.2$ & $114.7 \pm 0.7$ & $82.9 \pm 0.3$ & $83.1 \pm 0.3$  \\ 
2010-02-06 & 2455233.6647 & 3600 & -5.303 & $182 \pm 3$ & $-26.9 \pm 1.1$ & $11.9 \pm 1.0$ & $185 \pm 3$ & $60.7 \pm 0.2$ & $107.1 \pm 0.2$  \\ 
2010-02-12 & 2455239.5524 & 3600 & -7.641 & $169.8 \pm 1.6$ & $-12.8 \pm 1.5$ & $85  \pm 3$ & $84 \pm 3$ & $49.10 \pm 0.10$ & $117.3 \pm 0.2$  \\ 
2010-03-15 & 2455270.7057 & 3600 & -19.679 & $174.1 \pm 1.3$ & $-24.7 \pm 1.0$ & $13.6 \pm 0.3$ & $178 \pm 4$ & $54.9 \pm 0.7$ & $111.6 \pm 0.7$ \\
        \bottomrule
    \end{tabular}
    \begin{tablenotes}
        \item[$\chi$] Epoch labeled EP-V4 for BF plot of V4 in Figure~\ref{fig:bfs}.
        \item[\textdagger] Epoch excluded for V4.
        \item[\textdaggerdbl] At this epoch for V5 the uncertainties were obtained by fitting a single profile as the peaks were completely overlapping.
    \end{tablenotes}    
    \end{threeparttable}

    \label{tab:spec.V4}
\end{table*}

The spectroscopic data for V4 were reduced by the UVES data reduction pipeline described in \citet{art:ballester2000}, and for the GIRAFFE spectra we received the reduced data products from ESO on DVDs. The FIES spectra of V2032 were reduced using the instrument data reduction pipeline FIEStool (v. 1.3.2), developed in \python by E. Stempels and maintained and provided by the staff at NOT. Before each observing night, calibration frames were produced from a standard data set of 7 bias and 21 halogen flats and each object exposure was preceded by a Th-Ar lamp exposure for optimal wavelength calibration. 

\begin{table*}
    \centering
    \footnotesize
    \caption{The 19 spectra of V2032 (2 of which have been excluded due to low flux) taken with the FIES spectrograph at NOT, La Palma, Spain. Shown are the dates, exposure times, barycentric velocity corrections, and the radial velocities of the primary and secondary component.}
    \begin{threeparttable}
    
    \begin{tabular}{c c c c c c}
        
        \toprule
        YYYY-MM-DD & BJD & Exp. time (s) & BVC (km/s) & $v_{\rm rad}^{\rm p}$ (km/s) & $v_{\rm rad}^{\rm s}$ (km/s) \\
        \midrule
\multicolumn{6}{c}{FIES} \\ 
\cmidrule(r){3-4} 
2012-11-11 & 2456242.7270 & 2000 & 25.049 & $48.9 \pm 0.2$ & $117.8 \pm 0.3$ \\
2012-11-20 & 2456251.6688 & 2100 & 23.825 & $152.37 \pm 0.14$ & $13.7 \pm 0.2$ \\
2012-11-20 & 2456251.6993\tnote{$\chi$} & 3000 & 23.745 & $152.09 \pm 0.12$ & $14.41 \pm 0.15$ \\
2012-11-21 & 2456252.7693 & 1740 & 23.366 & $133.0 \pm 0.3$ & $33.5 \pm 0.6$ \\
2012-12-10 & 2456271.6468 & 1800 & 18.806 & $48.69 \pm 0.12$ & $118.10 \pm 0.16$ \\
2012-12-17 & 2456278.7121 & 1800 & 16.233 & $168.20 \pm 0.16$ & $-2.7 \pm 0.2$ \\
2012-12-17 & 2456278.7339 & 1800 & 16.177 & $167.69 \pm 0.13$ & $-2.3 \pm 0.2$ \\
2012-12-19 & 2456280.7451\tnote{$\dagger$} & 1800 & 15.419 & - & - \\
2013-01-15 & 2456307.6237 & 1800 & 4.403 & $148.2 \pm 0.2$ & $17.9 \pm 0.4$ \\
2013-01-16 & 2456308.6145 & 1800 & 3.971 & $130.9 \pm 0.2$ & $35.4 \pm 0.4$ \\
2013-01-27 & 2456319.5413 & 1800 & -0.871 & $62.98 \pm 0.18$ & $103.41 \pm 0.16$ \\
2013-01-28 & 2456320.5394 & 1800 & -1.324 & $60.6 \pm 0.2$ & $106.5 \pm 0.3$ \\
2013-01-28 & 2456320.5612 & 1800 & -1.388 & $60.39 \pm 0.19$ & $105.9 \pm 0.3$ \\
2013-02-04 & 2456328.4944\tnote{$\dagger$} & 1800 & -4.837 & - & -\\
2013-10-10 & 2456575.7223 & 2400 & 24.804 & $50.5 \pm 0.3$ & $115.9 \pm 0.3$ \\
2013-10-10 & 2456575.7510 & 2400 & 24.748 & $50.90 \pm 0.18$ & $115.8 \pm 0.2$ \\
2013-10-12 & 2456577.7167 & 2400 & 25.057 & $48.7 \pm 0.2$ & $118.2 \pm 0.3$ \\
2013-10-12 & 2456577.7454 & 2400 & 25.000 & $49.11 \pm 0.16$ & $118.2 \pm 0.3$ \\
2015-02-03 & 2457057.4462 & 2100 & -4.012 & $139.18 \pm 0.19$ & $25.0 \pm 0.4$ \\
        \bottomrule
    \end{tabular}
    \begin{tablenotes}
        \item[$\chi$] Epoch labeled EP-V2032.
        \item[$\dagger$] Excluded due to low flux. 
    \end{tablenotes}
    \end{threeparttable}

    \label{tab:spec.V2032}
\end{table*}

To extract the RVs of the components in all DEBs, a \python implementation of the broadening function (BF) formalism formulated by \citet{art:rucinski1999} was utilized. RVs were obtained by matching the spectra to appropriate model atmospheres from \citet{art:coelho2005}. As the spectroscopic data have been acquired with different telescopes with quite different instruments, the approach differs from instrument to instrument.

For the GIRAFFE spectra, which cover a single order, the procedure is straightforward; each spectrum was normalized and a BF was calculated giving an estimate for the RV. The UVES/FLAMES setup gives two measurements for each of the spectra listed in the upper part of Table~\ref{tab:spec.V4}. The divided spectra were normalized and the BF was calculated individually for each, yielding two RV measurements for a given epoch. The mean of the two then constituted the first estimate for the RV, however, at a later point in our analysis (see Section \ref{sec:orb.anal}), anti-correlations showed up in the residuals of the RVs between the primary and secondary component for V4. Therefore, we omitted RVs derived from spectra from the upper CCD of the FLAMES/UVES setup due to the absence of prominent lines in this part of the spectrum and only used the measurements from the lower part. We thus took the RV stemming from the lower CCD as our value. The error was estimated by dividing this part of the spectrum into three parts, where we calculated the BF for each, then calculated the standard deviation of those three. This was also the approach for the GIRAFFE spectra.

With FIES at the NOT a spectrum is divided into 78 orders. Each order for a given epoch in Table~\ref{tab:spec.V2032} was processed individually, i.e., each order was normalized and for this part of the spectrum, the BF was calculated. Therefore, for each spectrum in Table~\ref{tab:spec.V2032}, 78 estimates for the RVs of the components are available. However, seeing as some of the orders at shorter wavelenghts do not have a lot of flux and some of the redder orders contain telluric lines, not all orders are equally good. Therefore, orders we deemed bad were omitted. The RV estimate from a given epoch is then the mean of the RVs obtained from all the good orders and the corresponding error is the standard deviation of the measurements from these orders. Example BFs for V4 and V2032 can be seen in Figure~\ref{fig:bfs}. Note that the primary component of V4 is rotating rapidly, resulting in a broad peak and a lower signal-to-noise ratio. The peak from the primary component in the BF for V5 was quite prominent, whereas the peak from the secondary component was harder to locate for some epochs and we had to constrain the fit to a certain interval.

\begin{table}
    \centering
    \caption{Orbital output parameters from \sbop, which serve as initialization input for the models calculated in Section~\ref{sec:orb.anal}.}
    \begin{tabular}{c c c c}
        \toprule
        & V2032 & V4 & V5 \\
        \midrule
        $K^\mathrm{p}$ ($\rm km/s$) & $62.00 \pm 0.15$ & $96.5 \pm 0.5$ & $71.9 \pm 1.3$  \\
        $K^\mathrm{s}$ ($\rm km/s$) & $62.55 \pm 0.17$ & $114.0 \pm 0.2$ & $96.1 \pm 1.3$ \\
        $e$ & \SI{0.5858 \pm 0.0016}{} & \SI{0.187 \pm 0.003}{} & \SI{0.003 \pm 0.011}{} \\
        $\omega$ ($^\circ$) & $319.0 \pm 0.2$ & $272.3 \pm 0.6$ & $110 \pm 5$ \\
        $P$ (days) & \SI{27.8677 \pm 0.0004}{} & \SI{2.867630 \pm 0.000005}{} & \SI{3.3570 \pm 0.0014}{} \\
        $\gamma$ ($\rm km/s$) & $83.26 \pm 0.05$ & $85.03 \pm 0.15$ & $84.9 \pm 0.7$ \\
        \bottomrule
    \end{tabular}

    \label{tab:sbop}
\end{table}

With the RVs in hand, we could then create the RV curves. We used a \python implementation of the program Spectroscopic Binary Orbit Program \citep[\sbop;][]{SBOP} to obtain estimates of the spectroscopic orbital parameters for each system, which will be used as initial guesses for the further analysis. The starting orbital parameters from SBOP for all the DEBs are listed in Table~\ref{tab:sbop}. Here we fit for the velocity semi-amplitudes, eccentricity ($e$), argument of periastron ($\omega$), period ($P$), systemic velocity ($\gamma$), and the time of periastron passage $T_{\rm peri}$.

Evidently, V2032 is a very eccentric system with a rather long period and, interestingly, the radial velocity amplitudes, $K^\mathrm{p}$ and $K^\mathrm{s}$, are very similar suggesting that the masses of the components are almost identical. The superscripts p and s will denote quantities for the primary and secondary, respectively, throughout (and in the case for V4 t denotes the tertiary component).

\subsection{Luminosity ratios}\label{sec:spec.lrat}

The calculated broadening functions do not only hold information about the radial velocities of the components in the binary system, but are also an estimate for their luminosity ratio, $L^\mathrm{s}/L^\mathrm{p}$. When the stars belong to the same spectral type, then the luminosity ratio is simply the ratio of the areas under the peaks. An external constraint on the luminosity ratio for the further analysis is in general advantageous and proved to be necessary to obtain precise results for our binary systems.

\begin{figure}
    \centering
    \includegraphics[width=\columnwidth]{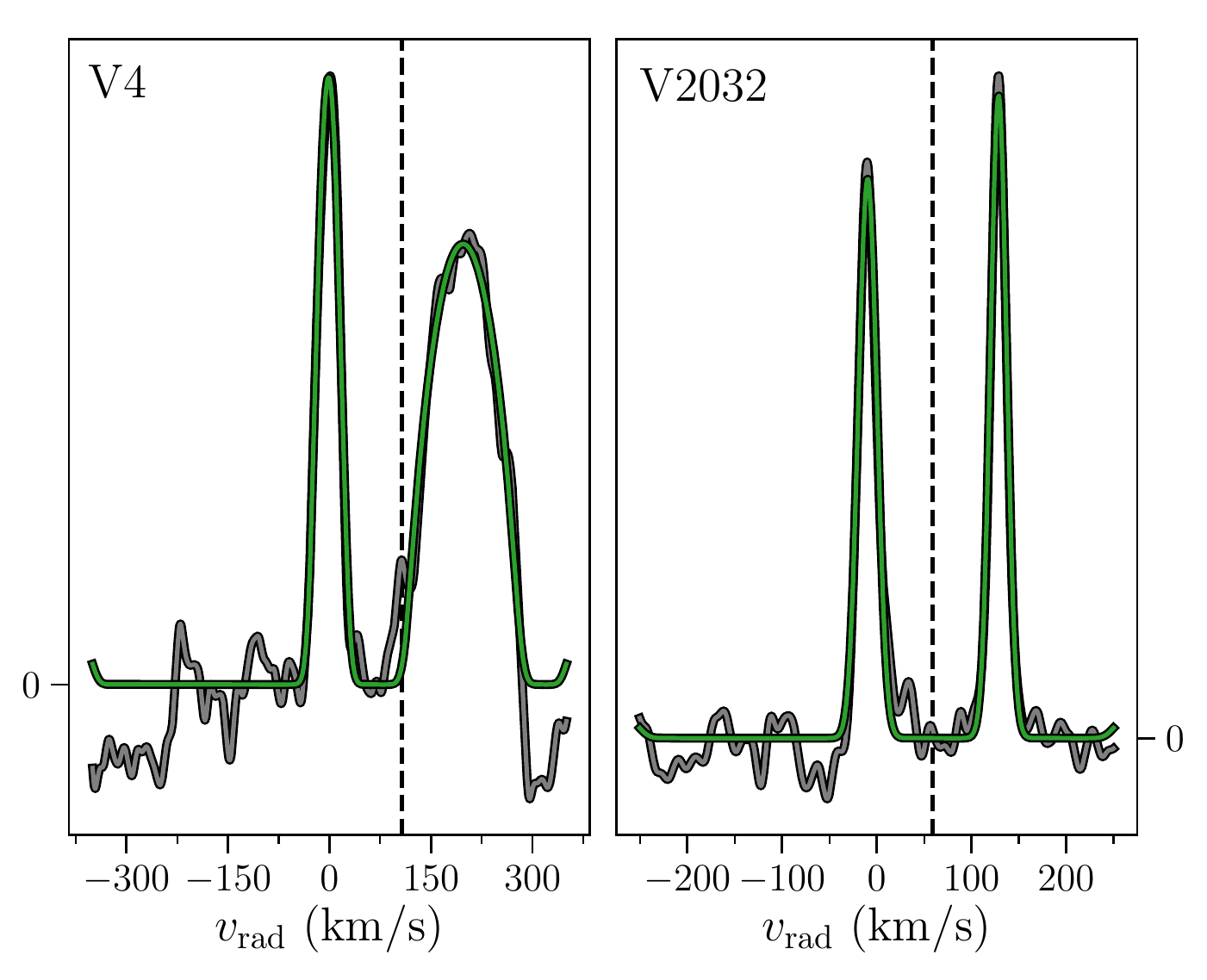}
    \caption{Example BF for V4 is shown to the left calculated from EP-V4 in Table~\ref{tab:spec.V4}. Shown to the right is an example BF for V2032 calculated from EP-V2032 in Table~\ref{tab:spec.V2032}. The grey lines in both figures are the smoothed calculated BFs and the green lines are the fitted rotational profile (see \citet{art:kaluzny2006} for details). The systemic velocity, $\gamma{\sim}83$~km~s$^{-1}$, corrected for the BVCs for the given epochs is marked with dashed lines. The y-axis is given in arbitrary units.}
    \label{fig:bfs}
\end{figure}

The ratio is easiest to calculate when the BF peaks are well separated (as is the case in Figure~\ref{fig:bfs}), so only epochs where the components have a large difference in RV were chosen from Tables~\ref{tab:spec.V4} and \ref{tab:spec.V2032}. As mentioned in Section~\ref{sec:spec.rad} due to the absence of lines in the part of the spectra imaged onto the upper CCD from the FLAMES/UVES setup, we only calculated the luminosity ratio for spectra stemming from the lower CCD. This yielded a value of $L^\mathrm{s}/L^\mathrm{p}= 0.40 \pm 0.02$ for V4. Because of the wavelength covered by this CCD this value corresponds to the luminosity ratio in $V$. We translated this ratio to corresponding values in $I$ and $B$ using filter transmission curves\footnote{Filter transmission curves from NOT: \url{http://www.not.iac.es/instruments/filters/filters.php}.} and obtained $0.39 \pm 0.02$ and $0.40 \pm 0.02$, respectively, corresponding to all available light curves for V4. We also calculated the luminosity ratio from the BFs for V5 using our GIRAFFE spectra and obtained a value of $0.36 \pm 0.03$ in $V$.

\begin{figure}
    \centering
    \includegraphics[width=\columnwidth]{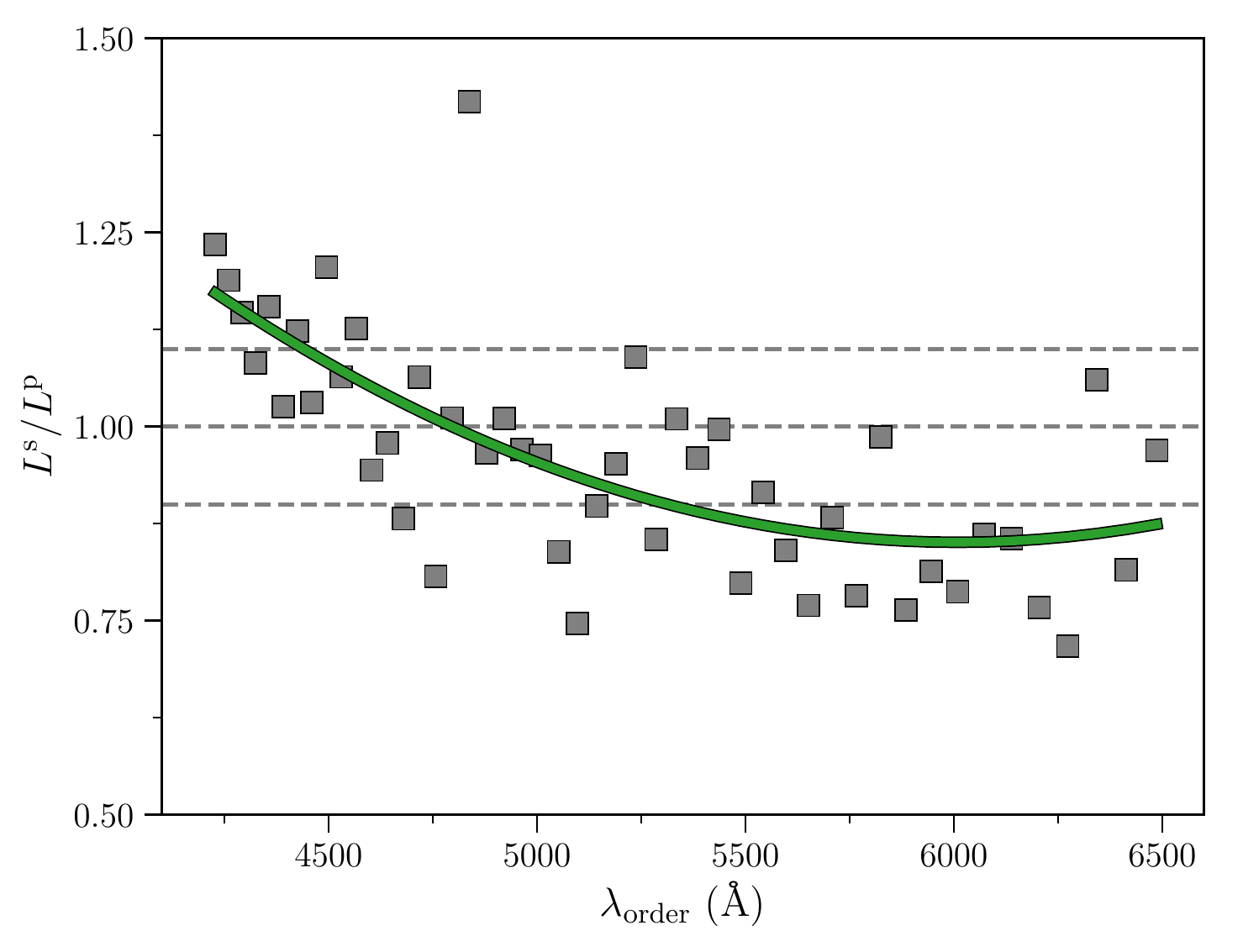}
    \caption{Luminosity ratio of V2032 as a function of order (here $\lambda_\mathrm{order}$ designates the midpoint of the wavelength interval for a given order). A grey square at a given order is the mean value of the luminosity ratio calculated from the BF for ``good'' epochs with well separated peaks (as in Figure~\ref{fig:bfs}). The green curve is a second order polynomial fit to these points, which is used to elucidate the trend.}
    \label{fig:lrat}
\end{figure}

For V2032 we used the FIES spectra to calculate the luminosity ratio, we again only used epochs where the peaks were well separated and again we only used the orders that we deemed suitable. The procedure was to, for a given order, calculate $L^\mathrm{s}/L^\mathrm{p}$ for all the spectra with well separated peaks and use the mean value of these as the value for this order. This was then repeated for all the good orders. This is shown with grey squares in Figure~\ref{fig:lrat}. Many of the measurements for the luminosity ratio of V2032 are very close to 1 and the overall value is $0.95 \pm 0.05$, however, a small trend is apparent when the values obtained for $L^\mathrm{s}/L^\mathrm{p}$ are plotted against the orders. The trend suggests that the secondary component is slightly more luminous at shorter wavelengths compared to the primary component meaning that $T_\mathrm{eff}^\mathrm{s}>T_\mathrm{eff}^\mathrm{p}$. 

The luminosity ratios are used in the subsequent analysis (Section~\ref{sec:orb.anal}) to help constrain the radii of the components. Specifically, for V2032 where we have photometric data in $V$ and $I$ as well as from TESS (Section~\ref{sec:phot}), which has a photometric passband similar to that of $I$, we derived luminosity ratios corresponding to these passbands. For $V$ this was done by simply selecting measurements of the BF from Figure~\ref{fig:lrat} in the range $4100-6100$~\AA\, and calculate the robust mean and standard deviation of these. This resulted in a value of $0.89 \pm 0.02$. For $I$ (TESS) we utilized the same scheme as for V4 to obtain a value of $0.84 \pm 0.02$.

\subsection{The Spectral Energy Distribution of V4}\label{sec:sed}

We examined the SED of V4 to confirm the value of the luminosity ratio we have obtained from spectroscopy (see Section~\ref{sec:spec.lrat}), but also to see if we can learn more about the fainter, third companion. A benefit of the binary's membership in a cluster is that it should be possible to describe the binary's light as the sum of the light of two single cluster stars. To that end, we compiled a database of photometric measurements from V4 and from likely single main sequence stars in NGC~2506, and sought a combination of stars whose summed fluxes most closely match the fluxes of the binary. For our sample of probable single stars, we selected likely members based on {\it Gaia} proper motions, parallaxes, and photometry. Likely binaries were rejected by restricting the sample to those with {\it Gaia} photometry placing them within about 0.03 mag of the blue edge of the main sequence band in the $G_{\rm BP}-G_{\rm RP}$. 

We briefly describe the photometric datasets and the conversions from magnitude to flux below. In the ultraviolet, \citet{siegel} presented photometry of more than 100 open clusters (including NGC~2506) using the UVOT telescope on the {\it Swift} satellite \citep{art:gehrels2004}. We used their magnitudes in the $uvw1$, $uvm2$, and $uvw2$ bands, and converted to fluxes.

\citet{art:twarog2016} and \citet{grundahl} presented narrow-band Str\"{o}mgren $uvby$ photometry for the cluster. We employed reference fluxes from \citet{gray-strom} to convert the magnitudes to fluxes. \citet{marconi} observed the cluster in 6 wide filters ($UBGVRI$). With the exception of the $G$ filter, the magnitudes were converted to fluxes using reference fluxes from \citet{bcp}, taking into account the known reversal of the zero point correction rows for the observed flux, $f_\lambda$ and $f_\nu$.

There are a couple of large ground-based optical surveys that provide calibrated broad-band photometric observations. The Pan-STARRS1 survey \citep{ps1desc} contains photometry in 5 filters ($grizy$), and we use their mean PSF magnitudes here. Zero points for its AB magnitude system are given in \citet{ps1}. The SkyMapper survey (Data Release 1; \citealt{skymap}) is a six filter ($uvgriz$) southern hemisphere study that provides PSF magnitudes on an AB system. In addition, {\it Gaia} has already produced high-precision photometry extending far down the main sequence of the cluster as part of \gaia DR2. We obtained the fluxes in the $G$, $G_{\rm BP}$, and $G_{\rm RP}$ bands from the {\it Gaia} Archive.
   
In the infrared, we have obtained Two-Micron All-Sky Survey (2MASS; \citealt{2mass}) photometry in $JHK_s$ from the All-Sky Point Source Catalog, and have converted these to fluxes using reference fluxes for zero magnitude from \citet{2masscal}. The stars were observed in $JK_s$ within the deeper VISTA survey \citep{vista}.  We also used PSF magnitudes in $iJ$ filters from the third data release of the DENIS database\footnote{{\tt cds.u-strasbg.fr/denis.html}}. 

Although we have strived to put the measurements on a consistent flux scale in order to construct spectral energy distributions, we emphasize that our procedure for decomposing the light from the two stars in a cluster binary does not depend on the exact calibration. What is important is that we are using measurements of a large number of cluster stars from uniform photometric studies, i.e., we are assuming the {\it relative} flux measurements are precise. The benefit of this procedure is that it is a {\it relative} comparison using other cluster stars with the same distance, age, and chemical composition, and not an absolute comparison. As such, it is independent of distance and reddening (as long as these are the same for the binary and comparison stars), the details of the filter transmission curves (as long as the same filter is used for observations of the different stars), and flux calibration of any of the filters (as long as the calibration is applied consistently). We can also avoid systematic errors associated with theoretical models or with the consistency of the different parts of empirical spectral energy distributions compiled from spectra.

We tested two ways of doing the decomposition of the binary's light: using well-measured NGC~2506 stars as proxies and checking all combinations of likely main sequence stars; and fitting all main sequence stars with photometry in a given filter as a function of {\it Gaia} $G$ magnitude. When using sums of real stars, we are somewhat at the mercy of the photometry that is available for each star (and the binary) and of the stellar sampling, i.e., the density of stars of the main sequence. The use of fits allows for finer examination of the main sequence, although there is some risk of diverging from the photometry of real stars.

To judge the degree to which a pair of stars reproduced the binary photometry, we looked for a minimum of a $\chi^2$-like parameter involving fractional flux differences in the different filter bands; $\sum_i [(F_{i,{\rm bin}} - (F_{i,1} + F_{i,2}))/(\sigma_{i,{\rm bin}} \cdot F_{i,{\rm bin}})]^2$, where $F_{i,{\rm bin}}$, $F_{i,1}$, and $F_{i,2}$ are the fluxes for respectively the binary and the two proxies, and $\sigma_{i,{\rm bin}} = 10^{-\sigma_{i,{\rm m}}/2.5} - 1$ with $\sigma_{i,{\rm m}}$ being the magnitude uncertainty in the $i$th filter band for the binary. The uncertainty was set to 0.02 mag for photometry without quoted errors or if the quoted uncertainty was below that value. This was done in order to deweight photometry with very low uncertainties (such as {\it Gaia}) that results partly from their very wide filter bandpasses.

The best fit combination of cluster star SEDs depends somewhat on the filters that were employed, to the point that the redder star could switch between the brighter and fainter star. The flux ratios were somewhat more stable, however, and the two stars cannot have temperatures that are too dissimilar. Our preferred set of photometry excluded DENIS $J$ and $K_s$, and WISE datasets due to low signal-to-noise, and had a goodness-of-fit value of 40.0 from measurements in 37 filters.

\begin{figure}
    \centering
    \includegraphics[width=\columnwidth]{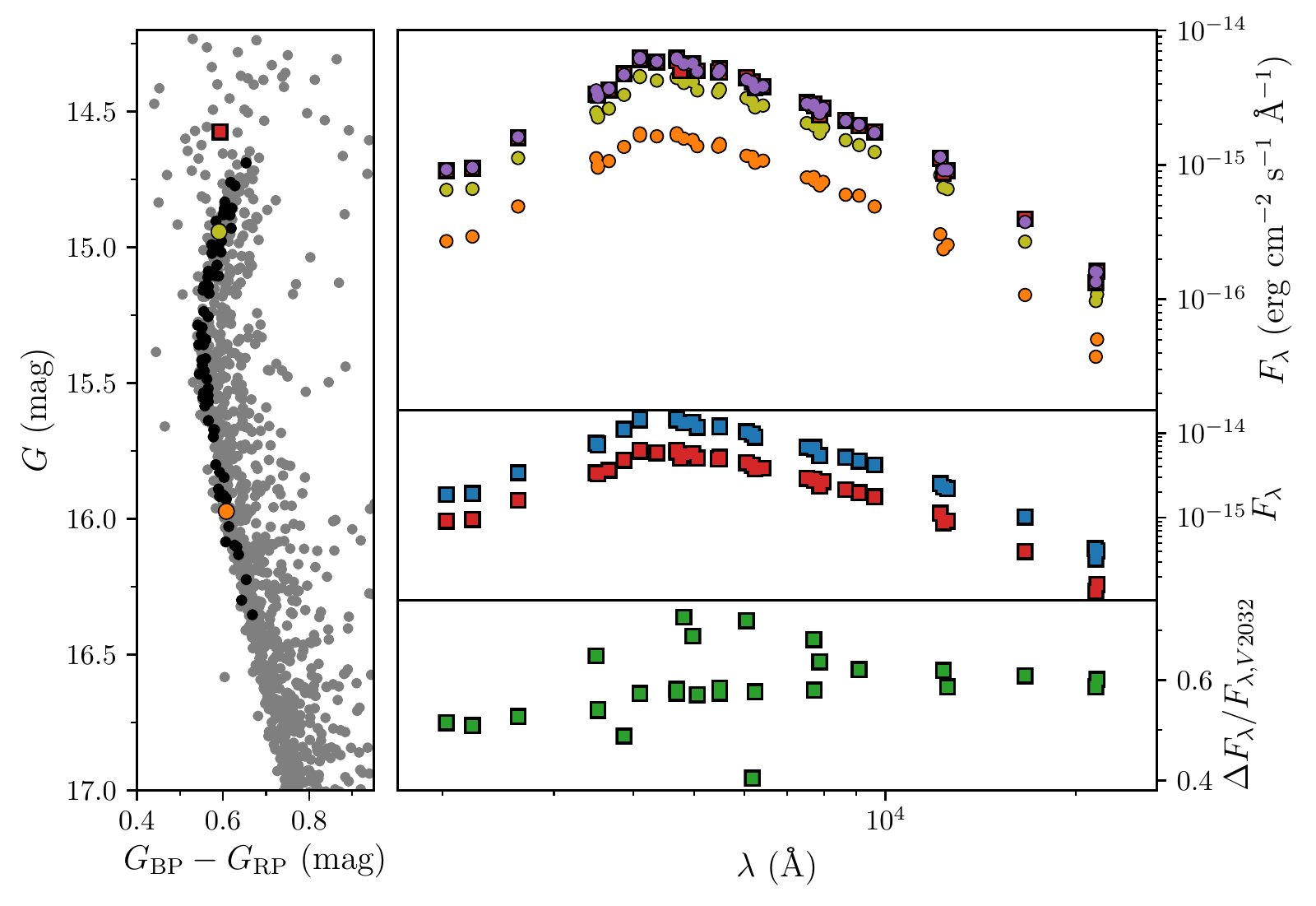}
    \caption{ Left: {\it Gaia} CMD for NGC~2506 cluster members in grey with the red square marking the combined photometry of V4. The yellow and orange points show the two stars identified as the best fit (MHT 772 and MHT 808, respectively). The black points are probable single cluster member stars that had photometry in all of the filter bands used in the SED fit. Top right: SEDs of V4 (red squares, which are mostly obscured by the purple points), MHT 772 (yellow points), MHT 808 (orange points), and the combined light of the two best-fitting stars (purple points). Middle right: SEDs of V4 (red squares) and V2032 (blue squares). Bottom right: Comparison of the SEDs for V4 and V2032, $(F_{\lambda,V2032} - F_{\lambda,V4})/F_{\lambda,V2032}$.}
    \label{fig:sed}
\end{figure}

Top panel of Figure~\ref{fig:sed} shows a comparison of the SED of V4 with the best fitting pair \citep[MHT 772 and 808 in][or WEBDA 4254 and 1247, respectively]{marconi}. A potential limiting factor is the stellar sampling available near the brighter star, but we have stars within 0.011 $G$~mag on the bright side and within 0.007~mag on the faint side. For the faint star, other stars in the sample fall within 0.06~mag. The resulting luminosity ratio in filters similar to $V$ (Str\"{o}mgren $y$, Sloan $g$, \citeauthor{marconi} $V$) was 0.39.

The main-sequence fitting procedure can be employed in any filter with a sufficient sample of stars covering the range of brightnesses for the binary's stars. In our case, this eliminates the DENIS $K_s$ and WISE filters from consideration in fitting V4. Our fit statistic had a minimum value of 47.6 for the selection of 38 filters.  We estimated the $2\sigma$ uncertainty in the fit based on where the goodness-of-fit statistic reached a value of 4 above the minimum value. For example, this returns $2\sigma(G_A) = 0.016$ and $2\sigma(G_B) = 0.05$. As expected, there is an anti-correlation between values for the primary and secondary stars because of the need to match the binary fluxes. For filters similar to $V$, the best-fit luminosity ratio comes out as $0.33\pm0.02$. Overall this fit is notably poorer than the cluster star fit in infrared $J$, $H$, and $K_s$ bands, with the computed fit being brighter than the observed binary. This appears to recommend the cluster star fit, with its slightly fainter primary star. 

\subsubsection{Effective temperatures for the components in V4}\label{sec:teff_v4}

We can attempt to get precise stellar temperatures for the components of V4 using the infrared flux method \citep[IRFM;][]{irfm}. With the available photometric databases for NGC~2506, we have measurements of fluxes covering the majority of the stellar energy emission. The IRFM relies on the difference in temperature sensitivity between the bolometric flux and monochromatic fluxes in the infrared on the Rayleigh-Jeans portion of the spectrum. The ratio of the bolometric and infrared fluxes can be compared to theoretical values: \[ \frac{F_{\rm bol}(\mbox{Earth})}{F_{\lambda_\mathrm{IR}}(\mbox{Earth})} = \frac{\sigma T_{\rm eff}^4}{F_{\lambda_\mathrm{IR}}(\mbox{model})} \] We used the 2MASS flux calibration of \citet{art:casagrande2010} in our implementation, in part because it produced greater consistency between the temperatures derived in the three bands. VISTA $J$ and $K_s$ filters returned $T_{\rm eff}$ estimates that were within the scatter of the 2MASS values, so we considered this corroboration. Starting from a solar-metallicity ATLAS9 model that produced a good fit by eye, we adjusted the temperature of the synthetic spectrum until it matched the average IRFM temperature from the three 2MASS bands. The model surface gravity was chosen from the eclipsing binary results (Section~\ref{sec:orb.anal}), although the results had little sensitivity to the gravity.

For MHT 772, which was identified as the best cluster representative of the primary star of V4, we found $T_{\rm eff} = 6830$ K, with a full range of 110 K for the estimates from different 2MASS bands. Thus, we estimate the uncertainty to be approximately 55 K. For comparison, we calculated the temperature for V4 itself, i.e., the combined light --- the two stars in our SED decomposition appear to have very similar colours. We found $6820 \pm 100$ K (with the uncertainty estimate from half of the full range in the 2MASS measurements).

\subsubsection{Effective temperatures for the components in V2032}\label{sec:teff_v2032}

We were unable to decompose the light of the V2032 binary in the same way we did for V4 because the component stars appear to reside in a part of the CMD where there is rapid evolution and few single stars to be found. However, the colour of the binary's combined light is very similar to that of V4, so we compared the SEDs of the two binaries to seek information about the component temperatures. The comparison (bottom panel Figure~\ref{fig:sed}) showed that V4 clearly has a larger fraction of its flux in the ultraviolet, which leads us to the conclusion that the primary (more massive) star of V2032 is cooler than the stars of V4. Employing the IRFM on the SED of V2032 gives $T_{\rm eff} = 6560 \pm 30$ K, although this should not be considered a direct measurement of the primary star's temperature. It is, however, fairly good evidence that the primary star is evolving towards the red -- if it is cooler but more luminous than the secondary star, expectations from normal single-star stellar evolution tracks would require it to be on the subgiant branch. The relative temperature difference between the components of V2032 is consistent with the results from the broadening functions in different spectral orders (see Section~\ref{sec:spec.lrat}).

Even though we could not get a good estimate of the effective temperature of the secondary component from the SED, we can still get a good measure for this value given that we have estimated the effective temperature of the primary component of V2032, and we have measured the metallicity and reddening, we can calculate the effective temperature of the secondary component. This was done by performing a Monte Carlo simulation, where we drew from Gaussian distributions in the sense $\mathcal{N}(\mu,\sigma)$ for the following parameters $T_{\rm eff}^{\rm p} = 6560 \pm 100$~K (where the 100~K is to account for any potential difference between the proxy and the primary), E$(b-y) = 0.057 \pm 0.004$~mag, $\rm [Fe/H] = -0.36 \pm 0.10$~dex, and the colour of the combined light of V2032 $(b-y) = 0.290 \pm 0.002$~mag. 

For each draw we found the colour for the primary, $(b-y)^{\rm p}$, that minimizes the difference between $T_{\rm eff}^{\rm p}$ estimated from the SED and the value resulting from using the temperature-colour-metallicity calibration in \citet{art:casagrande2010} given E$(b-y)$ and $\rm [Fe/H]$. From this it is possible to calculate the colour of the secondary component, $(b-y)^{\rm s}$, since $(b-y)=k^{\rm p}(b-y)^{\rm p} + k^{\rm s}(b-y)^{\rm s}$, where $k^{\rm p, s}$ is the fractional amount of light a component contributes to the system. We calculated this by drawing normally distributed values from the calculated luminosity ratio of $0.95 \pm 0.05$. A measure for $(b-y)^{\rm s}$ then yields a value for the effective temperature of the secondary component. From 5,000 draws this yielded a value of $T_{\rm eff}^{\rm s} = 7100 \pm 100$~K.

We caution that this is not a direct measure of the effective temperatures, rather it is a good estimate, which yields consistent results later in our analysis.



\section{Photometric observations}\label{sec:phot}

As V4 has been known to be an eclipsing binary for quite some time \citep[see, e.g.,][]{art:kim2001,art:arentoft2007}, a lot of data have been collected through the years with the earliest stemming from 2005 and the most recent from 2017. In contrast we only recently identified V2032 as being an eclipsing binary and as such only the most recent (ground-based) photometry contains light curves of this system. Common for both systems is that the (ground-based) photometry is CCD observations in the Johnson system. Table~\ref{tab:phot} displays all the ground-based photometric data available for the two binaries -- from the oldest taken with the Danish 1.54-metre to the latest stemming from the NOT. The observations made at the NOT using ALFOSC comprise all the photometric data available for V2032. The photometric data for V5 was obtained together with the earliest data for V4. All the photometric data were analyzed using the program DAOPHOT \citep{art:stetson1987} following the same procedure as in \citet{art:grundahl2008}. The Str\"omgren photometry presented here is the same as used in \citet{art:arentoft2007} stemming from \citet{art:grundahl2000}. Additionally we have obtained much more recent photometric data from the {\it Transiting Exoplanet Survey Satellite} \citep[TESS;][]{art:ricker2015}.

\subsection{Light curves}\label{sec:phot.light}

\begin{figure*}
    \centering
    \includegraphics[width=\textwidth]{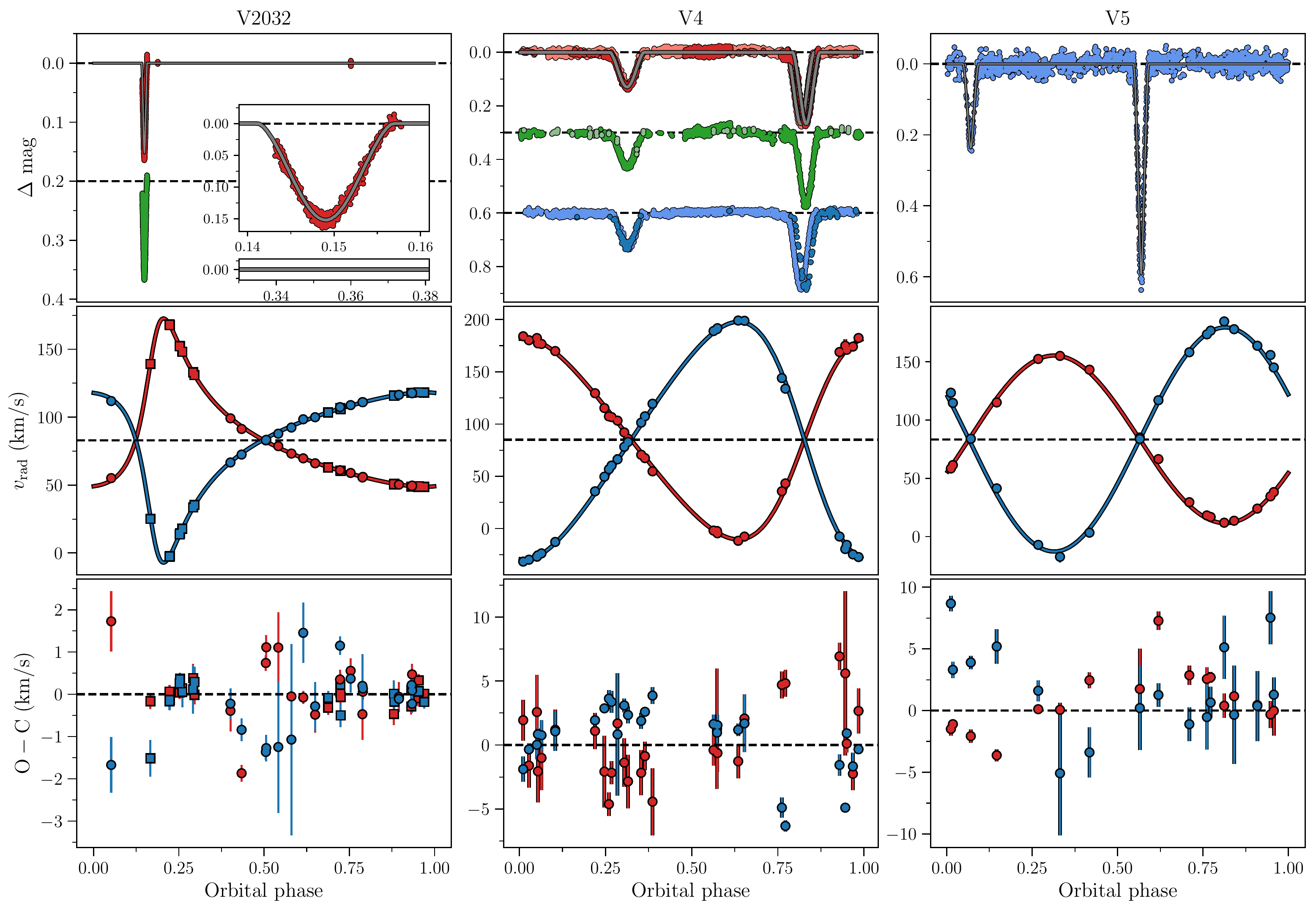}
    \caption{Top: Phase folded light curves of V2032, V4, and V5. For V4 we have light curves in $I$, $V$, and $B$ marked with respectively red, green, and blue points. The green points are shifted by $0.3$~mag and the blue points by $0.6$~mag. Shown in the left panel are light curves for V2032 in $I$ marked with red points and shifted by $0.2$~mag in green is $V$. The insets show a close-up of the eclipse in $I$ and the phase for the next conjunction, i.e., where we would expect a secondary eclipse if it was visible. In grey we have displayed a light curve model to show that our models suggest that there is only one eclipse in the system. The points in lighter colours (only for V4 and V5) are from the Danish 1.54-metre and the Flemish Mercator (see Table~\ref{tab:phot}) and the darker points are from the other telescopes. For V5 we only show observations in $B$ that we use in the analysis. Middle: Radial velocity curves for V2032, V4, and V5. The primary component is in all cases shown in red and the secondary in blue. The horizontal dashed lines denote the systemic velocity, $\gamma \sim 83$~km/s. Bottom: The calculated radial velocities subtracted from the observed ones.}
    \label{fig:lcs_rvs}
\end{figure*}

For the case of V4 with observations from many different telescopes, the photometry has to be brought to match by eliminating instrumental differences between the telescopes as well as night-to-night variations, which also apply to the observations of V2032 and V5. This was done by taking the mean of out-of-eclipse observations for a given night and subtract this value from the rest of the observations made that night. For observations where this was not possible (when all data points were obtained during an eclipse), the points were matched by eye. Figure \ref{fig:lcs_rvs} shows the phase folded light curves of V2032, V4, and V5. Evidently, the light curves of V4 and V5 are well-covered due to the amount of data available covering the entire phase in each, whereas the amount of observations of V2032 are much more sparse because of the more recent discovery of an eclipse in this system. 

\begin{table*}
	\centering
	\caption{Table showing dates, BJDs, and filters (Johnson) for the photometric data of the binaries. The data acquired with NOT comprises all the photometric data available for V2032. The data for V5 is from the Danish 1.54-metre and the Mercator telescope \citep{art:arentoft2007}. Note that no observations were made in $V$ and $B$ with the Flemish Mercator and the NOT, respectively.}
	\begin{threeparttable}
			\begin{tabular}{c c c c c c c}
				\toprule
				yyyy-mm-dd & BJD & Filter & BJD & Filter & BJD & Filter \\
				\midrule				
\multicolumn{7}{c}{Danish 1.54-metre\tnote{a}} \\
\cmidrule(r){3-5}
2005-01-05 & 2453375.6051 & $I$ & 2453381.6149 & $V$ & 2453375.5935 & $B$ \\
2005-01-05 & 2453375.6071 & $I$ & 2453381.6158 & $V$ & 2453375.5988 & $B$ \\
2005-01-05 & 2453375.6090 & $I$ & 2453381.6168 &  $V$ & 2453375.6021 & $B$ \\
\vdots & \vdots & \vdots & \vdots & \vdots & \vdots & \vdots \\
2006-02-14 & 2453780.6435 & $I$ & 2453759.8572 & $V$ & 2453780.6407 & $B$ \\
2006-02-14 & 2453780.6492 & $I$ & 2453759.8629 & $V$ & 2453780.6461 & $B$ \\
2006-02-14 & 2453780.6547 & $I$ & 2453759.8699 & $V$ & 2453780.6516 & $B$ \\ \hdashline
\multicolumn{7}{c}{Mercator\tnote{b}} \\
\cmidrule(r){3-5}
2005-01-08 & 2453378.5966 & $I$ & - & $V$ & 2453378.6008 & $B$ \\
2005-01-08 & 2453378.6058 & $I$ & - & $V$ & 2453378.6034 & $B$ \\
2005-01-08 & 2453378.6107 & $I$ & - &  $V$ & 2453378.6131 & $B$ \\
\vdots & \vdots & \vdots & \vdots & \vdots & \vdots & \vdots \\
2005-04-07 & 2453468.4169 & $I$ & - & $V$ & 2453468.4192 & $B$ \\
2005-04-07 & 2453468.4215 & $I$ & - & $V$ & 2453474.3746 & $B$ \\
2005-04-13 & 2453474.3770 & $I$ & - & $V$ & 2453474.3794 & $B$ \\ \hdashline				
\multicolumn{7}{c}{IAC-80\tnote{c}} \\
\cmidrule(r){3-5}
2013-01-17 & 2456310.4386 & $I$ & 2455580.4556 & $V$ & 2456311.4825 & $B$ \\
2013-01-17 & 2456310.4497 & $I$ & 2455580.4593 & $V$ & 2456311.4835 & $B$ \\
2013-01-17 & 2456310.4607 & $I$ & 2455580.4631 &  $V$ & 2456311.4844 & $B$ \\
\vdots & \vdots & \vdots & \vdots & \vdots & \vdots & \vdots \\
2016-01-08 & 2457395.6629 & $I$ & 2457397.7326 & $V$ & 2457395.6619 & $B$ \\
2016-01-08 & 2457395.6655 & $I$ & 2457397.7343 & $V$ & 2457395.6645 & $B$ \\
2016-01-08 & 2457395.7410 & $I$ & 2457397.7360 & $V$ & 2457395.6670 & $B$ \\ \hdashline
\multicolumn{7}{c}{LCOGT\tnote{d}} \\
\cmidrule(r){3-5}
2016-01-08 & 2457395.7173 & $I$ & 2457392.3383 & $V$ & 2457392.3400 & $B$ \\
2016-01-08 & 2457395.7252 & $I$ & 2457392.3411 & $V$ & 2457392.3423 & $B$ \\
2016-01-08 & 2457395.7311 & $I$ & 2457392.3438 &  $V$ & 2457392.3451 & $B$ \\
\vdots & \vdots & \vdots & \vdots & \vdots & \vdots & \vdots \\
2016-01-20 & 2457407.8129 & $I$ & 2457447.1025 & $V$ & 2457447.0996 & $B$ \\
2016-01-20 & 2457407.8211 & $I$ & 2457447.1076 & $V$ & 2457447.1046 & $B$ \\
2016-01-20 & 2457407.8292 & $I$ & 2457447.1125 & $V$ & 2457447.1146 & $B$ \\ \hdashline
\multicolumn{7}{c}{NOT\tnote{e}} \\
\cmidrule(r){3-5}
2016-12-30 & 2457753.4820 & $I$ & 2457753.4816 & $V$ & - & $B$ \\
2016-12-30 & 2457753.4829 & $I$ & 2457753.4825 & $V$ & - & $B$ \\
2016-12-30 & 2457753.4844 & $I$ & 2457753.4839 & $V$ & - & $B$ \\
\vdots & \vdots & \vdots & \vdots & \vdots & \vdots & \vdots \\
2017-02-25 & 2457810.4797 & $I$ & 2457810.4816 & $V$ & - & $B$ \\
2017-02-25 & 2457810.4800 & $I$ & 2457810.4819 & $V$ & - & $B$ \\
2017-02-25 & 2457810.4804 & $I$ & 2457810.4823 & $V$ & - & $B$ \\
				\bottomrule
		\end{tabular}
		\begin{tablenotes}
            \item[a] The Danish 1.54-metre, La Silla, Chile.
			\item[b] The Flemish Mercator, La Palma, Canary Islands, Spain. No observations were made in $V$.
			\item[c] The IAC-80, Tenerife, Canary Islands, Spain.
			\item[d] The Las Cumbres Observatory Global Telescope Network located at multiple sites around the world.
			\item[e] Nordic Optical Telescope, La Palma, Canary Islands, Spain.
		\end{tablenotes}
	\end{threeparttable}

	\label{tab:phot}
\end{table*}

Something quite peculiar can be seen in the panel for V4 in Figure~\ref{fig:lcs_rvs}. Evidently, the primary eclipse as observed by the Danish 1.54-Metre and the Flemish Mercator \citep[published data from][listed in the upper part of Table~\ref{tab:phot} and marked with lighter colours in Figure~\ref{fig:lcs_rvs}]{art:arentoft2007} is shifted from the more recent observations made with the IAC-80, LCOGT, and the NOT (darker points). These eclipse-timing variations (ETVs) are most likely caused by a third, but dimmer, companion in the V4 system. Indications for a third body can also be seen in the BFs for V4, where a small additional hump appeared around the systemic velocity for some epochs as in Figure~\ref{fig:bfs}, however, this is a somewhat more dubious indication.

\subsection{TESS data}\label{sec:phot.tess}

\begin{figure}
    \centering
    \includegraphics[width=\columnwidth]{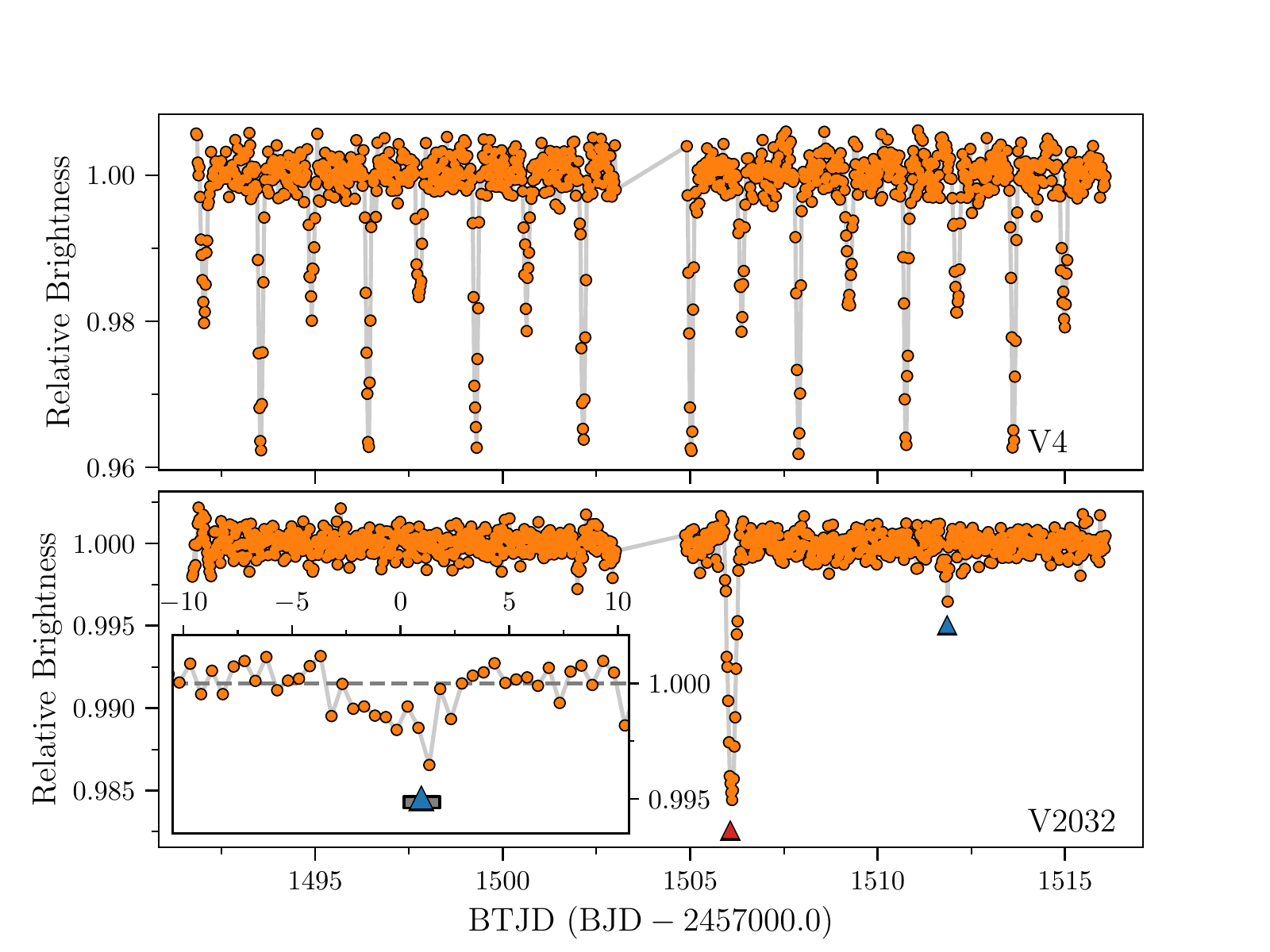}
    \caption{Light curves for V4 (top) and V2032 (bottom) extracted from the TESS FFIs (see Figure~\ref{fig:tessffi}). V4 is seen to eclipse multiple times as expected, given its $\sim$2.9~d period, whereas V2032 eclipses only once, consistent with this system having an orbital period of $\sim$27.9~d, which coincidentally is very close to that of TESS' orbit. The red triangle in the panel for V2032 shows the time for the primary eclipse and the blue triangle shows the expected time for the secondary eclipse (if visible) calculated from Equation~\eqref{eq:sec_ecl}. The inset is a zoomed view around the decrease in flux with the x-axis given in hours from the observed midpoint. The grey bar represents the smear in $T_0^{\rm s}$ (see Section~\ref{sec:signi}).}
    \label{fig:tesslc}
\end{figure}

During our analysis of this cluster it was observed by TESS. NGC~2506 was observed in TESS' Sector 7 and can be found in the $30$~min. cadence full-frame images (FFIs) displayed in Figure~\ref{fig:tessffi}. From the FFIs we were able to recover the signals from V4, V2032, and V5 by making use of the \lk package \citep{pack:lk}. In Figure~\ref{fig:tesslc} we display the light curves for V2032 and V4. V5 is not shown, since we do not use the TESS light curve in our analysis. 

For V4 we see multiple eclipses in Figure~\ref{fig:tesslc} and as expected V2032 only eclipses once due to the longer period. What is evident from Figure~\ref{fig:tessffi}, but also quite apparent when the depths seen in Figure~\ref{fig:tesslc} are compared to Figure~\ref{fig:lcs_rvs}, is how contaminated the signals are owing to the large pixel size of the TESS images \citep[approximately 21 arcseconds per pixel;][]{art:ricker2015}. Naturally, this is something we need to account for when these light curves are used to derive stellar parameters related to the depth of the eclipses. 

We tried estimating the time of the secondary eclipse in V2032 as we were unsure whether this would actually be visible due to the orientation of the system. Given that the orbit of V2032 is very eccentric (see Table~\ref{tab:sbop}) the time for the secondary eclipse, $T_\mathrm{0}^\mathrm{s}$, is not just found half a period after the time for the primary eclipse, $T_\mathrm{0}^\mathrm{p}$, but can be found from \citep{art:sterne1940}

\begin{equation}
    T_\mathrm{0}^\mathrm{s} - T_\mathrm{0}^\mathrm{p} = \frac{P}{\pi} \left ( \frac{h (1-e^2)^{1/2}}{1 - g^2} + \tan^{-1} \frac{h}{(1-e^2)^{1/2}} \right ) + \frac{1}{2} P \, ,
    \label{eq:sec_ecl}
\end{equation}
where $h=e \cos \omega$, and $g=e \sin \omega$. In Figure~\ref{fig:tesslc} we mark $T_\mathrm{0}^\mathrm{p}$ with a red triangle and $T_\mathrm{0}^\mathrm{s}$ as calculated from Equation~\ref{eq:sec_ecl} with a blue triangle. The calculated value for $T_\mathrm{0}^\mathrm{s}$ seems to coincide with a decrease in flux.

\subsubsection{Signal Significance}\label{sec:signi}
To assess the significance of the decrease in flux around $T_0^{\rm s}$ (blue triangle Figure~\ref{fig:tesslc}) and a potential secondary eclipse in V2032, we first looked at the distribution of the data in Figure~\ref{fig:tesslc}, with the exclusion of in-eclipse data, i.e., times around $T_\mathrm{0}^\mathrm{s}$ and $T_\mathrm{0}^\mathrm{p}$, and tried to find a proper match. An Anderson-Darling test \citep{art:anderson1952} suggested that we could reject the null hypothesis of normality at a significance level of at least $1\%$, so clearly the data are not normally distributed. A distribution that accounts for the data much better is the Student's t distribution. Here we chose 18 degrees of freedom as this neatly captured the tails of our distribution. We then ran a Monte Carlo simulation of 5,000 draws from the Student's t distribution as a representation of our data to see how often we get a sequence of 12 (as in Figure~\ref{fig:tesslc}) or more consecutive points below $1.0$. This happens in around $15\%$ of the cases. For each case of these $15\%$ we estimated the median and created a Gaussian distribution from these. Here we find that at a $6.4\sigma$ level we can reject that these points would have a median equal to or below the median of the in-eclipse points in Figure~\ref{fig:tesslc}, meaning that it is highly unlikely that this is caused by statistical fluctuations.

Finally, we looked at the timing of the signal, i.e., how likely is it that a signal of this duration ($\sim 6.0$~h) would appear at $T_0^{\rm s}$. Here we included a "smear" in $T_0^{\rm s}$ by incorporating the uncertainties in $P$, $e$, $\omega$, and $T_0^{\rm p}$ (from the $I$ column) in Table~\ref{tab:V2032}. This amounted to a spread of 1.7~h around $T_0^{\rm s}$ shown as the grey bar in Figure~\ref{fig:tesslc}. Here we  used 0.1 and 99.9 percentiles to be conservative resulting in a spread of 5.2~h. We then conducted another Monte Carlo simulation, where we picked out times from the time series at random, placed our 5.2~h smear for $T_0^{\rm s}$ there, and checked if it overlapped with the observed 6.0~h signal. In 5,000 draws this happens in roughly $0.1\%$ of the draws. Clearly, this signal cannot be ascribed to statistical fluctuations and the timing is suspicious to say the least. However, the contamination from nearby sources is so large in TESS (due to the pixel size as seen in Figure~\ref{fig:tessffi}) that we refrain from concluding that the observed signal in Figure~\ref{fig:tesslc} is in fact a secondary eclipse in V2032, especially seeing as our model suggests that a secondary eclipse should not be visible in the system (see Figure~\ref{fig:lcs_rvs}). Only observations around $T_0^{\rm s}$ from an instrument with a finer spatial resolution can resolve this. We therefore carry out the analysis of the system without employing additional constraints to this part of the TESS light curve.



\subsubsection{Asteroseismology from TESS data}\label{sec:astero}
With the TESS data it was natural to look for solar-like oscillations in the RGB stars for which we have determined $\log g$ and $T_\mathrm{eff}$ through our spectroscopic analysis. Solar-like oscillations are standing acoustic waves stochastically driven by surface convection and are expected to be present in all cool stars with convective envelopes \citep{book:aerts2010}. The reason why solar-like oscillations are interesting in the context of stellar clusters is that the oscillations a star display are related to the physical properties of the star and are thus independent of distance, extinction, and chance alignment in space velocity making them a valuable tool for cluster membership determination \citep[e.g., as for NGC~6791, NGC~6819, and NGC~6811 in][]{art:stello2011}. Furthermore, the global seismic parameters, namely the frequency of maximum oscillation power, $\nu_\mathrm{max}$, and average large frequency separation, $\Delta \nu$, have been shown to scale with the mass and luminosity of the star \citep{art:kjeldsen1995} meaning that these quantities can be inferred without invoking modelling of the stellar interior. These so-called asteroseismic scaling relations are, however, derived empirically necessitating thorough testing of their accuracy. The only way to test the seismically inferred masses is to compare them to model-independent masses derived from DEBs. This can be done in star clusters, where masses derived from DEBs in the turn-off region can be extrapolated to the RGB and the red clump \citep[e.g.,][]{art:brogaard2012,art:brogaard2015,art:brogaard2016,art:handberg2017}. 

Although it should be possible to detect solar-like oscillations in the 30~min. cadence TESS FFIs for RGB stars \citep[e.g.,][]{art:campante2017} at a magnitude of $y \sim 13.6$~mag these stars are, unfortunately, too faint. The amplitude would therefore not exceed the noise level \citep{art:huber2011,art:tasoc} and indeed we found no evidence for solar-like oscillations in the RGB stars from the TESS FFIs.

For the classical pulsators, i.e., the $\delta$~Scuti and $\gamma$~Dor stars, for which amplitudes in general are expected to be much higher \citep[e.g.,][]{art:uytterhoeven2011} we detect clear evidence for pulsations. In fact we detected clear pulsation signals for all the $\delta$~Scuti stars reported in \citet{art:arentoft2007} as well for roughly half of the $\gamma$~Dor stars. The $\gamma$~Dor stars for which we did not detect a clear signal are mostly located towards the center of the cluster where the light is highly blended. In Table~\ref{tab:pulse} we list the frequency of maximum power, $\nu_\mathrm{max}$, as well as the corresponding number of cycles per day for these. Light curves and power spectra can be found in Figures~\ref{fig:dsstars} and \ref{fig:gdstars}.

\begin{figure}
    \centering
    \includegraphics[width=\columnwidth]{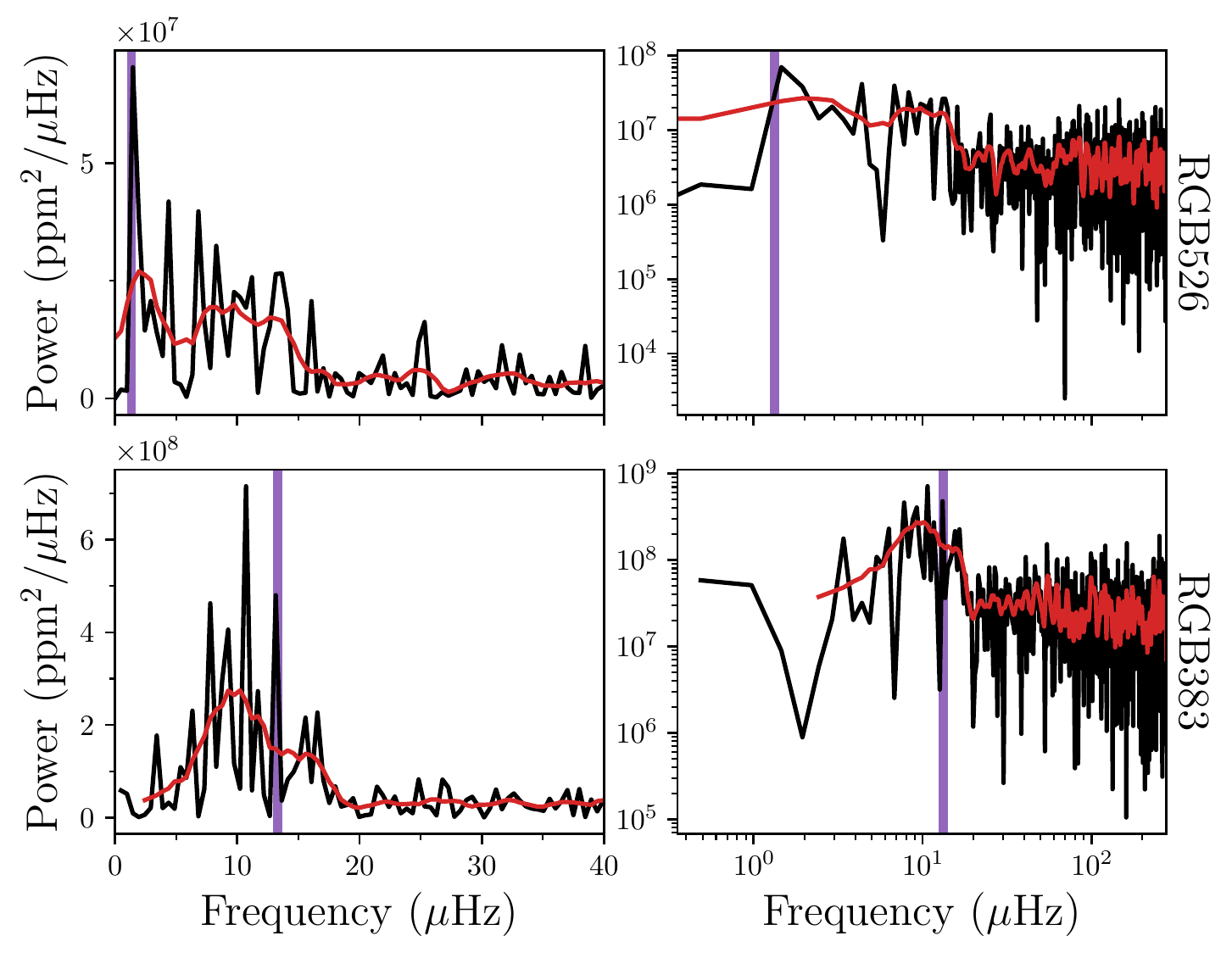}
    \caption{Power spectra for the two RGB stars marked with purple stars in Figure~\ref{fig:cmd} both in linear (left) and log-log plots (right). The black lines are the raw spectra and the smoothed spectra are shown in red. Top panels: The most luminous confirmed member, RGB526, of the cluster shows an excess of power at very low frequencies. Bottom panels: The third most luminous confirmed member, RGB383, shows a clear excess of power. The purple vertical lines denote $\nu_\mathrm{max}$ inferred from extracting $\log g$ and $T_\mathrm{eff}$ from an isochrone fitted to the CMD.}
    \label{fig:osc}
\end{figure}

As mentioned our spectroscopic RGB stars are too faint to detect solar-like oscillations using the TESS data. We therefore turned towards the more luminous part of the CMD and looked for solar-like oscillations in all the confirmed members brighter than the aforementioned RGB stars. In the power spectra for two of the stars we saw an excess of power close to their expected $\nu_\mathrm{max}$. The expected value for $\nu_\mathrm{max}$ is calculated by extracting stellar parameters from the isochrones in Figure~\ref{fig:G_sG} close to the stars' position in the CMD. These power spectra are displayed in Figure~\ref{fig:osc}. For the brighter of the two stars, RGB526, the expected as well as the observed $\nu_\mathrm{max}$ were at a very low frequency, which makes it difficult to assess the validity of this signal. We therefore report this as an indication for solar-like oscillations in this star. However, for RGB383 for which the observed and expected $\nu_\mathrm{max}$ is at a higher frequency, we were much more convinced that what can be seen are solar-like oscillations. If this is in fact solar-like oscillations, this would (to our knowledge) be the first detection of solar-like oscillations in a cluster observed with TESS.

\section{Orbital analysis: masses and radii}\label{sec:orb.anal}

The orbital analysis of V2032 and V5 was done differently from V4, given the difficulties arising from the probable third companion. To obtain masses and radii of V2032 and V5 we used the program {\tt ellc} \citep{art:maxted2016} to fit the light curves and the radial velocities. To obtain reliable estimates of the uncertainties we again used the program {\tt emcee} \citep{pack:foremanmackey2013} to do an MCMC sampling.

\begin{figure*}
    \centering
    \includegraphics[width=\textwidth]{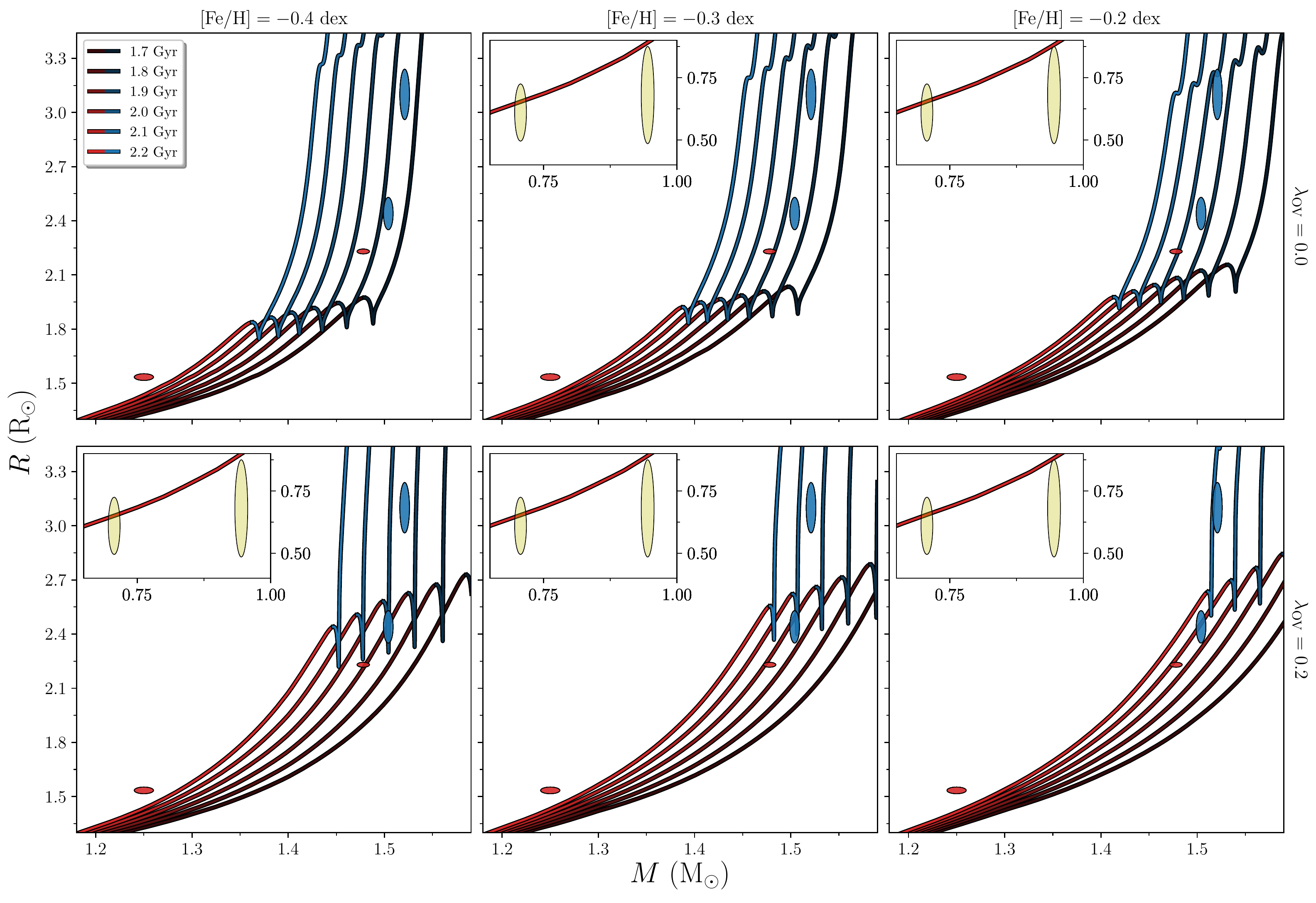}
    \caption{MR diagrams for the components in V2032, V4, and V5 marked with blue, red, and yellow 1$\sigma$ ellipses, respectively. The red (before the terminal age main sequence) and blue (after) coloured lines are BaSTI isochrones at different ages. Columns separate the isochrones in metallicity and rows are for different assumptions on model physics, where the isochrones in the bottom row take convective core-overshooting into account. Overshooting beyond the Schwarzschild boundary is parametrised in terms of the pressure scale height, $H_P$, as $\lambda_{OV}H_P$, where $\lambda_{OV}$ is set to 0.2 for models that include convective core overshooting. None of the models treat diffusion or mass-loss. $\rm [\alpha/Fe] = 0.0$~dex for all models. 
    }
    \label{fig:MR}
\end{figure*}

\subsection{V2032}\label{sec:V2032}

During our initial modelling of V2032 using the light curves in Figure~\ref{fig:lcs_rvs}, i.e., using the sparse Johnson photometry, it became evident that it was difficult to obtain consistent results for the radii between the two filters. We therefore also used the observations from TESS in Figure~\ref{fig:tesslc}, which covers both ingress and egress of the primary eclipse, to obtain estimates for the radii. As mentioned the light curve in Figure~\ref{fig:tesslc} is from a blended signal (not from a companion to the binary, but from the nearby sources entering the large pixels), which causes a decrease in the depth of the eclipse. We model this by including a contribution from a third (multiple) light(s) in {\tt ellc} as $\mathcal{F}^\mathrm{c} = l^\mathrm{c} (\mathcal{F}^\mathrm{p} + \mathcal{F}^\mathrm{s})$ with $\mathcal{F}^{p,s}$ being the flux from the primary or the secondary component \citep[see][]{art:maxted2016}. We estimated the contribution factor, $l^\mathrm{c}$, by comparing the difference in magnitude during an eclipse in the NOT data (Figure~\ref{fig:lcs_rvs}) to the fractional change in flux in the TESS data (Figure~\ref{fig:tesslc}). We found a value of $l^\mathrm{c}=7.6$ and we therefore adopted a Gaussian prior with this value and a width of $0.05$ for this parameter during our MCMC run of the TESS light curve.

Seeing as we do not cover ingress in the light curves of V2032 in the ground-based observations, it is somewhat difficult to constrain the semi-major axis, $a$. However, the orbital parameters derived from our spectroscopic measurements in Table~\ref{tab:sbop} constrain the product of the semi-major axis and inclination, $i$, through

\begin{equation}
    a \sin i = \frac{P (1-e^2)^{1/2}}{2 \pi} (K^\mathrm{p} + K^\mathrm{s}) \, .
    \label{eq:asini}
\end{equation}
We therefore used a Gaussian prior -- in the sense $\mathcal{N}(\mu = \mu (a \sin i),\sigma = \sigma (a \sin i))$ -- for this product in all cases ($V$, $I$, and TESS) created by drawing normally distributed samples from the parameters calculated by \sbop. Furthermore, we also incorporated Gaussian priors on the luminosity ratio of $0.89 \pm 0.02$ for $V$ and $0.84 \pm 0.02$ for $I$ and TESS (from the BF in Section~\ref{sec:spec.lrat}). We used our estimates of the effective temperatures in Table~\ref{tab:key} to estimate the surface brightness ratio, $J$. This was done by drawing normally distributed temperatures from these values, create corresponding Planck curves, which we multiplied by the filter transmission curves in $V$ and $I$, respectively, and take the ratio between the curves resulting from each star in a system to obtain values of $1.38 \pm 0.11$ and $1.25 \pm 0.07$. These values constituted our Gaussian priors for $J$, where we for each temperature draw then calculated $J$ in the same way. We used the same value for $I$ in the TESS fit due to the similarity in the passbands. The reason for adopting these constraints is that the light curves alone are not informative enough to yield fully consistent results.

For all light curves we adopted a quadratic limb darkening law with coefficients estimated using $\log g=3.7$~dex, and $\rm [Fe/H] = -0.3$~dex for both stars and $T_\mathrm{eff}^{\rm p}=6600$~K and $T_\mathrm{eff}^{\rm s}=7100$~K. We used $\xi = 2$~km/s for the micro turbulence. The linear, $c_1$, and quadratic, $c_2$, limb darkening coefficients were found from tables by \citet{art:claret2000,art:claret2016} for the Johnson and TESS filters, respectively, on which we placed Gaussian priors. We ran all our MCMCs with 100 walkers and for each of these we drew 20,000 times and applied a burn-in of 10,000, i.e., we rejected the first 10,000 steps of each walker. In Table~\ref{tab:key} we display our final results for the masses and radii of the components, which we have created by drawing from the posteriors of our MCMC for each passband in Table~\ref{tab:V2032} and created a joint posterior. 

\begin{table}
	\centering
    \caption{Key stellar parameters for the DEBs. The values for the masses and radii of V2032 and V5 are the medians and the uncertainties are from the highest posterior density (HPD) interval at a level of 68\% for V2032 and V5. The results for V4 are from our DE-MCMC analysis (see Section~\ref{sec:3V4}). The effective temperatures for the individual components of V2032 and V4 are calculated from the spectral energy distributions in Section~\ref{sec:sed}.}
    \begin{tabular}{c c c c}
    \toprule
    & V2032 & V4 & V5 \\
    \midrule
    $M^\mathrm{p}~(\mathrm{M}_\odot)$ & $1.521 \pm 0.005$ & $1.478^{+0.006}_{-0.007}$ & $0.945 \pm 0.012$ \\
    $M^\mathrm{s}~(\mathrm{M}_\odot)$ & $1.504 \pm 0.005$ & $1.250 \pm 0.010$ & $0.707^{+0.013}_{-0.009}$ \\
    $R^\mathrm{p}~(\mathrm{R}_\odot)$ & $3.10^{+0.07}_{-0.20}$ & $2.300^{+0.013}_{-0.014}$ & $0.68^{+0.22}_{-0.15}$ \\
    $R^\mathrm{s}~(\mathrm{R}_\odot)$ & $2.44^{+0.07}_{-0.10}$ & $1.534^{+0.019}_{-0.018}$ & $0.61^{+0.17}_{-0.06}$ \\ \hdashline
	$T_\mathrm{eff}^\mathrm{p}$~(K) & $6560 \pm 100$ & $6830 \pm 100$ & $5700 \pm 400$ \\
    $T_\mathrm{eff}^\mathrm{s}$~(K) & $7100 \pm 100$ & $6830 \pm 100$ & $4940^{+340}_{-190}$ \\
	\bottomrule
    \end{tabular}

	\label{tab:key}
\end{table}

\subsection{V5}\label{sec:V5}

For the modelling of V5 we employed the same strategy as for V2032 by using a prior on $a \sin i$ from Equation~\ref{eq:asini} and we used a Gaussian prior for the luminosity ratio, where we found $L^\mathrm{s}/L^\mathrm{p}=0.36 \pm 0.03$ (stemming from the BF in Section~\ref{sec:spec.lrat}). Again we adopted a quadratic limb darkening law using coefficients from the table in \citet{art:claret2000} and included them with Gaussian priors. As before the values and uncertainties are listed in Table~\ref{tab:key}. To get an estimate of the temperatures we drew uniformly distributed temperatures for both components, $\mathcal{U}(a,b)$ with $a=4200$~K and $b=6200$~K, which we then translated into a surface brightness ratio in $B$ again using a filter transmission curve. The results for V5 are summarised in Table~\ref{tab:V2032} with key parameters in Table~\ref{tab:key}.

\subsection{V4}\label{sec:V4}
As mentioned we strongly suspect a third body to be present in V4, which causes the shift we see in the eclipse times in Figure~\ref{fig:lcs_rvs}. We therefore dealt with this system in a manner different to that for V2032 and V5. First we tried dividing the data into different intervals in time so that we only used spectroscopic and photometric data obtained within a relatively short time of each other in our fits. This was done by combining photometric data from the Danish 1.54-metre and the Mercator telescope (Table~\ref{tab:phot}) with spectroscopic data from UVES only (Table~\ref{tab:spec.V4}), as well as a fit using the same photometric data with the inclusion of spectroscopic data from GIRAFFE. We also tried combinations that included all the spectroscopic data, but only included the photometry from IAC-80, LCOGT, and NOT as well as one that excluded the photometric data from NOT. All of these fits were performed using \jktebop \citep{art:southworth2013} and we invoked the constraints on the luminosity ratio of $0.40 \pm 0.02$, $0.39 \pm 0.02$ and $0.40 \pm 0.02$ for the fits using data in $V$, $I$, and $B$, respectively.

The reason for carrying out all of these different fits is that we wanted to see how consistent our results would be if we ignored the ETVs and treated the system as only being comprised of two bodies. We prefer the solutions that utilize as much of the data as possible, but still avoid including data with variations in the eclipse times. Therefore we report the results for two of the aforementioned fits that both made use of all the spectroscopic data; the one that only includes the newer photometry, i.e., from the IAC-80, LCOGT, and NOT, and the one using the older photometric data from the Danish 1.54-metre and the Mercator. The results for the masses and radii for these five different runs can be found in Table~\ref{tab:V4_johnson}. Our results here are in reasonable agreement, but they are not completely consistent and it would therefore be interesting to see what the consequences of not just treating the outer companion as a nuisance would be. 

\subsubsection{Three-body solution for V4}\label{sec:3V4}

Therefore, we did a full three-body solution of the system following the approach in \citet{art:orosz2019} using the ELC code \citep{art:orosz2000} to model the light and velocity curves. To sample the parameter space we used the Differential Evolution MCMC (DE-MCMC) algorithm \citep{art:terbraak2006}. Our first runs resulted in a radius for the secondary component that was significantly larger ($R^\mathrm{s} \sim 1.74$~$\mathrm{R}_\odot$) than that from our \jktebop runs.

It is not unusual to have an inflated secondary component in close-in binaries \citep[e.g.,][]{art:sandquist2016,art:brewer2016}, which can be explained by magnetic activity inhibiting convection. Given the smaller mass of the secondary component, it has a larger convective envelope, which generates strong magnetic fields. In turn these magnetic fields slow down the convective motion and thus make convection less effective. As a result the star has to expand to radiate away the excess heat that can not be transported by the inefficient convection, leading to radii increased by as much as 10\% above the expected theoretical value \citep{art:torres2006}. Radius inflation due to convective inhibition could therefore play a role in the secondary component of V4, however, it does not explain the discrepancy between the results presented above and those stemming from the three-body fits.

We were able to identify that the discrepancy between the results were caused by the limb darkening coefficients. In our \jktebop runs these were fixed, which underestimates systematic errors, whereas in our DE-MCMC runs we sampled for these coefficients using the formulation in \citet{art:kipping2013}, but with the result that they would wander into a physically unrealistic territory. We therefore made a range for the coefficients to sample from, limited by the values we found for $\log g$ ($\pm 0.05$~dex) and $T_{\rm eff}$ ($\pm 100$~K) in our previous runs and for $\rm [Fe/H]$ ($\pm 0.1$~dex) based on our analysis of the RGB stars. Again we used values from \citet{art:claret2000,art:claret2016} and invoked a constraint on the luminosity ratio of $L^\mathrm{s}/L^\mathrm{p}=0.40 \pm 0.02$. 

The results for the masses and radii from the DE-MCMC were $M^\mathrm{p}=1.478^{0.006}_{-0.007}$~$\mathrm{M}_\odot$, $M^\mathrm{s}=1.250 \pm 0.010$~$\mathrm{M}_\odot$, $R^\mathrm{p}=2.300^{+0.013}_{-0.014}$~$\mathrm{R}_\odot$, and $R^\mathrm{s}=1.534^{+0.019}_{-0.018}$~$\mathrm{R}_\odot$ for the primary and secondary component. Evidently, the secondary component is still slightly inflated compared to the results from \jktebop and compared to the theoretical models in Figure~\ref{fig:MR}, but overall the results are in much better agreement. Our final results for the masses and radii for the primary and secondary components of V4 are listed in Table~\ref{tab:key}. All other parameters form the fit can be found in Table~\ref{tab:V4_tess}.

\subsubsection{The outer companion in V4}\label{sec:V4t}
Our models suggest that the body orbiting the inner binary is in an eccentric ($e\sim0.5$) $443$~d orbit. From our modelling the mass of the third component is fairly well-determined, but as we have very little information of the radius, we are only able to place an upper limit of the amount of light this third body contributes to the system. This amounts to some 2\% of the total light. Given the mass suggested by our models, this can come about by having a very hot, compact object, i.e., a white dwarf, but it is also consistent with a main-sequence star similar to the components of V5. Therefore, for a given solution we imposed a $\chi^2$ penalty if the mass and radius of the third star fell outside the region in the mass-radius plane defined by BaSTI \citep[a Bag of Stellar Tracks and Isochrones;][]{art:hidalgo2018} isochrones for a main-sequence star. From this we find the mass to be $M^\mathrm{t}=0.74 \pm 0.03$~$\mathrm{M}_\odot$ and if the star were to be a (well-behaved) main-sequence star its radius would be similar to that of the components in V5.



\section{Cluster parameters}\label{sec:clpar}
To obtain cluster parameters for NGC~2506 we used the newly updated BaSTI isochrones. We compare these models to the masses and radii of the DEBs, the observed cluster sequence in Str\"omgren photometry, and the properties we derived for the spectroscopic RGB stars as well as the observed properties of the RGB stars potentially displaying solar-like oscillations.

\subsection{Mass-radius diagrams}\label{sec:mrdia}
In Figure~\ref{fig:MR} we compare our measurements of the masses and radii of the 6 stars in V2032, V4, and V5 listed in Table~\ref{tab:key} to the BaSTI isochrones. The models in the top row do not include convective core-overshooting, whereas the models in the bottom row do. We have colour-coded the isochrones so that blue corresponds to stars found after the terminal age main sequence (TAMS), where the components in V2032 are most likely found, and red denotes stars before the TAMS. None of the models treat atomic diffusion or mass loss (see \citet{art:hidalgo2018} for details regarding the input physics).

\begin{table}
	\centering
    \caption{Cluster parameters for NGC~2506. The age is determined from the binaries in Section~\ref{sec:mrdia}. The metallicity and $\alpha$-enhancement are based on the RGB stars in Section~\ref{sec:spec.rgb}, where we have calculated a weighted average and then added the systematic uncertainties ($0.1$~dex) in quadrature. Again using these stars we estimated the reddening in Section~\ref{sec:red}. The distance is estimated from the {\it Gaia} data in Section~\ref{sec:gaia}.}
    \begin{threeparttable}
    \begin{tabular}{c c}
    \toprule
    \multicolumn{2}{c}{NGC~2506}   \\
    \midrule
    $t$ & $2.01 \pm 0.10$~Gyr \\
    $\rm [Fe/H]$ & $-0.36 \pm 0.10$~dex \\
    $\rm [\alpha/Fe]$ & $0.10 \pm 0.10$~dex \\
    $\rm [M/H]$\tnote{\textdagger} & $-0.29 \pm 0.12$~dex \\
    $r$ & $3.101 \pm 0.017$~kpc \\
    E$(b-y)$ & $0.057 \pm 0.004$~mag \\
    E$(B-V)$ & $0.080^{+0.005}_{-0.006}$~mag \\
	\bottomrule
    \end{tabular}
	\begin{tablenotes}
	    \item[\textdagger] From Equation~\eqref{eq:met}.
	\end{tablenotes}
    \end{threeparttable}
    \label{tab:clust}
\end{table}

\begin{figure*}
    \centering
    \includegraphics[width=\textwidth]{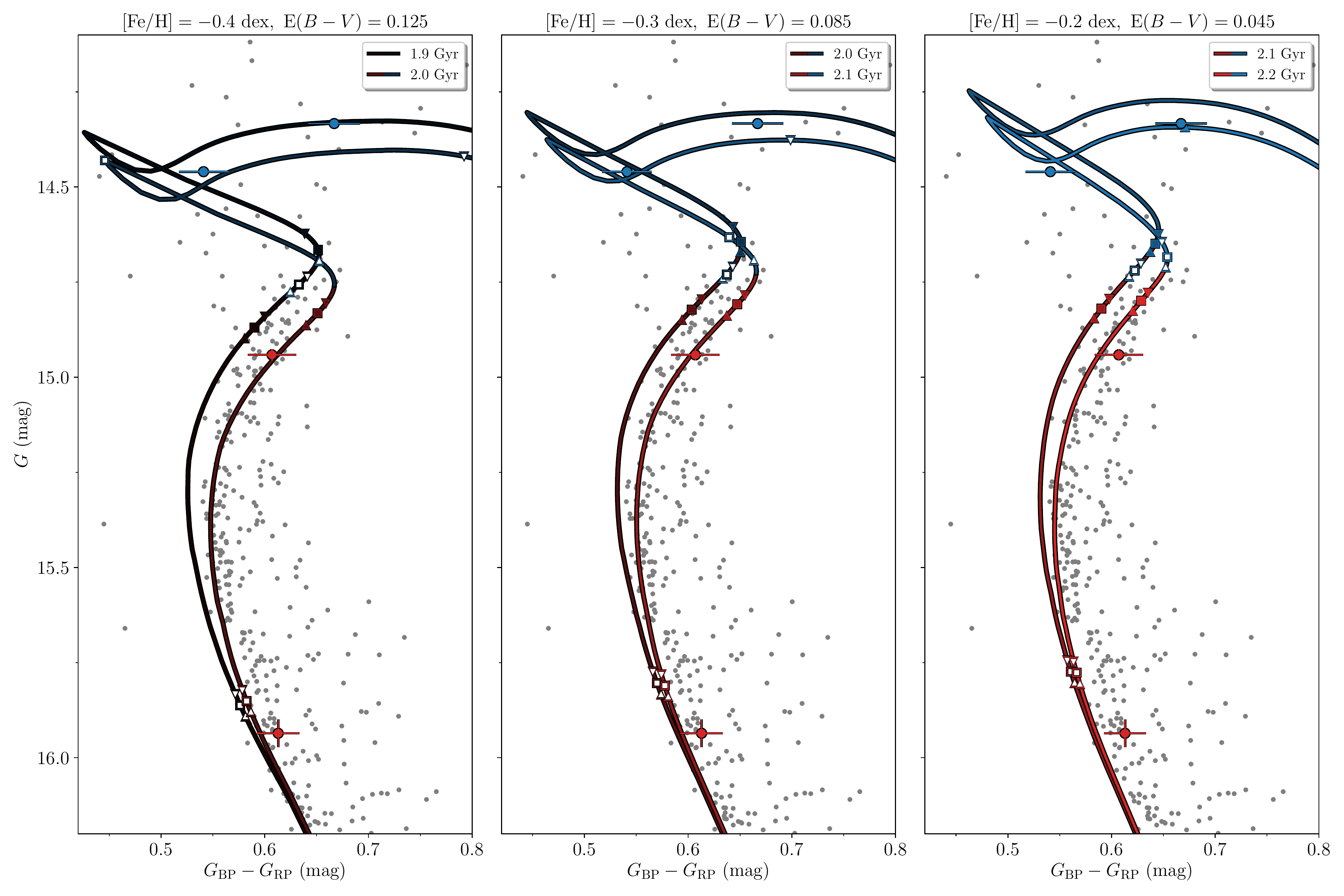}
    \caption{The CMD of NGC~2506 in \gaia colours compared to BaSTI isochrones at different metallicities and assuming different values for E$(B-V)$ and consequently slightly different values for $\mu$ for each metallicity. The colour coding for the isochrones is the same as in Figure~\ref{fig:MR}. The blue and red dots with errorbars are the components of V2032 and V4, respectively, which have been decomposed by firstly calculating the $G$ magnitudes based on the luminosity ratios in $V$ of $0.89 \pm 0.02$ respectively $0.40 \pm 0.02$. The colour for each star has been calculated from the bolometric corrections from \citet{art:casagrande2018a,art:casagrande2018b} using the radii and effective temperatures in Table~\ref{tab:key} and the colour excess from Table~\ref{tab:clust}. The squares denote interpolated values for the masses in the isochrones closest to those in Table~\ref{tab:key} and the upwards (downwards) facing triangles mark the lower (upper) 1$\sigma$ level. For the secondary components in both V2032 and V4 we have added white markers to distinguish these points from those corresponding to the primary components. For the isochrones where the interpolated mass corresponds to the observed evolutionary stage of V2032, we match them to these values (to the extent possible). For isochrones where this is not possible we match them to the primary component of V4. Note that for some isochrones the markers for the interpolated values for the primary and secondary components are not visible. They are either towards or on the RGB.}
    \label{fig:cmdiso}
\end{figure*}

Our analysis of the RGB stars suggested that the metallicity or more precisely the iron abundance is around $-0.40$~dex and with a value of $\rm [\alpha/Fe]=0.10$~dex, but since the isochrone grid we used does not include $\alpha$-enhanced isochrones, we accounted for this by making use of the formula for the actual metallicity in \citet{art:sharma2019}

\begin{equation}
    \mathrm{[M/H]} = \mathrm{[Fe/H]} + \log (0.694 \cdot 10^{\rm [\alpha/Fe]}  + 0.306) \, ,
    \label{eq:met}
\end{equation}
which was originally formulated by \citet{book:salaris2005}. In the present case the metallicity would be $\rm [M/H]=-0.29 \pm 0.12$~dex. We therefore used isochrones with an iron abundance close to this value to infer the age of the cluster, i.e., the middle panels in Figure~\ref{fig:MR}. Evidently, the inclusion of convective core-overshooting has significant impact on the evolutionary stage of the secondary component in V2032 and the primary component of V4. In the non-overshoot scenario the primary component of V4 is found at a post main-sequence evolutionary stage, but clearly the CMD in Figure~\ref{fig:cmdiso} suggests that the component is still on the main-sequence and therefore models including overshoot should be favoured.

Given that the stars in V2032 are at such an auspicious phase (as well as considering the difficulties for the radius of the secondary component in V4 and given the less informative stage of the components in V5) our age estimate is mostly hinged on this system and the primary component of V4. It is clear that these three components completely lock the isochrones, allowing for extremely precise age determination. It is also clear that if both components of V2032 are found after the TAMS, a smaller value than 0.2 is needed for $\lambda_{\rm OV}$, and as such V2032 and V4 can be used to not only distinguish between models with and without overshoot, but also assess the amount of overshoot needed quite precisely. However, the BaSTI isochrones only have the two options, 0.0 or 0.2.

Our age estimate is based on the isochrones in Figure~\ref{fig:MR} with a metallicity of $-0.3$~dex and which include convective core-overshooting. From these we estimate the age of the cluster to be $t=2.01 \pm 0.10$~Gyr, where the main source of error comes from the uncertainty of $0.1$~dex on $[\mathrm{Fe/H}]$. As argued the value for $\lambda_{\rm OV}$ should probably be a bit lower than 0.2 to bring the primary component of V4 and both components of V2032 to lie on the same isochrone. A crude estimate of how much smaller $\lambda_{\rm OV}$ should be is to consider the hooks on the isochrones in the middle panels ($[\mathrm{Fe/H}]=-0.3$~dex) of Figure~\ref{fig:MR}. The hook in the lower panel ($\lambda_{\rm OV}=0.0$) should be decreased by around $0.1$~R$_\odot$ to capture all three stars and the difference between the hook in the top panel and the bottom is about $0.5$~R$_\odot$, which means $\lambda_{\rm OV}$ should be decreased by about 20\%, i.e., to a value of around 0.16. There is roughly a $0.2$~Gyr difference in the age estimate between the two middle panels meaning that a change of 20\% in $\lambda_{\rm OV}$ would make the cluster around $0.04$~Gyr younger. The best age estimate of the cluster with core-overshoot adjusted to match both the primary star of V4 and both components of V2032 would thus be 2.01~Gyr, since the best fitting isochrone without such a correction (in the lower middle panel of Figure~\ref{fig:MR}) is 2.05~Gyr. Note the error for the age is the internal error and as such does not include deficiencies in the stellar models.

\subsection{The observed cluster sequence}
From the MR diagrams in Figure~\ref{fig:MR} it was clear that the primary component of V2032 should be found at a phase of rapid expansion and cooling. However, it is not as clear whether the secondary component is also found at this phase. Taking isochrones with $\rm [Fe/H]= -0.3$~dex as the ones most representative of the cluster metallicity, Figure~\ref{fig:MR} shows that the primary component of V2032 is definitely at a stage of rapid expansion, regardless of whether convective core-overshooting is included or not. The secondary could be located on either side of the TAMS depending on the inclusion of overshooting and also the value used for $\lambda_{\rm OV}$, even for the value available in the grid the secondary component could still be located before or after TAMS.

In the CMD in Figure~\ref{fig:cmdiso} we show the \gaia proper motion members (see Section~\ref{sec:gaia}) compared to the BaSTI isochrones, where we for each metallicity only show the two ages that best capture the components of V2032 and the primary component in V4 in Figure~\ref{fig:MR}. In the CMD we have decomposed the light from the binaries V2032 and V4. This was done by using the luminosity ratios in $V$ of $0.89 \pm 0.02$ and $0.40 \pm 0.02$, respectively, with the observed $G$ magnitudes and translate this into a $G$ magnitude for each component. The colours were calculated from the surface gravities (from the radii and masses) and effective temperatures in Table~\ref{tab:key} and the reddening and metallicity in Table~\ref{tab:clust} from which we calculated the bolometric corrections, ${\rm BC}_{G_{\rm RP}} - {\rm BC}_{G_{\rm BP}}= G_{\rm BP} - G_{\rm RP}$, from \citet{art:casagrande2018a,art:casagrande2018b}. The errors were created from drawing normally distributed values 500 times for each parameter that enters, then calculating the magnitude and colour, and subsequently measuring the spread of the resulting distributions. In Figure~\ref{fig:cmdiso} these are shown as blue and red dots with errorbars for V2032 and V4, respectively. As argued the primary component of V4 is clearly found on the main sequence in the CMD, which from the MR diagrams is only consistent with the inclusion of overshooting. Thus, we only consider those isochrones here. Adding to this is that the isochrones without overshooting clearly diverged from the observed cluster sequence.

We used the radii and effective temperatures for V2032 and V4 in Table~\ref{tab:key} to calculate the distance to the cluster. This was done by first calculating the total luminosity of the system, $L^{\rm tot}=L^{\rm p} + L^{\rm s}$, translating that to an absolute magnitude, M$_V$, to get the distance modulus, $\mu = {\rm m}_V - {\rm M}_V$, while again accounting for the extinction, A$_V$. We did a Monte Carlo simulation with 5,000 draws, where in each draw we drew normally distributed values (as in Section~\ref{sec:V2032}) for the effective temperatures, radii, apparent $V$-magnitude, and reddening. The resulting values for the distance was $2.92 \pm 0.12$~kpc and $3.17 \pm 0.08$~kpc for V2032 and V4, respectively.

We calculated and applied the true distance modulus, $\mu = 5 \log r - 5 + \mathrm{A}_V$ with $\mathrm{A}_V = 3.1 \cdot$E$(B-V)$ being the interstellar absorption and $r=3.04$~kpc being the mean of the values for the distance calculated from V2032 and V4. For each pair of isochrones in Figure~\ref{fig:cmdiso} we assumed values for E$(B-V)$ of 0.125, 0.085, and 0.045 and values for the metallicity of $-0.4$~dex, $-0.3$~dex, and $-0.2$~dex, respectively. On each of the isochrones we have highlighted the interpolated mass from Table~\ref{tab:key} with blue squares for the components in V2032 and red for those in V4. The upwards facing triangles denote the 1$\sigma$ lower limit and the downwards facing triangles mark the upper limit. To make it easier to distinguish between the components we have added smaller white markers on top of the symbols for both of the secondary components. 


\section{\gaia distance to the cluster}\label{sec:gaia}

With the \gaia Data Release 2 \citep[\gaia DR2;][]{art:gaia2018} data we can estimate the distance to the cluster with great precision. However, estimating the distance, $r$, to the cluster is not as simple as taking the inverse of the parallax, i.e., $r = 1 / \varpi$. This is because the measured parallax can be zero or even negative, while the distance is, of course, constrained to be positive \citep{art:luri2018}. Furthermore, the distance has a non-linear relationship to the measurement $1 / \varpi _\mathrm{True}$. To resolve this we therefore follow the approach recommended by \citet{art:luri2018}, which is to treat this as a Bayesian inference problem. 

\begin{figure}
    \centering
    \includegraphics[width=\columnwidth]{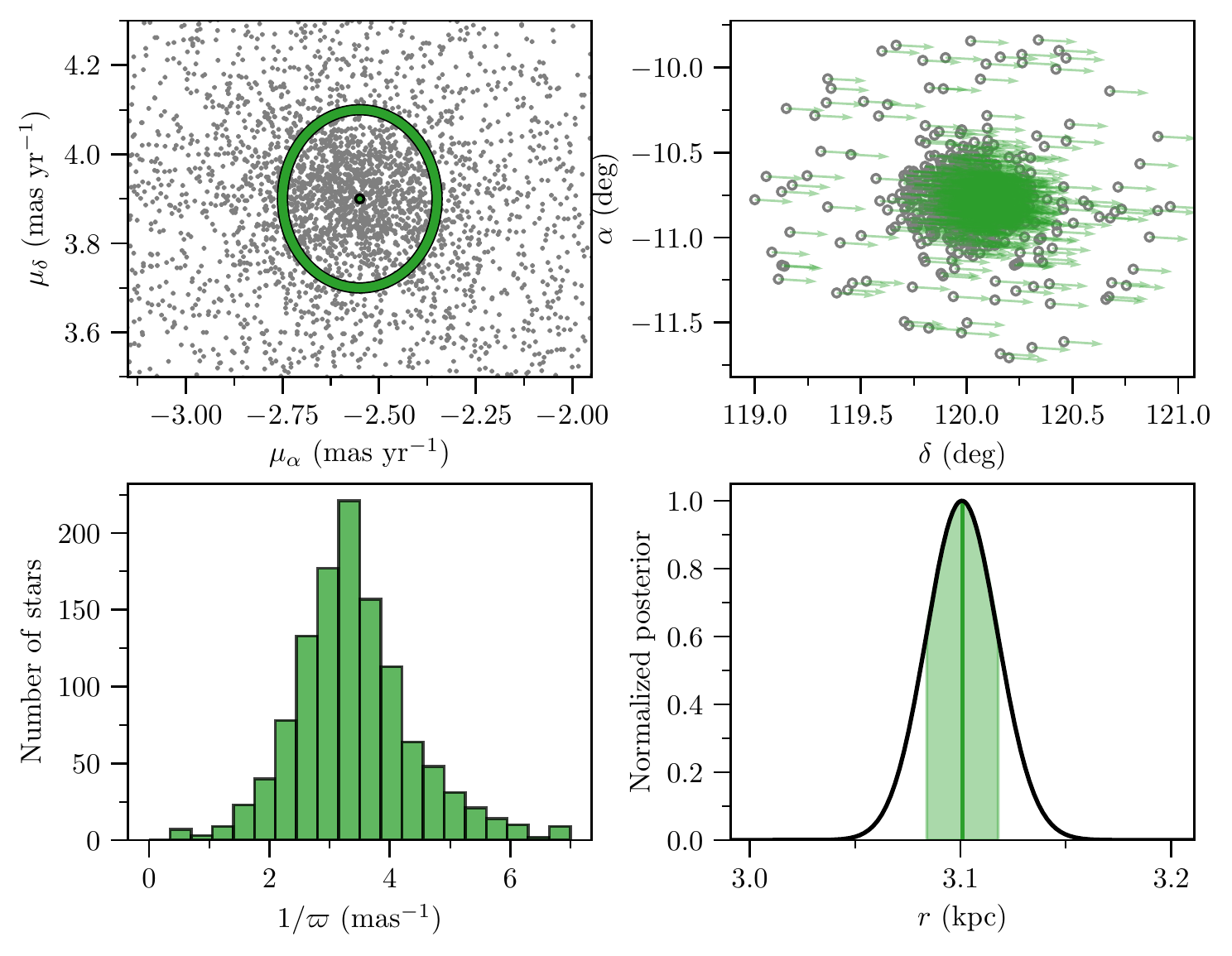}
    \caption{Top left: Stars in NGC~2506 as seen in proper motion space by {\it Gaia}, where the ring marks the stars included in the distance estimate. Top right: Stars in $\alpha,\delta$ with their proper motion vectors drawn (scaled for clarity). Bottom left: Histogram of $1/\varpi$ for the stars used in the distance estimate. Bottom right: The resulting posterior from Equation~\eqref{eq:post}.}
    \label{fig:gaia_dist}
\end{figure}

Firstly, we selected stars within a 1 degree radius of the cluster. We then located the cluster in proper motion space as shown in the top left corner of Figure~\ref{fig:gaia_dist}. Here we located the densest region, which should correspond to NGC~2506, and deemed stars within $0.2$~mas/yr of the center of this dense region to be members of NGC~2506 (as a sanity check we plot the selected stars in $\alpha,\delta$ in the top right corner with their proper motion vector scaled for clarity). From this sample we only included the stars with relatively well-determined parallaxes, i.e., $|\sigma_\varpi / \varpi| < 0.25$. These are displayed in the histogram of Figure~\ref{fig:gaia_dist}.

We adopt the exponentially decreasing space density prior in distance

\begin{equation}
    P (r|L) = 
\begin{cases}
\frac{1}{2 L^3} r^2 \exp (-r/L) & \text{if } r > 0\\
0 & \text{otherwise,}
\end{cases}
\label{eq:prior}
\end{equation}
where $L$ is a length scale to the cluster set to $3.55$~kpc \citep{art:twarog2016}. We estimate the likelihood as 

\begin{align}\label{eq:post}
    P(r_i | \{\varpi\},\{\sigma_\varpi\},L) =& \prod_{n=1}^{N} \int \frac{1}{\sqrt{2 \pi}\sigma_{\varpi_n}} \\\notag 
    & \times \exp \left [- \frac{(\varpi_n - \varpi_{\rm zp} - 1/r_i)}{2 \sigma_{\varpi_n}^2} \right]  \mathrm{d} r_i \, ,
\end{align}
where the subscript $n$ refers to the parallax and uncertainty in parallax of the $n$th star in the histogram of Figure~\ref{fig:gaia_dist} and $r_i$ is the proposed distance to the cluster, i.e., we created linearly spaced values for $r$ in the range 2 to 4.5~kpc. $\varpi_{\rm zp}$ is the global offset in parallax of $-0.029$~mas reported in \citet{art:bailer2018}, which we adopt. Here we have assumed that all $N$ parallax measurements are independent and exploited that the angular extent of the cluster is small. The resulting posterior can be seen in the lower right panel of Figure~\ref{fig:gaia_dist}, where we have displayed our result. The distance we found was $r = 3.101 \pm 0.017$~kpc. This value is in good agreement with the value of $3.04$~kpc we obtained from the binaries and in excellent agreement with the value of $3.112$~kpc reported in \citet{art:cantat2018}. Omitting the offset from \citet{art:bailer2018} in our analysis resulted in a distance of $r = 3.41 \pm 0.02$~kpc. An offset of around $-0.05$~mas was reported in \citet{art:khan2019} when comparing the \gaia distances to stars in the {\it Kepler} field with distances determined using asteroseismology. This means that in addition to the statistical error of 0.017~kpc that we report there is potentially a systematic error, which is significantly larger.

\begin{figure}
    \centering
    \includegraphics[width=\columnwidth]{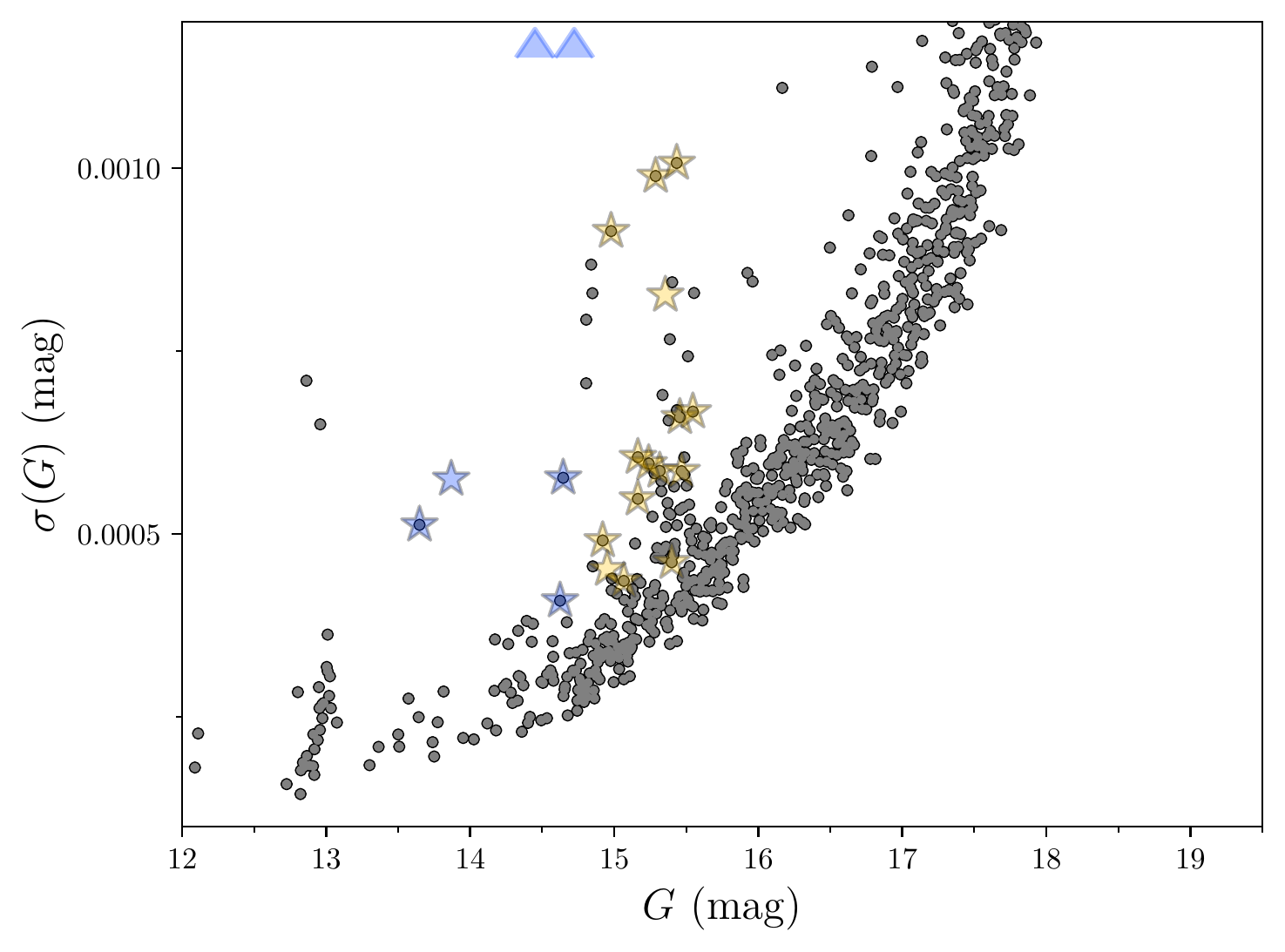}
    \caption{Cluster members (Section~\ref{sec:rv_members}) with the uncertainty on their {\it Gaia} $G$ magnitude against their $G$ magnitude. The $\delta$~Scuti and $\gamma$~Dor stars from \citet{art:arentoft2007} have been marked with respectively blue and yellow stars, but here they are transparent to make the underlying spread visible, which suggests that there are more of these type of stars in the cluster. The upwards pointing triangles denote the position of two $\delta$~Scuti stars at $(G,\sigma(G))=(14.722,0.002)$ and $(G,\sigma(G))=(14.450,0.003)$.}
    \label{fig:G_sG}
\end{figure}

Finally, we note that the {\it Gaia} data can be used to identify potential $\delta$~Scuti and $\gamma$~Dor stars in clusters. This is done by first identifying cluster members as in Section~\ref{sec:sed} and Figure~\ref{fig:gaia_dist}, and then by plotting their uncertainty in magnitude against their magnitude. This is shown for the {\it Gaia} $G$ magnitude in Figure~\ref{fig:G_sG}, where a clear spread in magnitude is seen at the place where these stars reside.


\section{Discussion}\label{sec:disc}
NGC~2506 is a very interesting open cluster, harboring a multitude of rare stellar systems. Over the years we have amassed a considerable amount of data for this cluster. Data stemming from many different telescopes and instruments, both ground-based and space-based. The spectroscopic data of the RGB stars allowed us to determine the metallicity of the cluster with high precision. This narrows the parameter space of the possible isochrones to choose from in the MR diagrams as well as in the CMD, enabling us to put a tight constraint on the age of the cluster.

Accurately determining the parameters of a cluster such as NGC~2506 is extremely valuable for several astrophysical reasons. First off, modelling stellar evolution is, of course, relying on having accurately determined parameters for a large number of stars to test against. Secondly, if the power excess seen in Figure~\ref{fig:osc} is indeed due to solar-like oscillations, NGC~2506 could help to test the asteroseismic scaling relations by comparing the results for the dynamically inferred properties from the binaries to those that can be inferred from asteroseismology. Furthermore, NGC~2506 can be used as a benchmark for modelling $\gamma$~Dor and $\delta$~Scuti stars, where again age and metallicity are key parameters, but here we would also have a firm grasp on the masses and radii of these stars. The power spectra for the $\delta$~Scuti stars in Figure~\ref{fig:dsstars} look very convincing in terms of detecting oscillations, whereas the power spectra for the solar-like oscillators in Figure~\ref{fig:osc} and for some of the $\gamma$~Dor stars in Figure~\ref{fig:gdstars} are a bit more dubious. This is why it would be interesting to see what could be achieved with difference imaging specifically designed for clusters in the TESS data \citep[e.g.,][]{art:bouma2019} as this might significantly enhance the signal for the variable stars.

\subsection{V4}\label{sec:disc.V4}

V4 is a testimony to the fact that sometimes acquiring more data can lead to unforeseen challenges and serendipitous discoveries. The exact nature of the third component of V4 is to some extent still uncertain. As mentioned the mass is constrained to be around $0.60$~$\rm M_\odot$, but we really have no constraints on its radius, except that our models suggested that the star should only contribute about 2\% to the total light of the system. Having a body that contributes about 2\% of the total light in the system is consistent with it either being a hot and compact object or a main-sequence star similar to the components of V5. If the third companion is a white dwarf its (final) mass suggests that the initial mass was around $3$~$\rm M_\odot$ \citep[e.g.,][]{art:cummings2018}. Given the cluster age of $2.05$~Gyr, a $3$~$\rm M_\odot$ star would have had sufficient time to evolve into a white dwarf \citep[e.g.,][]{book:kippenhahn2012}. Looking through a table of nearby white dwarfs by \citet{art:giammichele2012} with masses similar to that of the companion and with ages in the range $1.3$-$1.7$~Gyr, we find that if the star is a white dwarf it should have a temperature of around $8,000$~$\rm K$ (or hotter if the white dwarf is younger). This is significantly hotter than the components of the inner binary and could therefore be detected as an excess flux in UV. However, we did not detect such an excess (see Section~\ref{sec:sed}), which is not to say that a white dwarf can be ruled out, but it does speak in favour of the scenario with a V5-like component to the inner binary.

\begin{figure}
    \centering
    \includegraphics[width=\columnwidth]{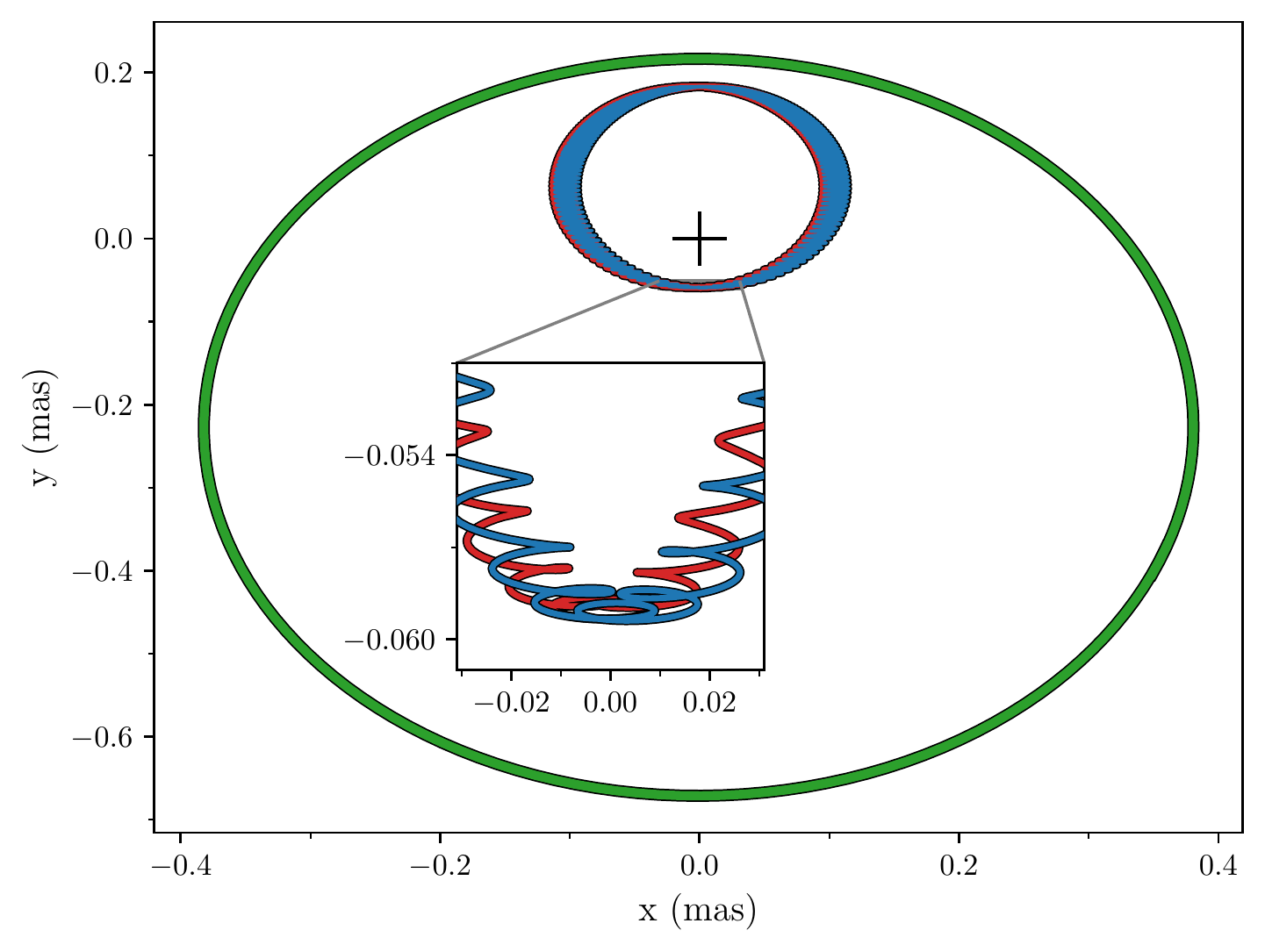}
    \caption{Orbit of the V4 system if the system was viewed face-on. The red and blue lines denote the orbits of the primary and secondary component, respectively, and the green line marks the orbit of the tertiary component. The orbit around the barycenter (black plus) of the inner binary has a diameter of around 0.3 milliarcseconds and should thus be easily detectable with {\it Gaia}. This figure is created using REBOUND \citep[with the IAS15 integrator;][]{art:rein2015} using the parameters in Table~\ref{tab:V4_tess}.}
    \label{fig:V4_orb}
\end{figure}

Regardless of the size of the third companion, it is massive enough to influence the orientation of the orbit of the inner binary. The wobble around the barycenter induced by the orbit of the third component to the inner binary is sufficiently large with a sufficiently short period that {\it Gaia} will be able to detect it in the full astrometric data release. The diameter of this orbit is around $0.3$~mas as seen in Figure~\ref{fig:V4_orb}. It is also interesting that given an inclination of around $90^\circ$ (Table~\ref{tab:V4_tess}) for the third body it could potentially at some point eclipse the stars in the inner binary. Observing this would be extremely valuable as this would yield the radius of this component, but it would also allow for a tighter constraint of the radii of the inner binary and ultimately the cluster parameters.

\subsection{V2032}\label{sec:disc.V2032}

For V2032 more photometry of the system would really help solidify the measurements of the radii, especially observations around ingress of the primary eclipse with pre-ingress well covered could make a significant improvement. What is perhaps even more interesting to investigate is the potential secondary eclipse seen in Figure~\ref{fig:tesslc}. We have already assessed that this decrease in flux can not be assigned to statistical fluctuations and the timing of the signal is striking. The signal is, of course, dependent on the aperture mask chosen and diminishes with certain choices, however, this signal seems to persistently follow the primary eclipse. As our current models and data suggest that the secondary eclipse should not be visible, it would be extremely interesting to observe this system around $T_0^{\rm s}$ with an instrument with a better resolution to see whether a secondary eclipse can be detected. This could alter the inclination somewhat, which in turn would affect our radii, but it should not have major implications for the masses and thus would not change the cluster parameters drastically.

Given the strong constraints presented in this paper on the cluster metallicity, membership, distance and precision masses and radii for three eclipsing systems an obvious next step would be to explore the model parameters in greater detail, i.e., calculate models which include alpha enhancement and has a finer grid in the overshoot parameter, which could potentially be stronger constrained in this way.

\subsection{Future TESS observations of NGC~2506}\label{sec:disc.TESS}

In the extended TESS mission the plan is for the spacecraft to revisit many of the already observed sectors and NGC~2506 should be observed again in TESS' Sector 34\footnote{\url{https://heasarc.gsfc.nasa.gov/cgi-bin/tess/webtess/wtv.py}} (primo 2021). This is extremely interesting for several reasons; firstly, we would acquire even more ephemerides for the V4 system, and might be able to place tighter constraints on the third body and we might be able to see if the potential secondary eclipse in V2032 persists (although as we have discussed we would probably require validation from instruments with higher spatial resolution). Secondly, the cadence of the FFIs in the extended mission will be changed from 30~min to 10~min, which could be of major importance for the detectability of solar-like oscillations further down the RGB (again the stars might be too faint), but a finer sampling will also aid in capturing the shape of the eclipses for the binaries. In addition a new 20~s cadence mode will be opened for selected targets (as opposed to the current 2~min cadence mode).



\section{Conclusions}\label{sec:conc}

In this paper we presented spectroscopic and photometric data of three detached eclipsing binaries -- V2032, V4, and V5 -- as well as spectroscopic data of four RGB stars; RGB231, RGB433, RGB913, and RGB2358. All of these stars are members of the open cluster NGC~2506 and we used the parameters derived from the data to determine the age and metallicity of the cluster. The spectrosopic data of the RGB stars allowed us to determine the metallicity of the cluster to be $\mathrm{[Fe/H]}=-0.36 \pm 0.10$~dex with $[\alpha/\mathrm{Fe}]=0.10 \pm 0.10$~dex. A value we used with our results for the masses and radii of the binaries to determine the age of the cluster to be $t = 2.01 \pm 0.10$~Gyr when we compared these results to the BaSTI isochrones. To properly model the cluster it is necessary to use models which include convective core-overshooting, although the value for the overshooting parameter of 0.2 available in the grid we used seems to be a bit too large. It should thus be possible to really quantify the value for the overshooting parameter in NGC~2506 using models specifically tailored to this cluster.

We found these values to be consistent with what is observed in the CMD of the cluster, which we have cleaned to only contain cluster members using {\it Gaia} DR2 data and additional spectroscopic observations. We find a very nice agreement between the distance to the cluster determined by {\it Gaia} and the distance we get from calculating the luminosity of the binaries V2032 and V4. We therefore conclude that the distance to the cluster is $r = 3.101 \pm 0.017$~kpc. Using the effective temperature of the RGB stars, we estimated the colour excess of the cluster to be E$(b-y)=0.057 \pm 0.004$~mag, which is in good agreement with the values required to fit the model isochrones to the observed sequence.

We furthermore report on the possible detection of solar-like oscillations in two of the most luminous members of the cluster using data from TESS. Namely, the RGB stars we have dubbed RGB526 and RGB383, with the latter showing quite prominent features in the power spectra in Figure~\ref{fig:osc} around the expected $\nu_\mathrm{max}$. If this detection is confirmed, it would to our knowledge be the first detection of solar-like oscillations in an open cluster detected by TESS. Much more prominent oscillations are seen in the power spectra of the $\delta$~Scuti stars (Figure~\ref{fig:dsstars}) and for some of the $\gamma$~Dor stars (Figure~\ref{fig:gdstars}).

%
\section*{Acknowledgements}
We thank the anonymous referee for useful comments and suggestions that helped improve the manuscript.
Funding for the Stellar Astrophysics Centre is provided by The Danish National Research Foundation (Grant agreement no.: DNRF106).
ELS gratefully acknowledges support from the (U.S.) National Science Foundation under grant AST 1817217.
This work has made use of data from the European Space Agency (ESA) mission {\it Gaia} (\url{https://www.cosmos.esa.int/gaia}), processed by the {\it Gaia} Data Processing and Analysis Consortium (DPAC, \url{https://www.cosmos.esa.int/web/gaia/dpac/consortium}). Funding for the DPAC has been provided by national institutions, in particular the institutions participating in the {\it Gaia} Multilateral Agreement.
This research has made use of the VizieR catalogue access tool, CDS, Strasbourg, France. The original description of the VizieR service was published in A\&AS 143, 23.
This paper includes data collected with the TESS mission, obtained from the MAST data archive at the Space Telescope Science Institute (STScI). Funding for the TESS mission is provided by the NASA Explorer Program. STScI is operated by the Association of Universities for Research in Astronomy, Inc., under NASA contract NAS 5$-$26555.
This research made use of Lightkurve, a \python package for {\it Kepler} and TESS data analysis \citep{pack:lk}.
This research made use of Astropy,\footnote{http://www.astropy.org} a community-developed core Python package for Astronomy \citep{astropy:2013, astropy:2018}.
FIEStool makes use of the packages NumPy \citep{art:vanderWalt2011} and PyFits \citep[now the Astropy fits sub-package;][]{astropy:2013} and to perform order tracing, extraction and wavelength calibration IRAF tasks from the echelle package \citep{art:tody1986} are used.
Based on observations made with the Nordic Optical Telescope, operated by the Nordic Optical Telescope Scientific Association at the Observatorio del Roque de los Muchachos, La Palma, Spain, of the Instituto de Astrofisica de Canarias.
\\
{\it Facilities}: TESS, {\it Gaia}, NOT, LCOGT, VLT, IAC-80, Danish 1.54-metre, Mercator.
\\
{\it Software}: Astroquery \citep{astroquery}, TESScut \citep[i.e., Astrocut;][]{tesscut}, celerite \citep{celerite}, REBOUND \url{http://github.com/hannorein/rebound}, FIEStool \url{http://www.not.iac.es/instruments/fies/fiestool/}, SciPy \citep{prog:scipy}.

\section*{Data availability}
Some of the data underlying this article are available in the ESO Data Portal at \url{http://archive.eso.org/cms/data-portal.html}, the NOT data server at \url{http://www.not.iac.es/archive/}, and the LCO Science Archive at \url{https://archive.lco.global/}.


\bibliographystyle{mnras}
\bibliography{biblio,library} 

\appendix

\section{Figures and tables}
\begin{table}
    \centering
    \caption{Extracted frequency of maximum power and the number of cycles per day at this frequency for the $\delta$~Scuti and $\gamma$~Dor stars reported in \citet{art:arentoft2007}. IDs refer to the labels therein.}
\begin{threeparttable}
    \begin{tabular}{c c c c}
         \toprule
         Type & ID & $\nu _\mathrm{max}$ ($\mu \mathrm{Hz}$) & Frequency (c/d) \\
         \midrule
         \multirow{6}{*}{$\delta$~Scuti} & V1 & 157.9 & 13.6 \\
          & V2 & 125.4 & 10.8 \\
          & V3 & 142.0 & 12.3 \\
          & V6 & 124.5 & 10.8 \\
          & V7 & 122.2 & 10.6 \\
          & V8 & 124.7 & 10.8 \\
          \hdashline
          \multirow{8}{*}{$\gamma$~Dor} & V11 & 14.4 & 1.2 \\
          & V12\textdagger & 3.3 & 0.3 \\
          & V13\textdagger & 4.3 & 0.4 \\
          & V14\textdagger & 4.3 & 0.4 \\
          & V15\textdagger & 8.1 & 0.7 \\
          & V16\textdagger & 9.1 & 0.8 \\
          & V17 & 3.7 & 0.3 \\
          & V18\textdagger & 4.3 & 0.4 \\
          & V19\textdagger & 6.4 & 0.6 \\
          & V21 & 6.9 & 0.6 \\
          & V22 & 4.8 & 0.4 \\
          & V23 & 14.4 & 1.2 \\
          & V24 & 10.7 & 0.9 \\
          & V25 & 13.9 & 1.2 \\
         \bottomrule
    \end{tabular}
		\begin{tablenotes}
        	\item[\textdagger] Very blended signal/minor detection.
		\end{tablenotes}
\end{threeparttable}
    \label{tab:pulse}
\end{table}


\begin{figure*}
    \centering
    \includegraphics[width=\textwidth]{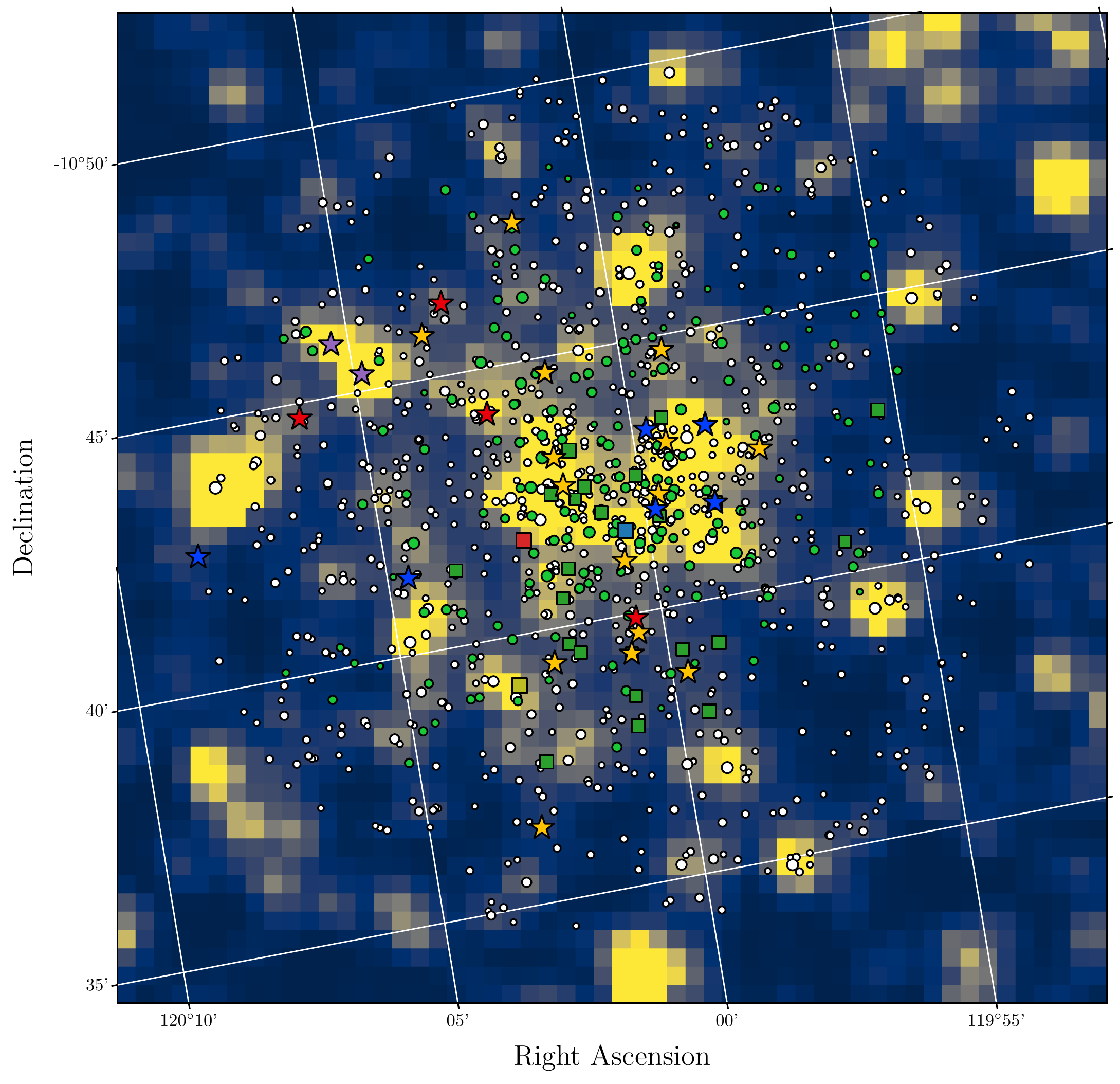}
    \caption{NGC~2506 as seen in the FFIs of TESS. As in Figure~\ref{fig:cmd} yellow and blue stars denote the $\gamma$~Dor stars and BSs/$\delta$~Scuti stars, respectively. V2032, V4, and V5 are respectively marked with blue, red, and yellow squares. Red and purple stars mark the position of the RGB stars for which we have respectively performed a spectroscopic analysis or possibly detected solar-like oscillations. Again the green squares and dots denote the binaries and single members, respectively, but this time they have been scaled according to their magnitude (the brighter the bigger). The white dots are {\it Gaia} sources brighter than $G<17$ within a $0.13^{\circ}$ radius of the cluster center -- these have also been scaled according to their magnitude.}
    \label{fig:tessffi}
\end{figure*}


\clearpage
\captionsetup[longtable]{labelfont=bf}
{\onecolumn

\begin{center}

    \begin{longtable}{c c c c c c c c c c c c c}
    \caption{Table containing the results of our spectroscopic membership determination in Section~\ref{sec:rv_members}. The naming is the WEBDA identification number found from cross-matching with \citet{art:twarog2016}, except for names starting with K, where we could not find a match and they therefore refer to their index in our Str\"omgren photometry. The last column is firstly the probability for membership based on \gaia proper motions and parallaxes derived by \citet{art:cantat2019} and secondly the membership class (M - probable RV member; NM - non-member; MB - probable binary member; MN - RV member with proper motion membership below 50\%; B - probable binary for which the radial velocity deviates significantly from the mean; BNM - binary with deviant RV, but proper motion probability below 50\%) determined in \citet{art:anthony-twarog2018} -- a dash denotes that we could not find a class. This table is available online with magnitudes for all Str\"omgren filters ($uvby$) with associated uncertainties.}
    \label{tab:rv_members}\\
    \toprule
    Name & $\alpha_{2000}$ &  $\delta_{2000}$ & $y$ & $b-y$ & $\langle v_{\rm rad} \rangle$ & sd$(v_{\rm rad})$ & $\sigma (v_{\rm rad})$ & Variable & Member & Probability \\
    & & & (mag) & (mag) & (km/s) & (km/s) & (km/s) & & \\
    \midrule
    \endfirsthead
    \multicolumn{11}{c}%
    {\tablename\ \thetable\ -- \textit{Continued from previous page}} \\
    \bottomrule
    Name & $\alpha_{2000}$ &  $\delta_{2000}$ & $y$ & $b-y$ & $\langle v_{\rm rad} \rangle$ & sd$(v_{\rm rad})$ & $\sigma (v_{\rm rad})$ & Variable & Member & Probability \\
    & & & (mag) & (mag) & (km/s) & (km/s) & (km/s) & & \\
    \bottomrule
    \endhead
    \bottomrule \multicolumn{11}{r}{\textit{Continued on next page}} \\
    \endfoot
    \bottomrule
    \multicolumn{11}{l}{$^{\chi}$ Excluded.} \\
    \endlastfoot
K1031 & 08 00 09.0 & -10 47 41.4 & 15.466 & 0.279 & 85.803 & 0.621 & 15.760 & 0 & 1 & 0.0/-- \\ 
3318 & 07 59 56.4 & -10 50 32.4 & 15.196 & 0.255 & 81.301 & 2.387 & 36.610 & 0 & 1 & 1.0/MN \\ 
2249 & 08 00 07.3 & -10 47 34.5 & 14.949 & 0.291 & 79.625 & 6.299 & 65.115 & 0 & 1 & 1.0/M \\ 
2215 & 08 00 10.9 & -10 46 27.8 & 15.386 & 0.262 & 83.843 & 2.131 & 31.293 & 0 & 1 & 1.0/MN \\ 
3329 & 07 59 56.5 & -10 49 55.6 & 15.396 & 0.268 & 82.459 & 6.388 & 55.785 & 1 & 1 & 1.0/-- \\ 
7054 & 07 59 47.9 & -10 52 02.4 & 14.161 & 0.594 & 19.111 & 0.131 & 11.502 & 0 & 0 & 0.0/NM \\ 
K1238 & 08 00 24.1 & -10 43 04.6 & 15.398 & 0.272 & 44.952 & 0.807 & 20.219 & 0 & 0 & 0.0/-- \\ 
K1245 & 08 00 16.5 & -10 44 55.6 & 14.930 & 0.296 & 87.018 & 2.833 & 46.335 & 0 & 1 & 1.0/-- \\ 
5011 & 08 00 10.9 & -10 46 13.8 & 15.013 & 0.397 & 49.272 & 0.139 & 10.462 & 0 & 0 & 0.0/-- \\ 
1268 & 08 00 13.6 & -10 45 32.7 & 15.184 & 0.280 & 95.125 & 11.285 & 12.900 & 1 & 1 & 1.0/-- \\ 
3328 & 07 59 55.2 & -10 50 00.1 & 14.982 & 0.272 & 80.008 & 4.632 & 60.865 & 0 & 1 & 1.0/M \\ 
2108 & 08 00 07.8 & -10 46 46.0 & 15.398 & 0.260 & 75.536 & 4.403 & 58.378 & 0 & 1 & 1.0/-- \\ 
K1315 & 08 00 22.2 & -10 43 12.6 & 14.409 & 0.445 & -5.119 & 0.128 & 12.061 & 0 & 0 & 0.0/-- \\ 
K1320 & 08 00 05.9 & -10 47 13.5 & 11.708 & 0.693 & 82.061 & 0.061 & 11.697 & 0 & 1 & 1.0/-- \\ 
2210 & 08 00 09.3 & -10 46 17.1 & 15.477 & 0.259 & 82.824 & 3.684 & 36.616 & 1 & 1 & 1.0/M \\ 
3217$^{\chi}$ & 07 59 58.5 & -10 48 57.2 & 18.767 & 0.594 & 87.266 & 189.361 & 35.554 & 1 & 1 & 1.0/M \\ 
1379 & 08 00 14.8 & -10 44 53.5 & 15.158 & 0.259 & 84.935 & 1.172 & 27.310 & 0 & 1 & 1.0/M \\ 
3206 & 08 00 01.4 & -10 48 11.8 & 14.847 & 0.298 & 92.171 & 6.814 & 51.763 & 1 & 1 & 1.0/M \\ 
K1390 & 07 59 48.0 & -10 51 21.0 & 15.097 & 0.397 & 21.480 & 0.165 & 10.950 & 0 & 0 & 0.0/-- \\ 
1241 & 08 00 13.7 & -10 44 46.2 & 14.835 & 0.283 & 86.131 & 1.254 & 28.857 & 0 & 1 & 1.0/-- \\ 
K1451 & 08 00 05.8 & -10 46 43.1 & 11.549 & 0.380 & 40.829 & 0.086 & 11.833 & 0 & 0 & 0.0/-- \\ 
2102 & 08 00 07.3 & -10 46 13.9 & 14.948 & 0.284 & 81.629 & 1.281 & 44.017 & 0 & 1 & 1.0/M \\ 
K1512 & 07 59 46.6 & -10 51 12.6 & 15.114 & 0.254 & 79.395 & 5.602 & 63.261 & 0 & 1 & 1.0/-- \\ 
1359 & 08 00 17.0 & -10 43 39.1 & 15.373 & 0.265 & 82.622 & 4.005 & 35.139 & 1 & 1 & 1.0/MB \\ 
K153 & 08 00 21.6 & -10 49 57.0 & 14.436 & 0.193 & 86.826 & 0.392 & 12.584 & 0 & 1 & 1.0/-- \\ 
K1536 & 08 00 20.2 & -10 42 48.0 & 13.195 & 0.593 & 75.117 & 2.481 & 11.453 & 1 & 1 & 1.0/-- \\ 
K1632 & 08 00 00.8 & -10 47 12.0 & 13.916 & 0.500 & 83.612 & 0.140 & 10.683 & 0 & 1 & 1.0/-- \\ 
3231 & 07 59 55.9 & -10 48 21.4 & 13.117 & 0.594 & 85.198 & 0.077 & 11.241 & 0 & 1 & 1.0/M \\ 
K1669 & 08 00 18.0 & -10 42 49.9 & 15.446 & 0.261 & 85.079 & 0.935 & 17.363 & 0 & 1 & 1.0/-- \\ 
5104 & 08 00 06.0 & -10 45 43.2 & 15.123 & 0.260 & 85.155 & 1.981 & 33.195 & 0 & 1 & 1.0/-- \\ 
1235 & 08 00 11.2 & -10 44 25.1 & 15.349 & 0.257 & 77.219 & 7.786 & 62.367 & 1 & 1 & 1.0/M \\ 
K1709 & 08 00 19.4 & -10 42 19.4 & 16.213 & 0.184 & 84.863 & 0.544 & 18.763 & 0 & 1 & 1.0/-- \\ 
K1733 & 08 00 15.8 & -10 43 09.8 & 15.398 & 0.384 & 39.552 & 0.234 & 10.465 & 0 & 0 & 0.0/-- \\ 
K1836 & 07 59 56.6 & -10 47 29.8 & 14.938 & 0.278 & 59.229 & 18.444 & 23.061 & 1 & 1 & 1.0/-- \\ 
3134 & 07 59 58.5 & -10 47 01.0 & 15.065 & 0.503 & -8.995 & 0.121 & 11.751 & 0 & 0 & 0.0/NM \\ 
1112 & 08 00 03.3 & -10 45 44.1 & 12.968 & 0.602 & 83.940 & 0.087 & 11.518 & 0 & 1 & 1.0/M \\ 
1354 & 08 00 13.4 & -10 43 13.4 & 15.353 & 0.253 & 85.046 & 0.626 & 12.685 & 0 & 1 & 1.0/M \\ 
2402 & 08 00 20.1 & -10 49 59.5 & 12.422 & 0.722 & 85.513 & 0.063 & 11.618 & 0 & 1 & 1.0/M \\ 
3111 & 08 00 00.1 & -10 46 23.2 & 14.641 & 0.311 & 86.334 & 0.408 & 20.162 & 0 & 1 & 0.0/-- \\ 
3152 & 07 59 59.2 & -10 46 34.0 & 14.928 & 0.296 & 90.102 & 9.246 & 26.304 & 1 & 1 & 1.0/-- \\ 
K1910 & 07 59 55.0 & -10 47 34.0 & 13.790 & 0.050 & 91.556 & 6.831 & 30.575 & 1 & 1 & 1.0/-- \\ 
7052 & 07 59 45.0 & -10 49 53.9 & 14.724 & 0.274 & 91.844 & 3.656 & 49.298 & 0 & 1 & 1.0/B \\ 
1224 & 08 00 06.0 & -10 44 40.1 & 14.911 & 0.293 & 81.071 & 5.934 & 17.585 & 1 & 1 & 0.0/-- \\ 
3143 & 07 59 56.5 & -10 46 46.7 & 15.039 & 0.284 & 84.127 & 2.500 & 37.509 & 0 & 1 & 1.0/-- \\ 
K2019 & 08 00 10.3 & -10 43 17.1 & 17.430 & 0.456 & 90.204 & 63.671 & 13.558 & 1 & 1 & 0.0/-- \\ 
4132 & 08 00 00.6 & -10 45 38.1 & 13.719 & 0.290 & 83.045 & 1.943 & 21.076 & 0 & 1 & 1.0/-- \\ 
3243$^{\chi}$ & 07 59 51.1 & -10 47 54.2 & 19.917 & 0.699 & 83.845 & 0.148 & 11.249 & 0 & 1 & 1.0/MN \\ 
K2071 & 08 00 10.6 & -10 43 01.0 & 15.411 & 0.271 & 84.416 & 0.908 & 15.267 & 0 & 1 & 1.0/-- \\ 
7082 & 08 00 09.8 & -10 52 18.1 & 14.908 & 0.445 & 83.922 & 0.153 & 12.475 & 0 & 1 & 1.0/M \\ 
K2138 & 08 00 03.0 & -10 44 33.3 & 15.212 & 0.270 & 82.741 & 2.802 & 35.608 & 0 & 1 & 1.0/-- \\ 
4109 & 07 59 56.8 & -10 46 04.3 & 13.773 & 0.549 & 84.028 & 0.081 & 11.064 & 0 & 1 & 1.0/M \\ 
4118 & 07 59 57.9 & -10 45 45.9 & 14.995 & 0.275 & 88.692 & 8.940 & 25.665 & 1 & 1 & 1.0/-- \\ 
1343 & 08 00 11.7 & -10 42 13.6 & 13.232 & 0.617 & 23.945 & 0.056 & 11.324 & 0 & 0 & 0.0/NM \\ 
3378 & 07 59 46.6 & -10 48 18.1 & 15.217 & 0.261 & 81.244 & 4.471 & 52.780 & 0 & 1 & 1.0/MB \\ 
1214 & 08 00 05.4 & -10 43 38.9 & 15.289 & 0.268 & 27.427 & 96.024 & 55.791 & 1 & 1 & 1.0/-- \\ 
4127 & 07 59 57.9 & -10 45 29.1 & 15.209 & 0.261 & 82.302 & 1.921 & 27.446 & 0 & 1 & 1.0/-- \\ 
K2216 & 08 00 10.2 & -10 42 25.6 & 17.464 & 0.344 & 25.590 & 0.222 & 10.128 & 0 & 0 & 0.0/-- \\ 
3367 & 07 59 47.4 & -10 48 00.5 & 15.341 & 0.254 & 84.244 & 1.783 & 24.978 & 0 & 1 & 1.0/M \\ 
K2309 & 07 59 38.8 & -10 49 48.1 & 15.108 & 0.259 & 86.216 & 0.570 & 17.089 & 0 & 1 & 1.0/-- \\ 
3392 & 07 59 48.1 & -10 47 15.2 & 13.139 & 0.575 & 82.968 & 0.080 & 11.126 & 0 & 1 & 1.0/MN \\ 
3260 & 07 59 49.8 & -10 46 49.9 & 15.006 & 0.359 & 85.554 & 0.147 & 10.617 & 0 & 1 & 0.0/-- \\ 
K2352 & 08 00 02.4 & -10 43 43.2 & 15.390 & 0.630 & 1.936 & 0.207 & 14.573 & 0 & 0 & 0.0/-- \\ 
7042 & 07 59 39.7 & -10 49 14.9 & 15.032 & 0.270 & 84.500 & 1.122 & 26.144 & 0 & 1 & 1.0/M \\ 
1328 & 08 00 09.2 & -10 41 50.7 & 14.715 & 0.276 & 104.159 & 8.098 & 18.561 & 1 & 1 & 1.0/NM \\ 
4272 & 08 00 00.8 & -10 43 47.0 & 15.138 & 0.264 & 81.931 & 3.660 & 42.904 & 0 & 1 & 1.0/-- \\ 
4230 & 07 59 55.4 & -10 45 03.4 & 12.841 & 0.401 & 58.770 & 0.104 & 11.264 & 0 & 0 & 0.0/-- \\ 
7047 & 07 59 41.0 & -10 48 36.6 & 15.363 & 0.231 & 83.879 & 4.018 & 26.236 & 1 & 1 & 1.0/M \\ 
4223 & 07 59 53.7 & -10 45 25.4 & 14.710 & 0.429 & 87.741 & 0.267 & 15.753 & 0 & 1 & 1.0/M \\ 
4241 & 07 59 56.3 & -10 44 46.6 & 15.044 & 0.281 & 75.976 & 5.716 & 63.394 & 0 & 1 & 1.0/M \\ 
K2449$^{\chi}$ & 08 00 01.6 & -10 43 26.5 & 15.424 & 0.281 & -126.008 & 7.806 & 11.009 & 1 & 0 & 0.0/-- \\ 
4228 & 07 59 54.4 & -10 45 10.9 & 11.986 & 1.059 & 109.973 & 0.281 & 11.655 & 0 & 0 & 0.0/NM \\ 
1301 & 08 00 02.5 & -10 42 48.1 & 14.605 & 0.564 & 84.355 & 0.096 & 11.166 & 0 & 1 & 1.0/M \\ 
4254 & 07 59 56.1 & -10 44 12.6 & 15.021 & 0.266 & 85.186 & 3.746 & 62.959 & 0 & 1 & 1.0/M \\ 
1302 & 08 00 01.8 & -10 42 40.6 & 15.389 & 0.242 & 84.202 & 5.655 & 11.984 & 1 & 1 & 1.0/M \\ 
7038 & 07 59 38.6 & -10 48 13.0 & 14.792 & 0.276 & 85.636 & 1.802 & 41.465 & 0 & 1 & 1.0/M \\ 
K2640 & 07 59 43.6 & -10 46 56.9 & 15.179 & 0.442 & 69.796 & 0.218 & 11.396 & 0 & 1 & 0.0/-- \\ 
K2663 & 08 00 03.8 & -10 41 50.7 & 15.002 & 0.321 & 84.281 & 2.657 & 54.428 & 0 & 1 & 1.0/-- \\ 
5343 & 07 59 52.8 & -10 44 33.2 & 15.308 & 0.434 & 31.623 & 0.275 & 10.493 & 0 & 0 & 0.0/-- \\ 
K2688 & 07 59 59.0 & -10 42 56.7 & 14.815 & 0.314 & 83.298 & 0.262 & 11.190 & 0 & 1 & 1.0/-- \\ 
1305 & 08 00 02.0 & -10 42 08.6 & 14.290 & 0.227 & 81.751 & 8.572 & 19.922 & 1 & 1 & 1.0/M \\ 
4237 & 07 59 51.8 & -10 44 36.7 & 15.320 & 0.254 & 82.097 & 2.185 & 43.207 & 0 & 1 & 1.0/MN \\ 
4262 & 07 59 55.0 & -10 43 19.2 & 14.413 & 0.345 & 52.569 & 11.840 & 15.966 & 1 & 0 & 0.0/BNM \\ 
7044 & 07 59 40.5 & -10 46 50.2 & 14.652 & 0.399 & 88.431 & 18.382 & 11.416 & 1 & 1 & 1.0/NM \\ 
4353 & 07 59 50.9 & -10 43 56.8 & 15.066 & 0.236 & 83.393 & 2.705 & 44.461 & 0 & 1 & 1.0/M \\ 
4374 & 07 59 57.6 & -10 42 13.0 & 14.713 & 0.324 & 62.236 & 0.160 & 11.459 & 0 & 0 & 0.0/NM \\ 
4331 & 07 59 46.9 & -10 44 36.1 & 15.155 & 0.264 & 83.934 & 5.010 & 44.997 & 1 & 1 & 1.0/M \\ 
4372 & 07 59 56.6 & -10 42 08.4 & 14.592 & 0.295 & 68.051 & 9.391 & 28.923 & 1 & 1 & 1.0/M \\ 
K2944 & 07 59 55.2 & -10 42 03.7 & 15.315 & 0.506 & 80.424 & 24.500 & 11.894 & 1 & 1 & 0.0/-- \\ 
K2956 & 07 59 44.5 & -10 44 38.1 & 15.500 & 0.350 & 60.365 & 12.846 & 13.775 & 1 & 1 & 0.0/-- \\ 
4318 & 07 59 41.4 & -10 45 21.6 & 15.294 & 0.265 & 84.162 & 2.694 & 33.190 & 0 & 1 & 1.0/MN \\ 
4337 & 07 59 43.6 & -10 44 23.0 & 14.724 & 0.306 & 86.970 & 1.802 & 39.833 & 0 & 1 & 1.0/M \\ 
4338 & 07 59 44.2 & -10 44 09.7 & 14.832 & 0.306 & 82.653 & 2.703 & 53.593 & 0 & 1 & 1.0/M \\ 
K3043 & 07 59 46.4 & -10 43 34.3 & 13.546 & 0.344 & 69.162 & 0.164 & 10.653 & 0 & 0 & 0.0/-- \\ 
7078 & 08 00 05.3 & -10 52 39.8 & 15.820 & 0.362 & 5.055 & 0.050 & 11.625 & 0 & 0 & 0.0/NM \\ 
K342 & 08 00 18.2 & -10 49 21.2 & 11.077 & 0.975 & 84.848 & 0.089 & 12.073 & 0 & 1 & 1.0/-- \\ 
K368 & 08 00 16.8 & -10 49 32.1 & 14.934 & 0.267 & 85.224 & 1.151 & 32.701 & 0 & 1 & 1.0/-- \\ 
K402 & 08 00 26.9 & -10 46 48.3 & 15.448 & 0.326 & 33.934 & 0.167 & 10.401 & 0 & 0 & 0.0/-- \\ 
K418 & 08 00 25.0 & -10 47 06.3 & 15.326 & 0.345 & 58.074 & 0.456 & 11.366 & 0 & 0 & 0.0/-- \\ 
K423 & 08 00 06.5 & -10 51 36.4 & 14.864 & 0.360 & 35.570 & 6.310 & 10.641 & 1 & 0 & 0.0/-- \\ 
2371 & 08 00 13.3 & -10 49 48.6 & 15.299 & 0.263 & 85.771 & 1.067 & 18.975 & 0 & 1 & 1.0/M \\ 
K449 & 08 00 22.0 & -10 47 33.3 & 14.913 & 0.379 & 103.562 & 0.342 & 11.529 & 0 & 0 & 0.0/-- \\ 
7079$^{\chi}$ & 08 00 07.1 & -10 51 04.5 & 18.622 & 0.515 & 52.346 & 0.160 & 11.023 & 0 & 0 & 0.0/NM \\ 
2351 & 08 00 17.3 & -10 48 16.8 & 14.556 & 0.298 & 82.557 & 1.605 & 34.115 & 1 & 1 & 0.0/M \\ 
K573 & 08 00 05.4 & -10 50 59.5 & 19.955 & 0.851 & 81.673 & 3.774 & 59.412 & 0 & 1 & 1.0/-- \\ 
2324 & 08 00 19.7 & -10 47 03.4 & 15.164 & 0.253 & 85.594 & 4.805 & 44.490 & 0 & 1 & 1.0/-- \\ 
7073 & 08 00 01.0 & -10 51 34.1 & 14.607 & 0.324 & 49.189 & 0.253 & 12.272 & 0 & 0 & 0.0/NM \\ 
K682$^{\chi}$ & 07 59 55.7 & -10 52 46.8 & 11.079 & 0.872 & -8.433 & 80.066 & 8.426 & 1 & 0 & 0.0/-- \\ 
2387 & 08 00 07.8 & -10 49 41.5 & 15.076 & 0.549 & 88.058 & 2.265 & 12.842 & 1 & 1 & 0.0/NM \\ 
2363 & 08 00 11.8 & -10 48 35.1 & 15.333 & 0.290 & 79.896 & 2.649 & 39.888 & 0 & 1 & 1.0/-- \\ 
2401 & 08 00 05.4 & -10 50 07.4 & 13.193 & 0.590 & 83.542 & 0.073 & 11.294 & 0 & 1 & 1.0/M \\ 
2347 & 08 00 14.5 & -10 47 48.0 & 14.990 & 0.277 & 86.246 & 1.169 & 26.171 & 0 & 1 & 1.0/M \\ 
2405 & 08 00 03.5 & -10 50 22.1 & 15.167 & 0.268 & 80.085 & 3.320 & 56.752 & 0 & 1 & 1.0/M \\ 
K84 & 08 00 21.8 & -10 50 18.7 & 13.235 & 0.598 & 84.809 & 0.106 & 11.372 & 0 & 1 & 1.0/-- \\ 
2262 & 08 00 09.4 & -10 48 33.1 & 14.069 & 0.332 & 40.598 & 0.304 & 12.782 & 0 & 0 & 0.0/-- \\ 
3308 & 08 00 02.6 & -10 50 07.7 & 15.460 & 0.423 & 47.347 & 0.223 & 11.092 & 0 & 0 & 0.0/-- \\ 
7068 & 07 59 57.4 & -10 51 12.3 & 15.313 & 0.248 & 76.617 & 3.743 & 56.293 & 0 & 1 & 1.0/M \\ 
2276 & 08 00 05.9 & -10 49 03.4 & 14.890 & 0.520 & 73.402 & 1.685 & 11.059 & 1 & 1 & 1.0/MB \\ 
K928 & 08 00 19.9 & -10 45 31.2 & 15.481 & 0.543 & 41.945 & 0.110 & 11.186 & 0 & 0 & 0.0/-- \\ 
K93 & 08 00 24.6 & -10 49 33.8 & 13.692 & 0.366 & 7.049 & 0.084 & 10.611 & 0 & 0 & 0.0/-- \\ 
2311 & 08 00 16.4 & -10 46 11.0 & 13.079 & 0.609 & 83.735 & 0.079 & 11.296 & 0 & 1 & 1.0/M \\ 
K965$^{\chi}$ & 08 00 06.6 & -10 48 36.7 & 12.792 & 0.047 & 153.056 & 180.707 & 9.334 & 1 & 0 & 1.0/-- \\ 
7065 & 07 59 55.4 & -10 51 22.1 & 15.055 & 0.517 & 107.776 & 0.137 & 11.480 & 0 & 0 & 0.0/NM \\ 
K996 & 08 00 07.3 & -10 48 16.2 & 14.577 & 0.311 & 91.884 & 3.857 & 46.863 & 0 & 1 & 1.0/-- \\ 
\end{longtable}
\end{center}
}
\twocolumn

\begin{table*}
    \centering
    \caption{Results for V2032 (left of vertical dashed line) and V5 (right) resulting from an MCMC sampling of 20,000 steps with a burn-in of 10,000 for the different photometric data available. The parameter space was sampled using 100 walkers. The value is taken as the 50th percentile of the chain and the uncertainties are the 16th and 84th percentile.}
\begin{threeparttable}
    \begin{tabular}{cccc:c}
        \toprule
        & \multicolumn{3}{c:}{V2032} & V5 \\
         & $I$ & $V$ & TESS & $B$\\
         \midrule
         $K^\mathrm{p}$~(km/s) & $61.99^{+0.10}_{-0.09}$ & $62.01^{+0.10}_{-0.09}$ & $61.88^{+0.10}_{-0.09}$ & $71.96^{+0.18}_{-0.13}$ \\
         $K^\mathrm{s}$~(km/s) & $62.70 \pm 0.11$ & $62.71 \pm 0.11$ & $62.61 \pm 0.10$ & $96.18^{+0.12}_{-0.11}$ \\
         $\gamma_{\rm GIRAFFE}$~(km/s) & $83.05 \pm 0.04$ & $83.05 \pm 0.04$ & $83.04 \pm 0.04$ & $83.40^{+0.19}_{-0.13}$ \\
         $\gamma_{\rm FIES}$~(km/s) & $83.28 \pm 0.04$ & $83.29 \pm 0.04$ & $83.28 \pm 0.04$ & - \\
         $e$ & $0.5867 \pm 0.0010$ & $0.5868 \pm 0.0010$ & $0.5860 \pm 0.0010$ & $0.0016^{+0.0007}_{-0.0008}$\\
         $\omega$~($^\circ$) & $138.85 \pm 0.10$ & $138.84 \pm 0.10$ & $138.88 \pm 0.10$ & $109.9^{+0.3}_{-0.7}$ \\
         $P$~(days) & $27.86780 \pm 0.00015$ & $27.86788 \pm 0.00015$ & $27.86741 \pm 0.00016$ & $3.35852^{+0.00014}_{-0.00017}$ \\
         $T_\mathrm{peri}$~($\rm BJD-$2,450,000) & $7754.495 \pm 0.006$ & $7754.498 \pm 0.006$ & $7754.485^{+0.007}_{-0.006}$ & $3387.112^{+0.014}_{-0.025}$ \\ 
         $M^\mathrm{p}$~(M$_\odot$) & $1.522 \pm 0.004$ &  $1.519 \pm 0.004$ & $1.523 \pm 0.005$ & $0.945^{+0.004}_{-0.003}$ \\
         $M^\mathrm{s}$~(M$_\odot$) & $1.505 \pm 0.004$ & $1.501 \pm 0.004$ & $1.505 \pm 0.005$ & $0.707^{+0.004}_{-0.003}$ \\
         \hdashline
         $R^\mathrm{p}$~($\rm R_\odot$) & $3.11^{+0.04}_{-0.05}$ & $2.92 \pm 0.04$ & $3.19 \pm 0.11$ & $0.68 \pm 0.04$ \\
         $R^\mathrm{s}$~($\rm R_\odot$) & $2.44^{+0.08}_{-0.04}$ & $2.39^{+0.07}_{-0.05}$ & $2.50^{+0.10}_{-0.09}$ & $0.610^{+0.021}_{-0.016}$ \\
         $a$~($\rm R_\odot$)\tnote{$\chi$} & $56.01^{+0.15}_{-0.14}$ & $55.94^{+0.15}_{-0.14}$ & $56.05^{+0.15}_{-0.14}$ & $9.04^{+0.10}_{-0.05}$ \\
         $i$~($^\circ$)\tnote{$\chi$} & $83.47 \pm 0.10$ & $84.01^{+0.13}_{-0.17}$ & $83.0 \pm 0.3$ & $88.91^{+0.26}_{-0.17}$ \\
         $T_0^{\rm p}$~($\rm BJD-$2,450,000) & $7781.5157 \pm 0.0002$ & $7781.5173 \pm 0.0002$ & $7781.554^{+0.005}_{-0.006}$ & $3385.6608^{+0.0005}_{-0.0004}$ \\ \hdashline
         $T_\mathrm{eff}^\mathrm{p}$\tnote{\textdagger}~(K) & $6560^{+80}_{-70}$ & $6590^{+90}_{-80}$ & $6560^{+80}_{-70} \pm 70$ & $5690^{+140}_{-120}$ \\ 
         $T_\mathrm{eff}^\mathrm{s}$\tnote{\textdagger}~(K) & $7100 \pm 80$ & $7080 \pm 90$ & $7100 \pm 80$ & $4940^{+110}_{-60}$ \\
         $c_1^\mathrm{p}$\tnote{\textdaggerdbl} & $0.38^{+0.08}_{-0.09}$ & $0.24 \pm 0.09$ & $0.21 \pm 0.10$ & $0.49 \pm 0.09$ \\
         $c_2^\mathrm{p}$\tnote{\textdaggerdbl} & $0.10^{+0.10}_{-0.09}$ & $0.37 \pm 0.10$ & $0.32 \pm 0.10$ & $0.36^{+0.17}_{-0.13}$ \\
         $c_1^\mathrm{s}$\tnote{\textdaggerdbl} & $0.31 \pm 0.10$ & $0.30 \pm 0.10$ & $0.18^{+0.10}_{-0.09}$ & $0.38^{+0.10}_{-0.15}$ \\
         $c_2^\mathrm{s}$\tnote{\textdaggerdbl} & $0.13 \pm 0.10$ & $0.36 \pm 0.10$ & $0.33 \pm 0.10$ & $0.18^{+0.06}_{-0.08}$ \\
         $l^\mathrm{c}$\tnote{\textdaggerdbl} & - & - & $7.60 \pm 0.05$ & - \\
         \bottomrule
    \end{tabular}
		\begin{tablenotes}
		    \item[$\chi$] $a \sin i$ constrained by Equation~\ref{eq:asini}.
        	\item[\textdagger] Sampled using a Gaussian prior for V2032 and a uniform prior for V5: $\sigma(T_{\rm eff}) = 100$~K and $\mathcal{U}(4200 \mathrm{K}, 6200 \mathrm{K})$.
        	\item[\textdaggerdbl] Sampled using a Gaussian prior: $\sigma(c_i) = 0.1$ and $\sigma(l^\mathrm{c}) = 0.05$.
		\end{tablenotes}
\end{threeparttable}    
    \label{tab:V2032}
\end{table*}


\begin{table*}
    \centering
    \caption{10,000 Monte Carlo simulations for V4 using \jktebop.}
\begin{threeparttable}
    \begin{tabular}{cccc:cc}
         \toprule
         & \multicolumn{3}{c}{IAC-80, LCOGT, NOT} & \multicolumn{2}{c}{Danish 1.54-metre, Mercator} \\
         & $I$ & $V$ & $B$ & $I$ & $B$ \\
         \midrule
        $K^\mathrm{p}$~(km/s) & $96.0 \pm 1.0$ & $96.0 \pm 1.1$ & $96.1 \pm 1.1$ & $96.4 \pm 0.4$ & $96.5 \pm 1.1$ \\
        $K^\mathrm{s}$~(km/s) & $112.4 \pm 0.4$ & $112.5 \pm 0.4$ & $112.6 \pm 0.5$ & $113.8 \pm 0.5$ & $114.0 \pm 0.4$ \\
        $\gamma^\mathrm{p}$~(km/s) & $85.1 \pm 0.8$ & $85.1 \pm 0.8$ & $85.2 \pm 0.8$ & $85.4 \pm 0.7$ & $85.4 \pm 0.8$ \\
        $\gamma^\mathrm{s}$~(km/s) & $85.2 \pm 0.3$ & $85.2 \pm 0.3$ & $85.0 \pm 0.3$ & $85.1 \pm 0.3$ & $85.1 \pm 0.3$ \\
        $e$ & $0.182 \pm 0.003$ & $0.199 \pm 0.003$ & $0.182 \pm 0.004$ & $0.176 \pm 0.004$ & $0.182 \pm 0.004$ \\
        $\omega$~($^\circ$) & $281.9 \pm 0.3$ & $280.49 \pm 0.19$ & $281.5 \pm 0.3$ & $272.91 \pm 0.13$ & $272.64 \pm 0.14$ \\
        $P$~(days) & $2.8676350 \pm 0.0000013$ & $2.8676325 \pm 0.0000010$ & $2.8676383 \pm 0.0000016$ & $2.867632 \pm 0.000003$ & $2.867636 \pm 0.000003$ \\ 
        $M^\mathrm{p}$~(M$_\odot$) & $1.411 \pm 0.019$ & $1.430 \pm 0.019$ & $1.444 \pm 0.019$ & $1.490 \pm 0.019$ & $1.49 \pm 0.02$ \\
        $M^\mathrm{s}$~(M$_\odot$) & $1.23 \pm 0.03$ & $1.22 \pm 0.03$ & $1.23 \pm 0.03$ & $1.26 \pm 0.03$ & $1.26 \pm 0.03$ \\
        \hdashline
        $R^\mathrm{p}$~($R_\odot$) & $2.37 \pm 0.02$ & $2.30 \pm 0.02$ & $2.24 \pm 0.02$ & $2.36 \pm 0.02$ & $2.36 \pm 0.02$ \\
        $R^\mathrm{s}$~($R_\odot$) & $1.41 \pm 0.04$ & $1.48 \pm 0.03$ & $1.37 \pm 0.03$ & $1.43 \pm 0.03$ & $1.46 \pm 0.03$ \\
        $a$~($R_\odot$) & $11.78 \pm 0.07$ & $11.75 \pm 0.07$ & $11.79 \pm 0.07$ & $11.90 \pm 0.07$ & $11.90 \pm 0.07$ \\
        $i$~($^\circ$) & $80.25 \pm 0.12$ & $80.14 \pm 0.09$ & $80.56 \pm 0.12$ & $80.22 \pm 0.10$ & $80.20 \pm 0.11$ \\
        $T_0^{\rm p}$~($\rm BJD-$2,450,000) & $3396.2536 \pm 0.0019$ & $3396.2581 \pm 0.0014$ & $3396.250 \pm 0.002$ & $3396.2791 \pm 0.0006$ & $3396.2796 \pm 0.0007$ \\
        $J$ & $1.078 \pm 0.019$ & $1.003 \pm 0.014$ & $1.11 \pm 0.02$ & $1.120 \pm 0.018$ & $1.087 \pm 0.018$ \\ \hdashline
        $c_1^\mathrm{p}$\tnote{\textdagger} & $0.126$ & $0.2831$ & $0.4206$ & $0.126$ & $0.4206$ \\
        $c_2^\mathrm{p}$\tnote{\textdagger} & $0.379$ & $0.3802$ & $0.3399$ & $0.379$ & $0.3399$ \\
        $c_1^\mathrm{s}$\tnote{\textdagger} & $0.1273$ & $0.2743$ & $0.3969$ & $0.1273$ & $0.3969$ \\
        $c_2^\mathrm{s}$\tnote{\textdagger} & $0.3741$ & $0.3837$ & $0.3567$ & $0.3741$ & $0.3567$ \\
         \bottomrule
    \end{tabular}
		\begin{tablenotes}
        	\item[\textdagger] Fixed during fit.
		\end{tablenotes}
\end{threeparttable}  
    \label{tab:V4_johnson}
\end{table*}

\begin{table*}
    \centering
    \caption{Results from the DE-MCMC run of the V4 system in which all the photometric data are included simultaneously. Here we give the median and the upper/lower 1$\sigma$ result for a given parameter. $i_\mathrm{mutual}$ is the mutual inclination between the orbit of the third body and the binary orbit and $\Omega$ is the nodal angle.}
    \begin{tabular}{ccccc}
        \toprule
        & $I$ & $V$ & $B$ & TESS \\
        \midrule
        $M^\mathrm{p}$~(M$_\odot$) & \multicolumn{4}{c}{$1.478^{+0.006}_{-0.007}$}  \\
        $M^\mathrm{s}$~(M$_\odot$) & \multicolumn{4}{c}{$1.250 \pm 0.010$}  \\
        $K^\mathrm{p}$~(km/s) & \multicolumn{4}{c}{$96.3 \pm 0.4$}  \\
        $K^\mathrm{s}$~(km/s) & \multicolumn{4}{c}{$113.84 \pm 0.12$} \\
        $\gamma$~(km/s) & \multicolumn{4}{c}{$79.8 \pm 0.3$} \\
        $e$ & \multicolumn{4}{c}{$0.1891 \pm 0.0011$} \\
        $\omega$~($^\circ$) & \multicolumn{4}{c}{$272.62^{+0.09}_{-0.08}$} \\
        $P$~(days) & \multicolumn{4}{c}{$2.867623 \pm 0.000002$}  \\ \hdashline
        $R^\mathrm{p}$~($R_\odot$) & \multicolumn{4}{c}{$2.300^{+0.013}_{-0.014}$} \\
        $R^\mathrm{s}$~($R_\odot$) & \multicolumn{4}{c}{$1.534^{+0.019}_{-0.018}$} \\
        $a$~($R_\odot$) & \multicolumn{4}{c}{$11.87 \pm 0.02$} \\
        $i$~($^\circ$) & \multicolumn{4}{c}{$80.14 \pm 0.06$}  \\
        $T_0^{\rm p}$~($\rm BJD-$2,450,000) & \multicolumn{4}{c}{$3379.0738 \pm 0.0005$} \\
        $T_\mathrm{eff}^\mathrm{p}$~(K) & \multicolumn{4}{c}{$6690^{+140}_{-120}$} \\
        $T_\mathrm{eff}^\mathrm{s}/T_\mathrm{eff}^\mathrm{p}$ & \multicolumn{4}{c}{$1.0162 \pm 0.0019$} \\ \hdashline
        $c_1^{\rm p}$\tnote{\textdagger} & $0.235^{+0.004}_{-0.005}$ & $0.404^{+0.006}_{-0.004}$ & $0.548^{+0.005}_{-0.006}$ & $0.250^{+0.005}_{-0.006}$ \\
        $c_2^{\rm p}$\tnote{\textdagger} & $0.173^{+0.004}_{-0.007}$ & $0.242^{+0.005}_{-0.003}$ & $0.292 \pm 0.005$ &  $0.178^{+0.005}_{-0.006}$ \\
        $c_1^{\rm s}$\tnote{\textdagger} & $0.230^{+0.004}_{-0.003}$ & $0.414^{+0.003}_{-0.005}$ & $0.547^{+0.006}_{-0.005}$ & $0.251 \pm 0.005$ \\
        $c_2^{\rm p}$\tnote{\textdagger} & $0.174^{+0.004}_{-0.003}$ & $0.248^{+0.002}_{-0.004}$ & $0.292 \pm 0.005$ & $0.184 \pm 0.004$ \\
        $l^\mathrm{c}$ & \multicolumn{4}{c}{$0.8288 \pm 0.0015$} \\ \hdashline
        $M^\mathrm{t}$~(M$_\odot$) & \multicolumn{4}{c}{$0.74 \pm 0.03$} \\
        $R^\mathrm{t}$~(R$_\odot$) & \multicolumn{4}{c}{$0.68^{+0.03}_{-0.02}$} \\
        $P^\mathrm{t}$~(days) & \multicolumn{4}{c}{$443.4231^{+0.0017}_{-0.0022}$}  \\
        $a^\mathrm{t}$~($R_\odot$) & \multicolumn{4}{c}{$370.7 \pm 1.3$} \\
        $e^\mathrm{t}$ & \multicolumn{4}{c}{$0.512 \pm 0.014$} \\
        $\omega^\mathrm{t}$~($^\circ$) & \multicolumn{4}{c}{$221 \pm 3$} \\
        $i^\mathrm{t}$~($^\circ$) & \multicolumn{4}{c}{$89.59^{+0.03}_{-0.02}$}  \\
        $i_\mathrm{mutual}$~($^\circ$) & \multicolumn{4}{c}{$9.45^{+0.08}_{-0.07}$} \\
        $\Omega$~($^\circ$) & \multicolumn{4}{c}{$0.19 \pm 0.19$} \\
        $T_\mathrm{0}^\mathrm{t}$~($\rm BJD-$2,450,000) & \multicolumn{4}{c}{$3210 \pm 6$} \\
        $T_\mathrm{eff}^\mathrm{t}$~(K) & \multicolumn{4}{c}{$5500^{+200}_{-300}$} \\
        \bottomrule
    \end{tabular}
    
    \label{tab:V4_tess}
\end{table*}
\clearpage
{\onecolumn

\begin{figure}
    \centering
    \includegraphics[width=0.9\textwidth]{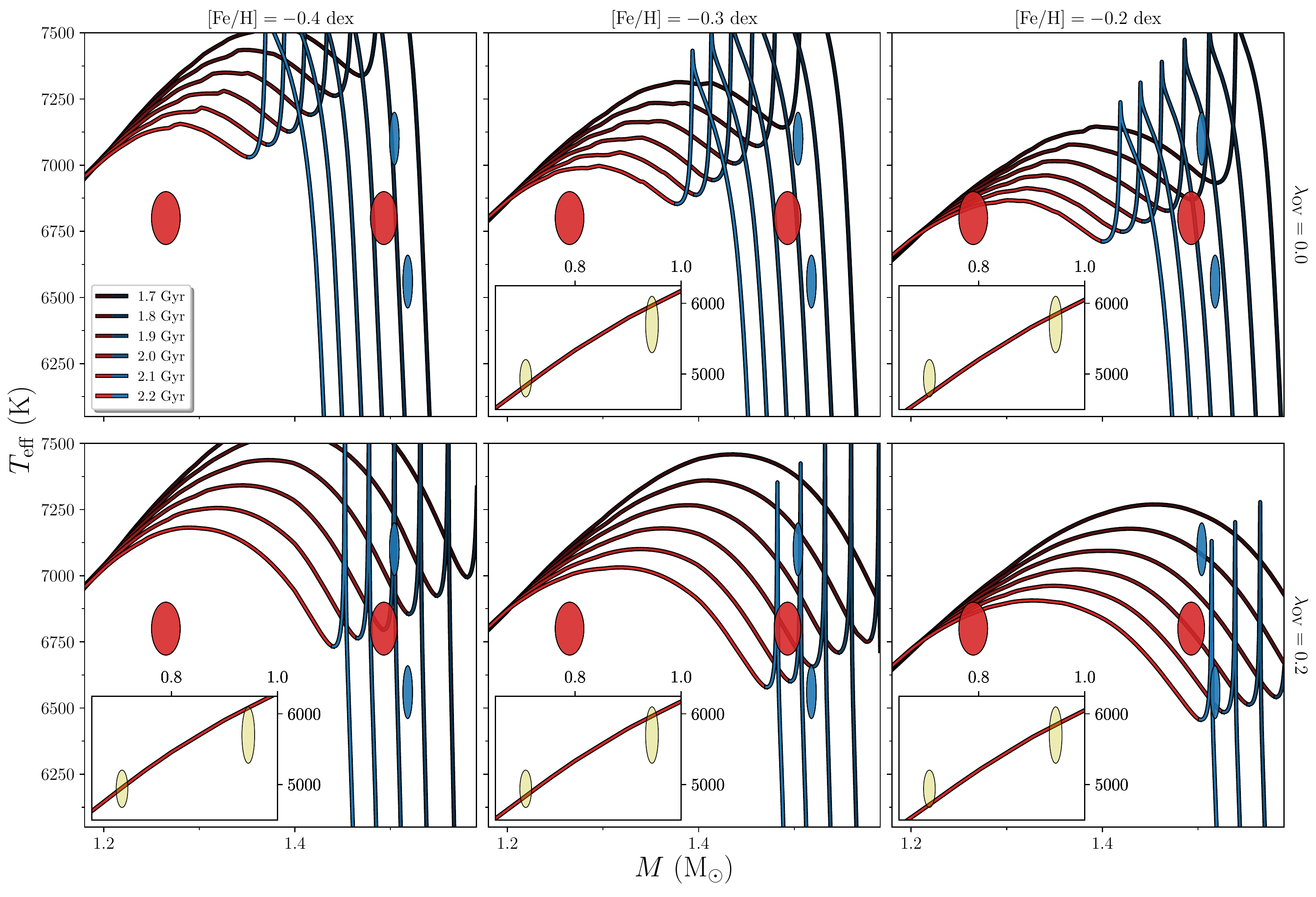}
    \caption{Mass-temperature diagram with the temperatures from Table~\ref{tab:key}, but otherwise the same as in Figure~\ref{fig:MR}.}
    \label{fig:app.MT}
\end{figure}
}


\begin{figure*}
    \centering
    \includegraphics[width=\textwidth]{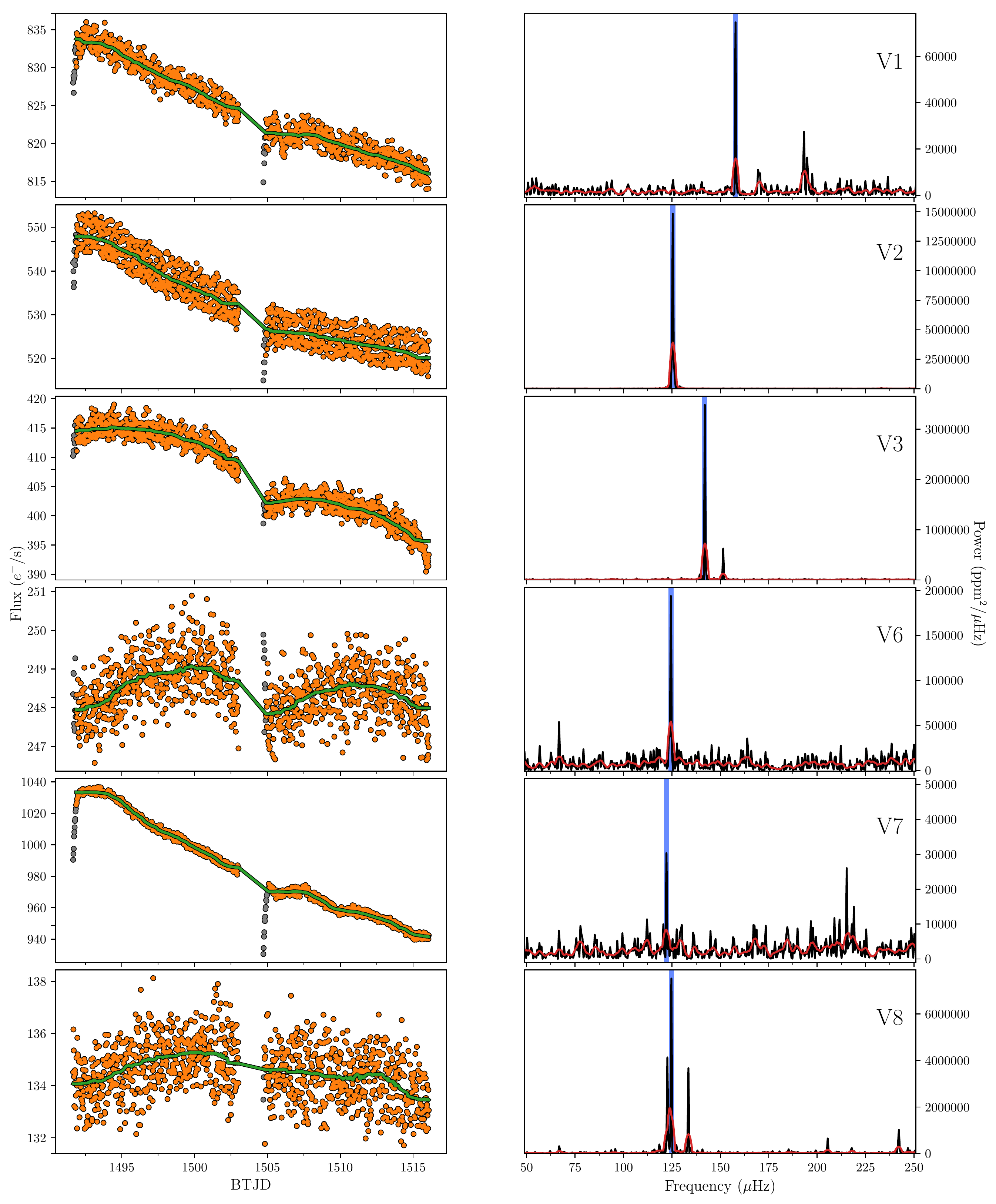}
    \caption{Light curves (left) and power spectra (right) for the $\delta$~Scuti stars in \citet{art:arentoft2007} created from the TESS FFIs. The grey points in the light curves show all the raw data points and orange points show the data used to create the power spectra. The green line is a running median used to normalize the data. The power spectra are plotted as black lines with a smoothed version in red. The vertical blue line marks the position for the frequency of maximum power.}
    \label{fig:dsstars}
\end{figure*}



\begin{figure*}
    \centering
    \includegraphics[width=\textwidth]{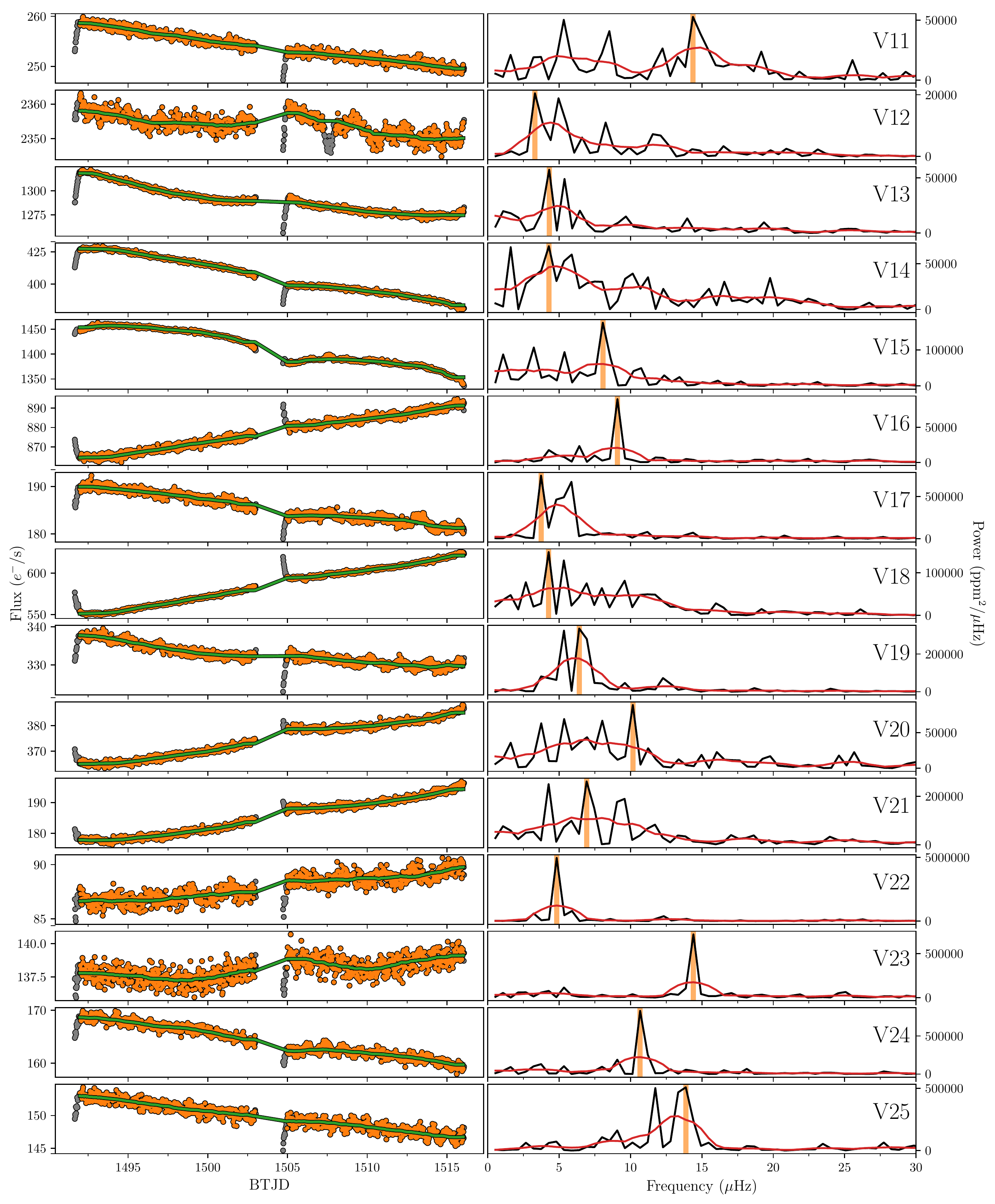}
    \caption{Light curves and power spectra for the $\gamma$~Dor stars in \citet{art:arentoft2007} created from the TESS FFIs. Here we have plotted the frequency of maximum power as an orange vertical line, otherwise colours mean the same as in Figure~\ref{fig:dsstars}.}
    \label{fig:gdstars}
\end{figure*}



\bsp	
\label{lastpage}
\end{document}